\documentclass[12pt]{article}
\usepackage[utf8]{inputenc}

\usepackage{times}

\RequirePackage{etex}

\usepackage{lineno}

\usepackage{pdfpages}
\usepackage{siunitx}

\usepackage{enumitem}

\usepackage{graphicx}
\usepackage[export]{adjustbox}
\usepackage{needspace}
\usepackage{tabularx}

\usepackage{wrapfig}
\usepackage{pdflscape}
\usepackage{threeparttable}
\usepackage[referable]{threeparttablex}

\usepackage{mathptmx}

\usepackage{amsmath}
\usepackage{amsfonts}
\usepackage{amssymb}

\usepackage{subcaption}
\usepackage{float}

\usepackage[labelfont = bf, justification = justified, singlelinecheck = false, font = footnotesize, figurename = FIGURE, tablename = TABLE]{caption}

\usepackage{bm}
\usepackage{color,soul}
\sethlcolor{lime}

\usepackage[titletoc,toc,title]{appendix}

\usepackage{setspace}

\usepackage{titlesec}

\titleformat*{\section}{\normalfont\bfseries}
\titleformat*{\subsection}{\normalfont\bfseries}
\titleformat*{\subsubsection}{\normalfont\itshape}

\usepackage{longtable}
\usepackage{multirow}
\usepackage{booktabs} 
\usepackage{afterpage}
\usepackage{relsize}
\usepackage{cprotect}
\usepackage{xcolor}

\newcounter{tablenote}[table]

\usepackage[margin=1.0in]{geometry}

\usepackage[round]{natbib} 

\usepackage{fancyhdr}

\usepackage{array}
\newcolumntype{P}[1]{>{\centering\arraybackslash}p{#1}}

\allowdisplaybreaks

\usepackage[colorlinks,citecolor=blue,urlcolor=blue,bookmarks=false,hypertexnames=true]{hyperref}

\newcommand{\ind}{\mathbf{1}}

\usepackage{lipsum}

\usepackage{autonum}

\begin{document}
    
    \onehalfspacing
    \frenchspacing
    
    {\Large\bf A robust mixed-effects quantile regression model using generalized Laplace mixtures to handle outliers and skewness}
    
    \vskip 0.0cm
    
    {\bf Divan A. Burger$^{1,2}$, Sean van der Merwe$^{2}$, and Emmanuel Lesaffre$^{3,4}$}
    {\it\small
        \vskip -0.7cm    
        \tabcolsep 0.0cm   
        \begin{threeparttable}
            \begin{longtable}{l p {\linewidth}}
                $^{1}$ & Syneos Health, Bloemfontein, Free State, South Africa \\
                $^{2}$ & Department of Mathematical Statistics and Actuarial Science, University of the Free State, Bloemfontein, South Africa \\
                $^{3}$ & I-BioStat, KU Leuven, Leuven, Belgium \\
                $^{4}$ & Department of Statistics and Actuarial Science, University of Stellenbosch, South Africa
            \end{longtable}
        \end{threeparttable}
    }
    
    \noindent\rule{\textwidth}{1pt}
    Address correspondence to Divan Burger, Syneos Health, Bloemfontein, South Africa 9301. Email: \verb|divanaburger@gmail.com|
    
    \newpage
    
    \section*{Abstract}
    
    Mixed-effects quantile regression models are widely used to capture heterogeneous responses in hierarchically structured data. The asymmetric Laplace (AL) distribution has traditionally served as the basis for quantile regression; however, its fixed skewness limits flexibility and renders it sensitive to outliers. In contrast, the generalized asymmetric Laplace (GAL) distribution enables more flexible modeling of skewness and heavy-tailed behavior, yet it remains vulnerable to extreme observations. In this paper, we extend the GAL distribution by introducing a contaminated GAL (cGAL) mixture model that incorporates a scale-inflated component to mitigate the impact of outliers without requiring explicit outlier identification or deletion. We apply this model within a Bayesian mixed-effects quantile regression framework to model HIV viral load decay over time. Our results demonstrate that the cGAL-based model more reliably captures the dynamics of HIV viral load decay, yielding more accurate parameter estimates compared to both AL and GAL approaches. Model diagnostics and comparison statistics confirm the cGAL model as the preferred choice. A simulation study further shows that the cGAL model is more robust to outliers than the GAL and exhibits favorable frequentist properties.

    \vskip -0.7cm
    
    \tabcolsep 0.0cm 
    \begin{threeparttable}
        \begin{longtable}{p {2.3cm} p {14.2cm}}
            {\bf Keywords}: & Bayesian; Generalized asymmetric Laplace distribution; Mixture distribution; Mixed-effects models; Robust quantile regression; Viral load
        \end{longtable}
    \end{threeparttable}
    
    \setcounter{table}{0}
    
    \newpage

    \section{Introduction}

    Mixed-effects quantile regression models have become increasingly important in statistical analyses because they account for the correlation structure inherent in repeated or clustered measurements, thus facilitating more accurate inference of the conditional distribution across various quantiles \citep{KoenkerHallock2001, GERACI2007}. Under a standard mixed-effects framework, this approach is well-suited to heteroscedastic settings, where variability can differ substantially across individuals or groups. In longitudinal clinical trials, for example, certain treatments may have a stronger effect on patients in the lower or upper tails of the response distribution than those near the mean. By capturing these diverse responses, quantile-based methods reveal treatment effects that mean-focused analyses might overlook. The asymmetric Laplace (AL) distribution has long been the favored choice for parametric mixed-effects quantile regression because its fixed skewness parameter directly targets the quantile of interest \citep{GERACI2007}.
    
    Despite its usefulness, the AL distribution in quantile regression has a key limitation: its skewness parameter is locked to a specific quantile. Rather than estimating skewness, one must pre-select it based on the quantile of interest. For instance, modeling the 25\textsuperscript{th} quantile (Q25) necessitates a right-skewed AL distribution, and this skewness grows more pronounced for lower quantiles, such as the 10\textsuperscript{th} percentile. Consequently, if the actual data shape deviates from the skewness imposed by the AL distribution, the model may simply provide a poor fit.
    
    To overcome the AL distribution's forced skewness constraint, \citet{yan2017new} proposed the generalized AL (GAL) distribution, which gives greater flexibility for asymmetric and heavy-tailed data. \citet{yu2023flexible} further extended this approach to mixed-effects models in a quantile regression framework. By adding an estimable shape parameter, the GAL distribution accommodates a wider range of skewness and kurtosis, making it more adaptable than the AL in many settings.
    
    However, while the GAL distribution is more flexible, it can still be vulnerable to extreme outliers, especially in heavy-tailed data. By construction, its location parameter is the $p_0$\textsuperscript{th} quantile. When $p_0$ is close to 0.5, both tails remain relatively heavy. But as $p_0$ moves away from 0.5 (i.e., $p_0 \to 0^+$ or $p_0 \to 1^-$), the skewness parameter ``lightens" one tail while weighting the other more heavily, leaving the lighter-tail side more susceptible to outliers. To address this, we propose a more robust approach: the contaminated GAL (cGAL) distribution. Constructed as a mixture of two GAL components, the cGAL inflates the variance of the second component to account for extreme observations, effectively mitigating outlier sensitivity. Crucially, it retains the property that the location parameter aligns with the targeted quantile, preserving interpretability while reducing sensitivity to skewness and outliers.
    
    Other relevant work includes \citet{barataexdqlm}, which examined dynamic quantile regression using the GAL distribution in Bayesian settings, though without extending it to mixed models or multivariate data with cluster-level random effects. \citet{florez2022quantile} applied quantile regression to multivariate and longitudinal data with the multivariate generalized hyperbolic distribution, which effectively models complex dependence but omits random effects and relies on the Bessel function, complicating computation. \citet{wichitaksorn2014generalized} explored flexible Bayesian quantile regression via scale mixtures of truncated normals. However, akin to AL, their methods lack asymmetry at middle quantiles and produce lighter tails at extremes, increasing outlier sensitivity. Moreover, these approaches do not extend to random effects. \citet{bernardi2018bayesian} employed the skew exponential power distribution for quantile regression, yielding more flexible tails than the AL but applying it only to fixed-effects models, thus excluding random and mixed model contexts.
    
    In this paper, we adopt a Bayesian approach to estimate parameters of the mixed-effects cGAL model with \verb|JAGS| \citep{PLUMMER2003}. This approach enables full posterior inference and does not rely on asymptotic approximations. Full posterior inference can be especially beneficial for smaller or more complex models. Because \verb|JAGS| supports flexible, user-defined model specifications, our conceptual approach can be translated directly into code, which conveniently facilitates the implementation of the proposed mixed-effects quantile regression model.
    
    We apply the proposed model to a human immunodeficiency virus (HIV) viral load dataset, using a nonlinear mixed-effects (NLME) model to capture patient-specific trajectories of viral load decay across various quantiles. 
    
    The remainder of this paper is organized as follows: \autoref{sec:HIV_DATASET} describes the ACTG~315 HIV clinical trial dataset that motivates our analysis. In \autoref{sec:DISTRIBUTIONS}, we present the quantile regression distributions used (AL, GAL, and cGAL). \autoref{sec:BAYES_MODEL} outlines the Bayesian mixed-effects model for quantile regression. \autoref{sec:APPLICATION} applies these methods to the HIV dataset to demonstrate their use in a real-world context. In \autoref{sec:SIMULATION}, we conduct a simulation study by sampling from the cGAL model and then fitting cGAL and GAL models to compare their performance under varying degrees of contamination and assess their sensitivity to outliers. Finally, \autoref{sec:DISCUSSION} presents concluding remarks and outlines possible directions for future research.

    \section{ACTG~315 HIV clinical trial} \label{sec:HIV_DATASET}
    
\begin{figure}
        \centering
        \includegraphics[width=0.8\textwidth]{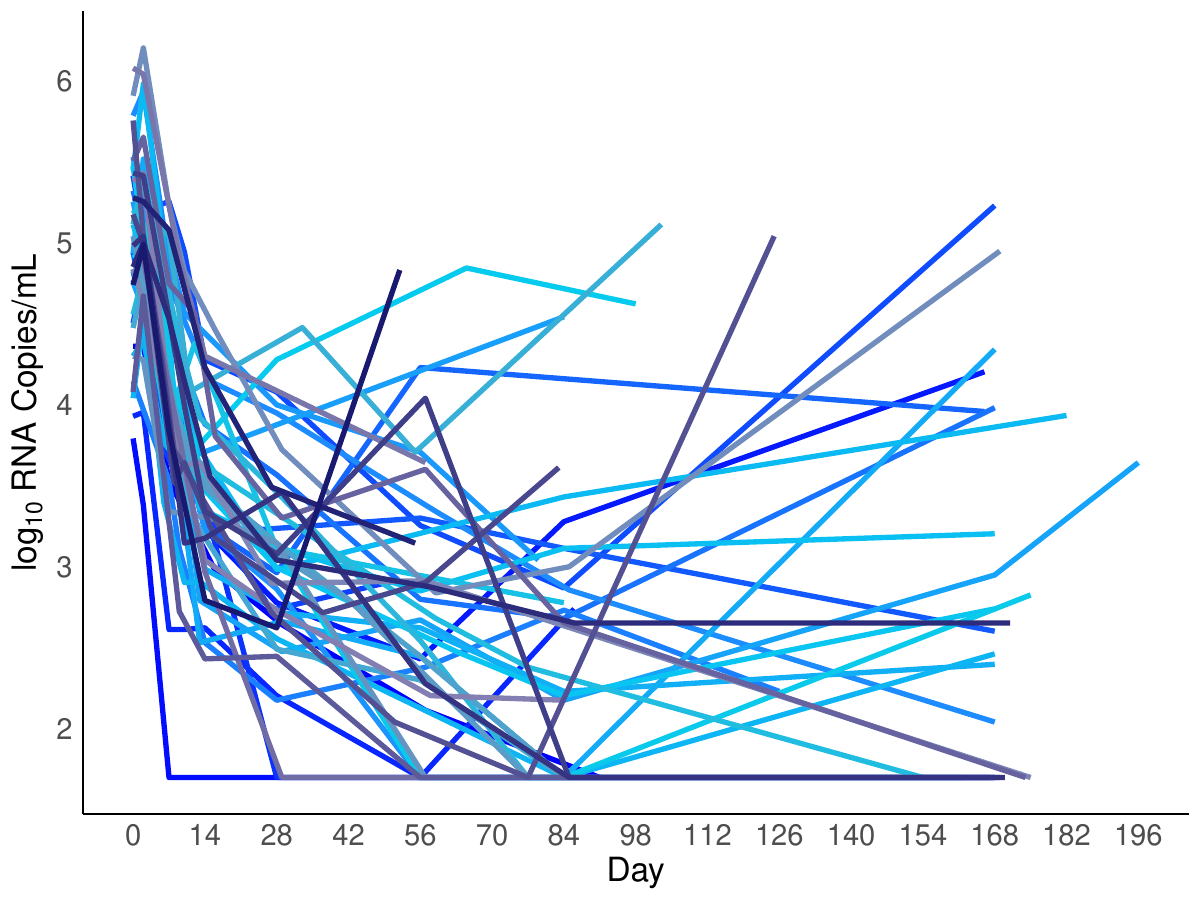}
        \caption{HIV viral load over time among patients in the ACTG~315 study. 
        Each line shows an individual patient's log$_{10}$ HIV viral load.}
        \label{fig:ACTG315_RNA}
    \end{figure}

    The ACTG~315 dataset, available in the \verb|ushr| package in \verb|R| \citep{MORRIS2020}, includes longitudinal measurements of HIV viral load (log$_{10}$ RNA copies/mL) over time. It features data on 46 patients, with the longest measurement recorded on Day~196 after baseline (Day~0). Models of HIV viral load dynamics typically show a biphasic decline: an initial rapid reduction in viral load due to effective antiretroviral therapy, followed by a slower stabilization phase as the virus is further suppressed \citep{perelson1996hiv}. Traditionally, many analyses (e.g., generalized linear models) have concentrated on the mean response, giving insights into average patient trajectories. Yet such approaches often overlook the considerable variability among patients, as shown in \autoref{fig:ACTG315_RNA}. Some patients experience sharper declines, while others have irregular fluctuations or slower viral suppression, indicating outliers or subgroups with distinct response patterns \citep{nelson2002mathematical}. A more flexible framework, such as quantile regression, is therefore needed to capture the full distribution of viral load trajectories and better understand HIV progression across diverse patient subgroups.
    
    We use an NLME quantile regression model that accounts for both the nonlinear dynamics of viral load and the presence of outliers, such as biologically extreme surges or measurement errors. In particular, we apply the cGAL model, which extends the GAL distribution through a contamination mechanism to address irregularities in typical viral load data. By comparing the cGAL model with the standard GAL model, we demonstrate that a robust quantile regression approach can yield more reliable predictions and insights into HIV progression, especially in the presence of outliers.

    \section{Laplace distributions for quantile regression} \label{sec:DISTRIBUTIONS}

    This section introduces the three Laplace distributions used in quantile regression in this paper: the AL distribution, the GAL distribution, and its robust extension, the cGAL distribution.

    \subsection{Asymmetric Laplace distribution}

    Consider a random variable $y \in \mathbb{R}$ following the AL distribution. The probability density function (PDF) of the AL distribution is given by:
    \begin{equation}
        f_\text{AL}\left(y \left| \mu, \sigma, p_0 \right.\right) = \frac{p_0\left(1 - p_0\right)}{\sigma} \exp\left\{-\rho_{p_0}\left(\frac{y - \mu}{\sigma}\right)\right\},
    \end{equation}
    where $\mu \in \mathbb{R}$ is a location parameter, representing the $p_0^\text{th}$ quantile, $\sigma > 0$ is a scale parameter, and $p_0 \in \left(0,1\right)$ is a skewness (quantile) parameter, fixed a {\it priori} (e.g., 0.25 for the first quartile). When $p_0 = 0.5$, the distribution is symmetric around $\mu$, corresponding to the standard Laplace distribution. The distribution is skewed for $p_0 \neq 0.5$, with skewness increasing as $p_0$ moves away from 0.5.
    
    The function $\rho_{p_0}\left(u\right)$ is defined as:
    \begin{equation}
        \rho_{p_0}\left(u\right) = u\left(p_0 - \ind\left(u < 0\right)\right),
    \end{equation}
    where the indicator function $\ind\left(u < 0\right)$ is defined as:
    \begin{equation}
        \ind\left(u < 0\right) =
        \begin{cases}
            1, & \text{if } u < 0, \\
            0, & \text{if } u \geq 0.
        \end{cases}
    \end{equation}
    The piecewise function $\rho_{p_0}\left(u\right)$ governs the exponential decay on either side of $\mu$, with asymmetry controlled by $p_0$. In quantile regression, $p_0$ thus specifies the distribution's focus (e.g., the 25\textsuperscript{th} vs. 75\textsuperscript{th} quantile).
    
    By construction, $\mu$ satisfies:
    \begin{equation}
        \int_{-\infty}^{\mu} f_\text{AL}\left(y \left| \mu, \sigma, p_0 \right.\right) \verb|d|y = p_0,
    \end{equation}
    confirming $\mu$ as the $p_0^\text{th}$ quantile.
    
    The AL distribution can also be viewed as a normal-exponential mixture (e.g., \citealp[]{kotz2001laplace}):
    \begin{equation} \label{eq:AL_MIXTURE}
        f_\text{AL}\left(y \left| \mu, \sigma, p_0 \right.\right) = \int_0^\infty \mathcal{N}\left(y \left| \mu + \theta_1 z, \sigma^2 \theta_2^2 z \right.\right) \text{Exp}\left(z \left| 1 \right.\right) \verb|d|z,
    \end{equation}
    where $ \mathcal{N}\left(y \left| \mu_N, \sigma_N^2 \right.\right) $ denotes the normal distribution with mean $ \mu_N $ and variance $ \sigma_N^2 $, and $ \text{Exp}\left(z \left| 1 \right.\right) $ is the exponential distribution with rate 1. The parameters $\theta_1$ and $\theta_2^2$ are functions of $p_0$ and are given by:
    \begin{equation}
        \theta_1 = \frac{1 - 2p_0}{p_0\left(1 - p_0\right)}, \quad \theta_2^2 = \frac{2}{p_0\left(1 - p_0\right)}.
    \end{equation}    
    The AL distribution is commonly used in quantile regression because its skewness parameter $p_0$ directly targets a chosen quantile. However, while its tails are heavier than the normal distribution (the standard Laplace has kurtosis 6), it can still be sensitive to very heavy-tailed or extreme observations, prompting the need for more flexible alternatives such as the GAL and cGAL distributions. These extensions are explored next.

    \subsection{Generalized asymmetric Laplace distribution}
    
    Next, consider the GAL distribution suggested by \citet{yan2017new}, which generalizes the AL. As with the AL distribution, we fix $\mu$ to be the $p_0$\textsuperscript{th} quantile so that
    \begin{equation}
        \int_{-\infty}^{\mu} f_\text{GAL}\left(y \left| \mu, \sigma, \gamma, p_0 \right.\right) \verb|d|y = p_0,
    \end{equation}
    where $\sigma > 0$ and $\gamma \neq 0$ control the scale and shape around the $p_0$\textsuperscript{th} quantile $\mu$.
    We then standardize $y$ by defining 
    \begin{equation}
        y^* = \frac{y - \mu}{\sigma}.
    \end{equation}
    Next, define the terms:
    \begin{equation} \nonumber
        p_{\gamma^+} = p - \ind\left(\gamma > 0\right), \quad p_{\gamma^-} = p - \ind\left(\gamma < 0\right),
    \end{equation}
    \begin{equation}
        g\left(\gamma\right) = 2 \Phi\left(-\left|\gamma\right|\right) \exp\left(\frac{\gamma^2}{2}\right), \quad p \equiv p\left(\gamma, p_0\right) = \ind\left(\gamma < 0\right) + \left\{\frac{p_0 - \ind\left(\gamma < 0\right)}{g\left(\gamma\right)}\right\},
    \end{equation}
    where $\Phi\left(\cdot\right)$ is the standard normal cumulative distribution function (CDF), and $\gamma$ lies in the interval $\left(L_\gamma, U_\gamma\right)$, defined by roots of $g\left(\gamma\right) = 1 - p_0$ and $g\left(\gamma\right)=p_0$.
    
    For $\gamma \neq 0$, the GAL PDF is:
    \begin{align}
        f_\text{GAL}\left(y \left| \mu, \sigma, \gamma, p_0 \right.\right) &= \frac{2p\left(1-p\right)}{\sigma} \Bigg[ \Bigg( \Phi\left(\frac{-p_{\gamma^+} y^*}{\left|\gamma\right|} + \frac{p_{\gamma^-}}{p_{\gamma^+}} \left|\gamma\right|\right) \nonumber \\
        & \quad - \Phi\left(\frac{p_{\gamma^-}}{p_{\gamma^+}} \left|\gamma\right|\right) \Bigg) \exp\left\{-p_{\gamma^-} y^* + \frac{\gamma^2}{2} \left(\frac{p_{\gamma^-}}{p_{\gamma^+}}\right)^2 \right\} \ind\left(\frac{y^*}{\gamma} > 0\right) \nonumber \\
        & \quad + \exp\left\{-p_{\gamma^+} y^* + \frac{\gamma^2}{2} \right\} \Phi\left(-\left|\gamma\right| + \frac{p_{\gamma^+} y^*}{\left|\gamma\right|} \ind\left(\frac{y^*}{\gamma} > 0\right)\right) \Bigg], \label{eq:PDF_GAL}
    \end{align}
    which reduces to the AL distribution when $\gamma \to 0$.

    The GAL distribution can also be expressed as a triple mixture of normal, exponential, and truncated normal variables \citep{yan2017new}:
    \begin{equation} \label{eq:GAL_MIXTURE}
        f_\text{GAL}\left(y \left| \mu, \sigma, \gamma, p_0 \right.\right) = \int_0^\infty \int_0^\infty \mathcal{N}\left(y \left| \mu + \sigma C \left| \gamma \right| s + \sigma A z, \sigma^2 B z \right.\right) \text{Exp}\left(z \left| 1 \right.\right) \mathcal{N}^+\left(s \left| 0, 1 \right.\right) \verb|d|z \verb|d|s,
    \end{equation}
    where $ \mathcal{N}^+\left(s \left| 0, 1 \right.\right) $ is a truncated normal on the positive real line, and 
    \begin{equation}
        A \equiv A\left(p\right) = \frac{1 - 2p}{p\left(1 - p\right)}, 
        \quad 
        B \equiv B\left(p\right) = \frac{2}{p\left(1-p\right)},
        \quad
        C \equiv C\left(\gamma,p\right) = \left[\mathbf{1}\{\gamma>0\} - p\right]^{-1}.
    \end{equation}
    While the GAL distribution presents greater flexibility in modeling skewness and accommodating heavier tails compared to the AL, as will be demonstrated in the next section, its kurtosis is capped, limiting its ability to capture extremely heavy-tailed behavior.
    
    \subsection{Contaminated generalized asymmetric Laplace distribution}

    To increase the robustness of a quantile model to outlying values, we introduce the cGAL distribution, formed as a mixture of two GAL distributions. Both components share the same location $\mu$ but differ in scale parameters: $\sigma$ for the main component and $\tau_0\sigma$ for the ``contaminated" component, where $\tau_0>1$. Let $\alpha \in \left(0,1\right)$ be the mixture weight. The cGAL PDF is:
    \begin{equation}
        f_{\text{cGAL}}\left(y \left| \mu, \sigma, \gamma, \alpha, p_0 \right.\right) = \left(1 - \alpha\right) f_{\text{GAL}}\left(y \left| \mu, \sigma, \gamma, p_0 \right.\right) + \alpha f_{\text{GAL}}\left(y \left| \mu, \tau_0 \sigma, \gamma, p_0 \right.\right).
    \end{equation}
    By inflating the scale in the second component, the cGAL distribution can better accommodate outliers. This two-component, scale-inflated mixture approach closely resembles the well-known contaminated normal distribution \citep{tukey1960survey, huber1992robust}.
    
    Because each component has the same location $\mu$, the mixture preserves $\mu$ as the $p_0^\text{th}$ quantile:
    \begin{equation}
        F_{\text{cGAL}}\left(y \left| \mu, \sigma, \gamma, \alpha, p_0 \right.\right) = \left(1 - \alpha\right) F_{\text{GAL}}\left(y \left| \mu, \sigma, \gamma, p_0 \right.\right) + \alpha F_{\text{GAL}}\left(y \left| \mu, \tau_0 \sigma, \gamma, p_0 \right.\right),
    \end{equation}
    where $F_{\text{GAL}}\left(y \left| \mu, \sigma, \gamma, p_0\right) \right.$ is the CDF of the GAL distribution with parameters $\mu$, $\sigma$, $\gamma$, and $p_0$. Evaluating at $y=\mu$ yields $F_{\text{cGAL}}(\mu)=p_0$. The parameter $\alpha$ controls the contamination degree: as $\alpha \to 0$, the model reverts to the standard GAL distribution; as $\alpha \to 1$, it degenerates to a single, inflated-scale GAL (with scale $\tau_0\sigma$), and is thus no longer a two-component contaminated distribution. Larger values of $\tau_0$ inflate the tails more strongly, improving robustness against extreme outliers.

\begin{figure}
        \centering
        \begin{subfigure}{0.50\textwidth}
            \centering
            \includegraphics[width=\linewidth]{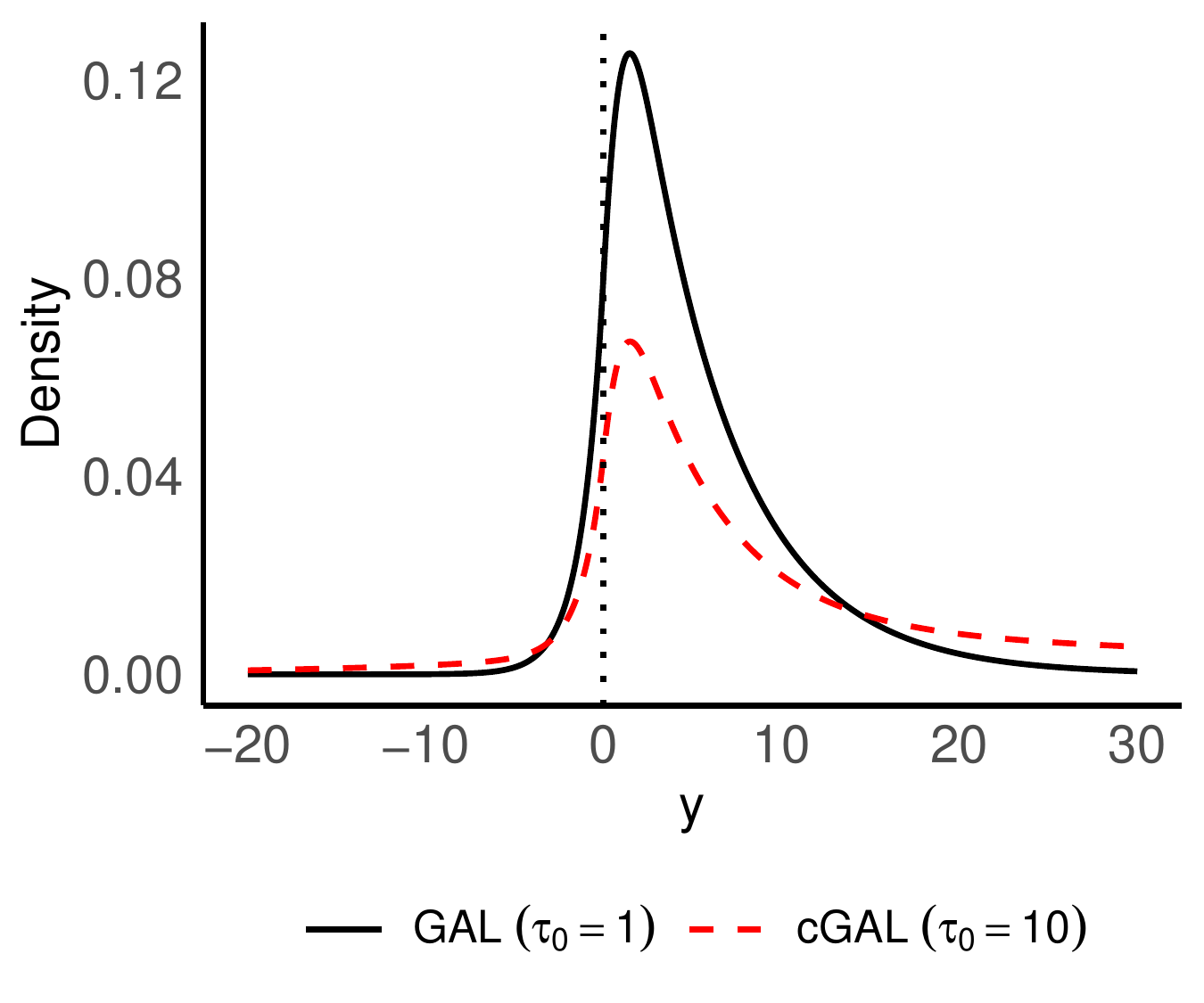}
            \caption{$p_0 = 0.1$}
        \end{subfigure}
        \hspace{-0.4cm}
        \begin{subfigure}{0.50\textwidth}
            \centering
            \includegraphics[width=\linewidth]{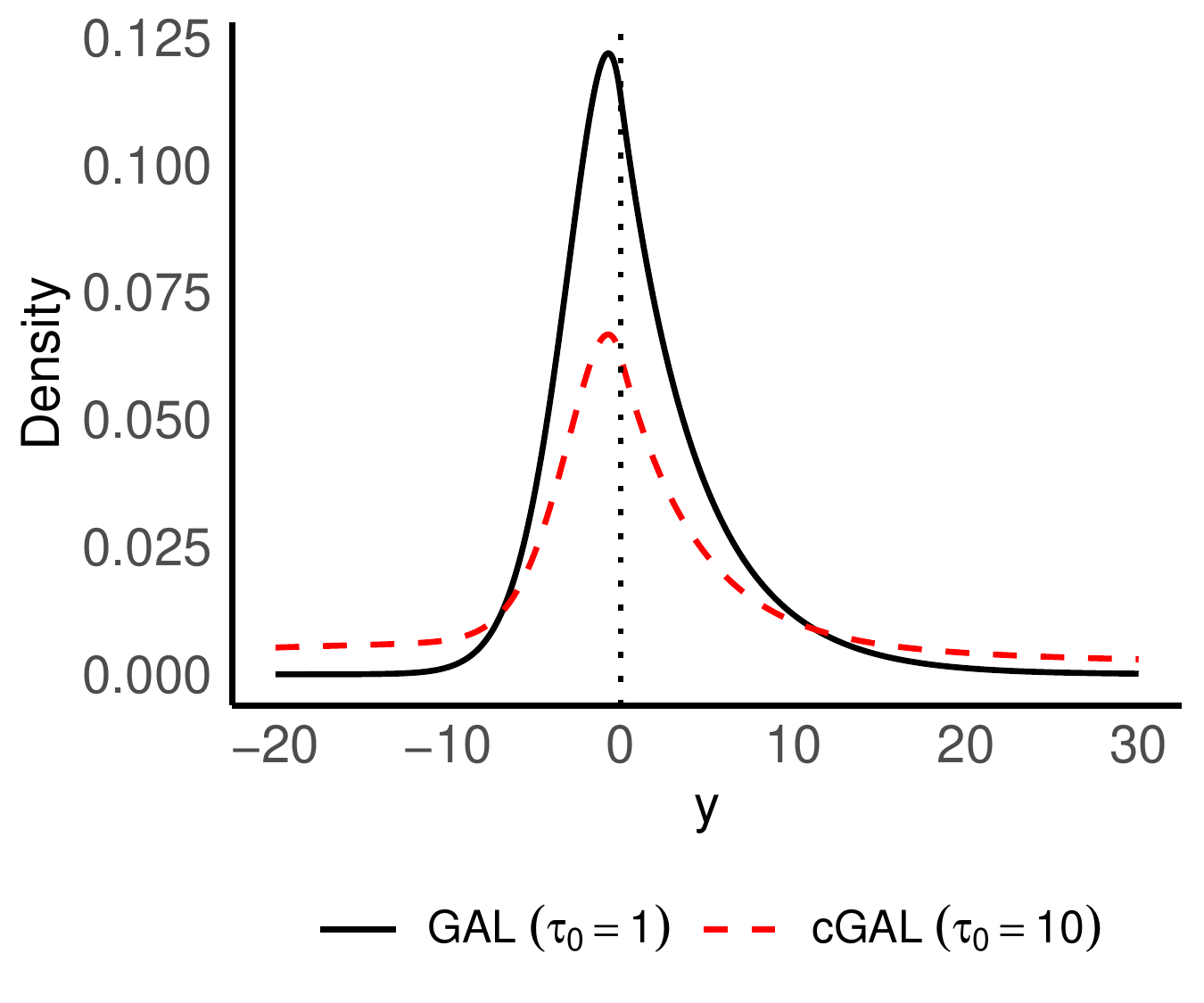}
            \caption{$p_0 = 0.5$}
        \end{subfigure}
        \caption{PDF comparison of GAL and cGAL distributions for $p_0 = 0.1$ with $\gamma = 1$ and $p_0 = 0.5$ with $\gamma = -0.625$, assuming $\mu = 0$ and $\sigma = 1$. The cGAL distribution is shown for $\tau_0 = 10$ and $\alpha = 0.5$, while the GAL corresponds to $\tau_0 = 1$. When $\tau_0 > 1$ and $\alpha \neq 0,1$, cGAL exhibits heavier tails, making it more robust to outliers. The black dotted line at $y = 0$ indicates the $p_0^\text{th}$ quantile.}
        \label{fig:CGAL_PDF}
    \end{figure}

    \autoref{fig:CGAL_PDF} compares the PDFs for the GAL and cGAL distributions at two quantiles, $p_0 = 0.1$ with $\gamma = 1$ and $p_0 = 0.5$ with $\gamma = -0.625$, assuming $\mu = 0$ and $\sigma = 1$. The cGAL distribution, shown for $\tau_0 = 10$ and $\alpha = 0.5$, exhibits heavier tails compared to the GAL distribution ($\tau_0 = 1$) when contamination occurs.

\begin{figure}
        \centering
        \begin{subfigure}[b]{0.45\textwidth}
            \centering
            \includegraphics[width=\textwidth]{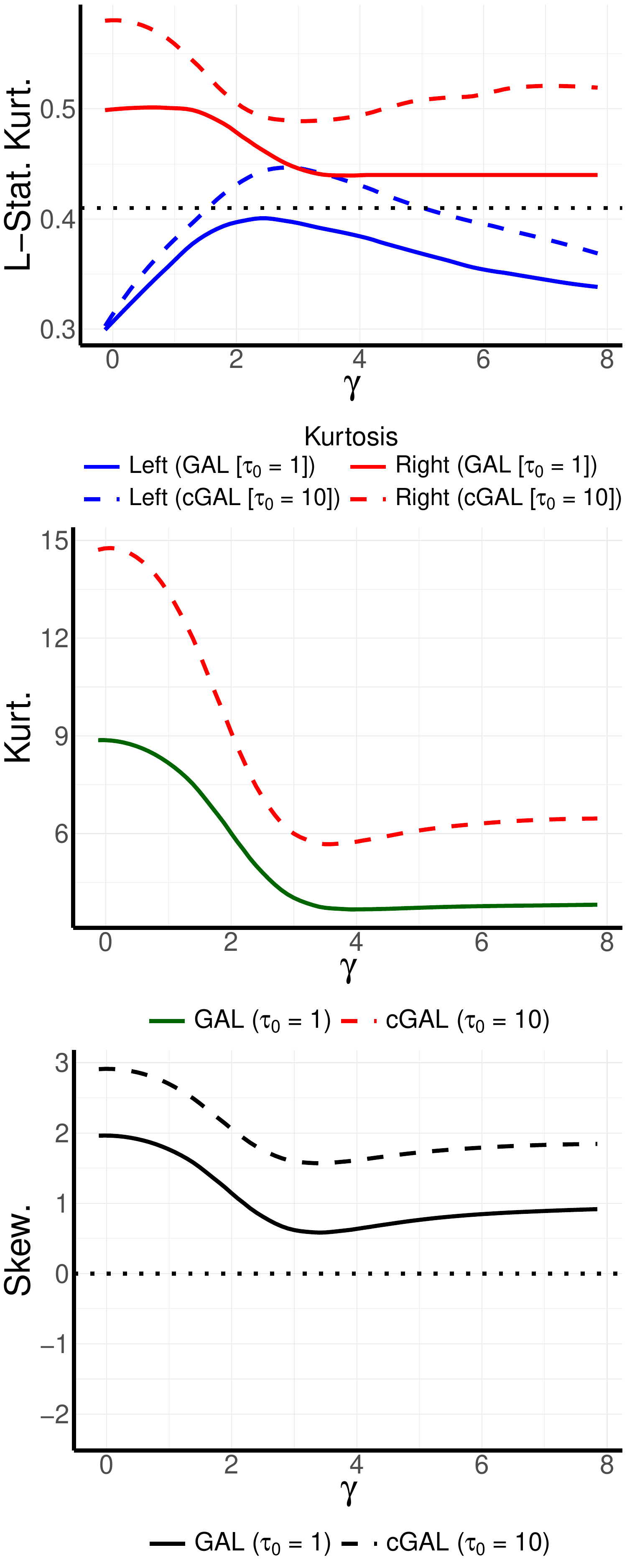}
            \caption{$p_0 = 0.1$}
        \end{subfigure}
        \hspace{-0.3cm}
        \begin{subfigure}[b]{0.45\textwidth}
            \centering
            \includegraphics[width=\textwidth]{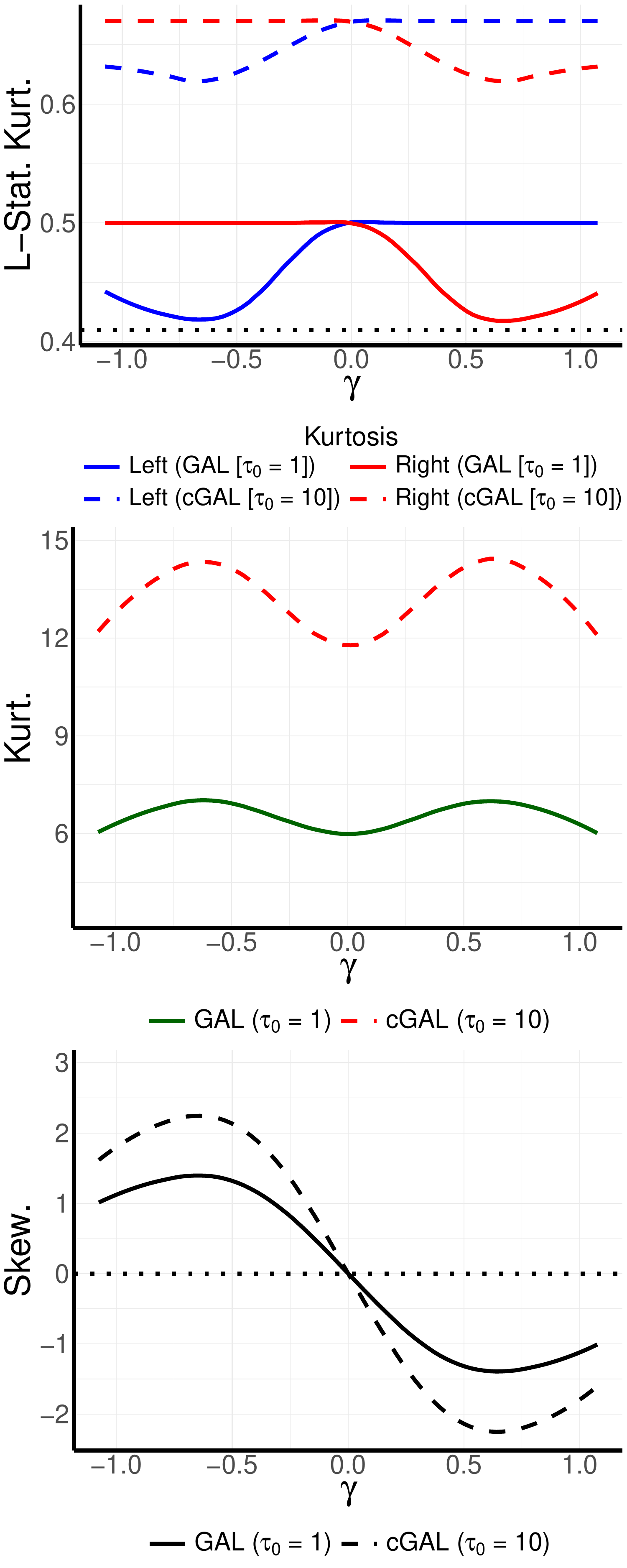}
            \caption{$p_0 = 0.5$}
        \end{subfigure}
        \caption{Kurtosis and skewness for GAL and cGAL ($\tau_0=10$, $\alpha=0.5$) over $\gamma$. Left and right L-statistic kurtosis \citep{fiori2014right} are shown in blue and red, respectively. A dotted black line marks the normal-reference kurtosis of $0.4142$. The overall kurtosis (green) and skewness (black) for GAL ($\tau_0=1$, solid) vs. cGAL ($\tau_0=10$, dashed) highlight cGAL's heavier tails and increased robustness. A dotted line at $0$ on the skewness plot indicates symmetry.}
    \label{fig:CGAL_KURT_SKEW}
    \end{figure}

    \autoref{fig:CGAL_KURT_SKEW} presents the sample kurtosis and skewness for the GAL and cGAL ($\tau_0 = 10$, $\alpha = 0.5$) distributions as functions of the parameter $\gamma$ to illustrate the impact of contamination in the cGAL distribution on both the tails and asymmetry of the distribution in comparison to the GAL model for two quantiles ($p_0 \in \left\{0.1, 0.5\right\}$). The left and right L-statistic kurtosis \citep{fiori2014right} are displayed alongside both distributions' overall kurtosis and skewness. Details on the calculation of the L-statistic kurtosis can be found in \autoref{sec:L_STAT_KUR}. For instance, the GAL distribution exhibits lighter-than-normal left tails at the lower decile ($p_0 = 0.1$). However, with the introduction of contamination in the cGAL distribution, the left tail becomes heavier, making it more robust to outliers, particularly for modeling at the lower decile ($p_0 = 0.1$).

    \section{Bayesian mixed-effects quantile regression model} \label{sec:BAYES_MODEL}

    This section outlines the general form of this study's mixed-effects quantile regression model and its Bayesian specification. We now embed the AL, GAL, and cGAL distributions into a hierarchical framework for repeated measures or clustered data.
    
    Let $y_{ij}$ denote the response variable for subject or cluster $i$ ($i = 1, \dots, N$) and observation $j$ ($j = 1, \dots, J_i$). We model $y_{ij}$ as:
    \begin{equation}
        y_{ij} = \mu_{ij}\left(p_0\right) + \epsilon_{ij}\left(p_0\right),
    \end{equation}
    where $\mu_{ij}\left(p_0\right)$ is the conditional $p_0^\text{th}$ quantile of $y_{ij}$ given the covariates and $\epsilon_{ij}\left(p_0\right)$ is the error term such that the $p_0^\text{th}$ quantile of $\epsilon_{ij}\left(p_0\right)$ is zero. Thus, the distribution of $\epsilon_{ij}\left(p_0\right)$ is centered at zero in the ``$\tau$-quantile" sense.
    
    The quantile $\mu_{ij}\left(p_0\right)$ conditional on random effects is modeled as a function of fixed effects, random effects, and covariates:
    \begin{equation}
        \mu_{ij}\left(p_0\right) = g\left({\bm x}_{ij}, {\bm\beta}, {\bm b}_i\right),
    \end{equation}
    with ${\bm x}_{ij}$ being the vector of covariates for observation $j$ of subject $i$, ${\bm\beta}$ the vector of unknown fixed-effect parameters, ${\bm b}_i \in \mathbf{\mathbb{R}}^d$ the vector of random effects for subject $i$, and $g$ a known link function, which can be linear or nonlinear (e.g., a compartmental viral load decay model).

    The random effects ${\bm b}_i$ are assumed to follow a multivariate normal distribution with mean $\bm{0}$ and covariance matrix $\bm{\Sigma}$:
    \begin{equation}
        {\bm b}_i \sim \text{MVN}\left({\bm0}, {\bm\Sigma}\right). \label{eq:RANDOM_B}
    \end{equation}
    The mixed-effects quantile model is formulated under the three different distributional frameworks in \autoref{sec:DISTRIBUTIONS}: the AL, GAL, and cGAL distributions. In the AL model, $y_{ij} \sim f_{\text{AL}}\left(y_{ij} \left| \mu_{ij}\left(p_0 \right), \sigma, p_0 \right.\right)$. The GAL model introduces a skewness parameter, giving $y_{ij} \sim f_\text{GAL}\left(y_{ij} \left| \mu_{ij}\left(p_0\right), \sigma, \gamma, p_0 \right.\right)$. Lastly, the cGAL model incorporates an additional contamination parameter to handle outliers, leading to $y_{ij} \sim f_{\text{cGAL}}\left(y_{ij} \left| \mu_{ij}\left(p_0\right), \sigma, \gamma, \alpha, p_0 \right.\right)$.

    In the cGAL setting, ``outliers" or extreme data points are partially explained by the contaminated component.

    For the fixed effects $\bm{\beta}$, we assume a multivariate normal prior with mean $\bm{0}$ and covariance matrix $s_\beta^{2} \bm{I}_p$:
    \begin{equation}
        {\bm\beta} \sim \text{MVN}\left({\bm 0}, s^2_\beta {\bm I}_p\right),
    \end{equation}
    where $\bm{I}_p$ is the $p \times p$ identity matrix.
    
    The error term's scale parameter $\sigma$ is modeled using a truncated Student's $t$-distribution:
    \begin{equation}
        \sigma \sim t_{+}\left(0, s_\sigma, \nu_\sigma\right),
    \end{equation}
    indicating a $t$-distribution with $\nu_\sigma$ degrees of freedom, centered at $0$, scale $s_\sigma$, and truncated to be positive.
    
    For the skewness parameter, we introduce an auxiliary proportion $B \in \left(0, 1\right)$ with a beta prior and then rescale it to the admissible interval, $\left(L_\gamma, U_\gamma\right)$:
    \begin{equation}
        B \sim \mathrm{Beta}\left(a_\gamma, b_\gamma\right), 
        \qquad
        \gamma = L_\gamma + \left(U_\gamma - L_\gamma\right) B,
    \end{equation}
    where 
    $E\left[B\right] = a_\gamma / \left(a_\gamma + b_\gamma\right)$ and variance 
    \begin{equation}
      \operatorname{Var}\left[B\right] 
      = \dfrac{a_\gamma b_\gamma}
              {\left(a_\gamma + b_\gamma\right)^{2} \left(a_\gamma + b_\gamma + 1\right)}.
    \end{equation}
    Setting $a_\gamma = b_\gamma = 1$ reproduces a uniform prior on $\gamma$; alternative choices can up- or down-weight the ends of the interval.

    We specify the matrix-generalized half-$t$ (MGH-$t$) prior distribution of \citet{HUANG2013} for the inverse covariance matrix $\bm{\Sigma}^{-1}$ as a hierarchical mixture of a Wishart distribution and gamma distributions for the diagonal elements of a scale matrix, $\bm{\Psi}$. Specifically, the diagonal elements of $\bm{\Psi}$, denoted as $\psi_{ii}$, follow independent gamma distributions with shape and scale parameters set to $0.5$ and $1/A_\psi^2$, respectively: $\psi_{ii} \sim \text{Gamma}\left(0.5, 1/A_\psi^2\right)$. The off-diagonal elements $\psi_{ij}$ for $i \neq j$ are set to zero. The inverse covariance matrix, $\bm{\Sigma}^{-1}$, follows a Wishart distribution with inverse scale matrix $2\nu_\Sigma\bm{\Psi}$ and degrees of freedom equal to $\nu_\Sigma + d - 1$. We write the MGH-$t$ prior for $\bm{\Sigma}^{-1}$, compactly, as:
    \begin{equation}
        \bm{\Sigma}^{-1} \sim \mathrm{MGH}\text{-}t\left(A_\psi,\nu_\Sigma\right).
    \end{equation}
    This hierarchical formulation corresponds to specifying a $t_{+}\left(0, A_\psi, \nu_\Sigma\right)$ prior for the standard deviation terms of $\bm{\Sigma}$.

    For the contamination weight, we assign a beta prior
    \begin{equation}
      \alpha \sim \text{Beta}\left(a_\alpha, b_\alpha\right),
    \end{equation}
    Choosing $a_\alpha \ll b_\alpha$ encodes the prior belief that only a small fraction of observations belong to the contaminated (heavier-tailed) component, while still allowing the likelihood to pull $\alpha$ upward when the data warrant it.

    Throughout, we treat $\tau_0$ as a tuning constant; any sufficiently large value provides heavy-tail protection without altering the model's quantile-centering property.

    The quantities $s_\beta^{2}$, $s_\sigma$, $\nu_\sigma$, $A_\psi$, $\nu_\Sigma$, $a_\gamma$, $b_\gamma$, $a_\alpha$, $b_\alpha$, and the scale-inflation constant $\tau_0$ can be chosen to reflect substantive prior beliefs or pragmatic weak-information defaults. \autoref{sec:APPLICATION} specifies the values adopted in the HIV analysis and presents sensitivity checks under markedly flatter alternatives.
    
    \section{Application to HIV dataset} \label{sec:APPLICATION}

    \subsection{Nonlinear mixed-effects model} \label{sec:BIPHASIC_MODEL}

    The NLME model we explore is a specific case of the general mixed-effects quantile regression framework introduced in \autoref{sec:BAYES_MODEL}, tailored for modeling the HIV viral load in the ACTG~315 dataset. In this case, the model is nonlinear, specified as a biphasic (two-phase) exponential decay process that captures both short-term and long-term dynamics of HIV viral load \citep{perelson1996hiv}. We model the observed viral load for patient $i$ at time $t_{ij}$ as a combination of two exponential decay processes to explain the biphasic viral decline. The model accounts for both short-term and long-term decay components and includes random effects to capture patient-specific variability.

    We model the observed viral load $y_{ij}$ as follows \citep{wu1999population}:
    \begin{equation}
        y_{ij} = \mu_{ij}\left(p_0\right) + \epsilon_{ij}\left(p_0\right),
    \end{equation}
    where $\mu_{ij}\left(p_0\right)$ represents the conditional $p_0^\text{th}$ quantile of the viral load for patient $i$ at time $t_{ij}$, and $\epsilon_{ij}\left(p_0\right)$ is an error term such that the $p_0^\text{th}$ quantile of $\epsilon_{ij}\left(p_0\right)$ is zero. The conditional quantile $\mu_{ij}\left(p_0\right)$ is specified as
    \begin{equation}
        \mu_{ij}\left(p_0\right) = \log_{10} \left(P_{1i} e^{-\lambda_{1i} t_{ij}} + P_{2i} e^{-\lambda_{2ij} t_{ij}}\right),
    \end{equation}
    where $P_{1i}$ and $P_{2i}$ are the initial viral load components for the two decay phases, and $\lambda_{1i}$ and $\lambda_{2ij}$ are the corresponding decay rates.
    
    The parameters $P_{1i}$, $P_{2i}$, $\lambda_{1i}$, and $\lambda_{2ij}$ vary across patients according to the following random-effects structure:
    \begin{align}
        \log\left(P_{1i}\right) &= \beta_1 + b_{i1}, \\ \nonumber
        \lambda_{1i} &= \beta_2 + b_{i2}, \\ \nonumber
        \log\left(P_{2i}\right) &= \beta_3 + b_{i3}, \\ \nonumber
        \lambda_{2ij} &= \beta_4 + \beta_5 \text{CD4}_{ij} + b_{i4}.
    \end{align}
    Here, $\beta_1$ to $\beta_5$ are fixed effects, capturing the population-level behavior, while $b_{i1}$ to $b_{i4}$ are random effects that account for patient-specific deviations from the population averages. In this formulation, CD4 count ($\text{CD4}_{ij}$), expressed in units of CD4 cells per 100~mm\textsuperscript{3}, is included as a covariate in the long-term decay rate, $\lambda_{2ij}$ \citep{yu2023flexible}, allowing immune status to influence the long-term phase of viral decay.

    The random effects vector $\bm{b}_i = \left(b_{i1}, b_{i2}, b_{i3}, b_{i4}\right)^\top$ is assumed to follow a multivariate normal distribution as in Equation~(\ref{eq:RANDOM_B}).

    The NLME model is explored under the three distributional frameworks outlined in \autoref{sec:BAYES_MODEL}. Hence, each observation $y_{ij}$ is assumed to follow an AL/GAL/cGAL distribution for which the $p_0$\textsuperscript{th} quantile is {\it centered} at $\mu_{ij}\left(p_0\right)$.
    
    The priors used in these models will follow those specified earlier in \autoref{sec:BAYES_MODEL}.

    The above NLME model, based on the GAL distribution, was also applied to the ACTG~315 dataset by \citet{yu2023flexible}. However, they implemented a custom sampler instead of \verb|JAGS| and adopted different prior specifications. For instance, they employed a Wishart prior for $\bm\Sigma$, whereas we use the matrix-generalized half-$t$ distribution.

    \subsection{Model implementation, adequacy, and comparison} \label{sec:IMPLEMENT}

    We investigated viral load dynamics at five quantiles, $p_0 = 0.05$, $0.25$, $0.5$, $0.75$, and $0.95$, to capture both lower and upper extremes along with the median. Clinically, examining the tails (e.g., $p_0=0.05$ and $p_0=0.95$) helps identify subgroups that either rapidly suppress their viral load or remain persistently high, thereby informing targeted interventions. This broad distribution coverage also allows us to assess both short-term and long-term decay phases across diverse patient trajectories.

    We adopt the following weakly informative specifications:
    \begin{equation}
        \begin{aligned}
            \bm\beta &\sim \mathrm{MVN}\left(\mathbf{0}, 1000\mathbf{I}_p\right), \\[2pt]
            \sigma &\sim t_{+}\left(0, \sqrt{10}, 3\right), \\[2pt]
            B &\sim \mathrm{Beta}\left(1, 1\right), \\[2pt]
            \gamma &=  L_\gamma + (U_\gamma - L_\gamma)B, \\[2pt]
            \bm\Sigma^{-1} &\sim \mathrm{MGH}\text{-}t\left(A_\psi = 50, \nu_\Sigma = 2\right), \\[2pt]
            \alpha &\sim \mathrm{Beta}\left(1, 9\right), \\[2pt]
            \tau_0 &= 10.
        \end{aligned}
    \end{equation}
    Here $\bm\beta = \left(\beta_1, \dots, \beta_5\right)^\top$ ($p = 5$) and $\bm\Sigma$ is a $d \times d$ random-effects covariance matrix with $d = 4$; thus the MGH-$t$ construction uses $\nu_\Sigma + d - 1 = 5$ degrees of freedom in the implied Wishart step, conditional on the diagonal scale matrix $\bm\Psi$.

    The prior for $\bm\beta$ implies the 95\% prior region covers approximately $\pm 60$ log$_{10}$ RNA copies/mL, which exceeds any clinically plausible range. Therefore, the likelihood dominates inference.

    The choice of $\alpha \sim \mathrm{Beta}\left(1, 9\right)$ implies an expected value of $E\left[\alpha\right] = 0.10$ and results in a strongly right-skewed distribution. This prior reflects our expectation that only a small fraction of observations arise from the contaminated component, while still permitting the data to suggest larger values of $\alpha$ if necessary. Similarly, setting $\tau_{0} = 10$ increases the scale of the contaminated component tenfold, large enough to capture heavy-tailed outliers without dominating the main GAL component.

    All models were implemented and run in \texttt{JAGS} \citep{PLUMMER2003} through the \texttt{runjags} package in \verb|R| \citep{DENWOOD2016}. The Bayesian models were executed using Markov chain Monte Carlo (MCMC) sampling, with multiple chains initiated to ensure a thorough exploration of the posterior distribution. Convergence diagnostics were checked for each model using the potential scale reduction factor ($\hat{R}$), with values below 1.05 indicating acceptable convergence across chains.

    For the AL and GAL models, we utilized their mixture representations in Equations~(\ref{eq:AL_MIXTURE}) and (\ref{eq:GAL_MIXTURE}) to streamline the MCMC sampling process. In contrast, for the cGAL model, we used the ``ones" trick to define its custom likelihood in \verb|JAGS|. This method uses binary indicator variables that follow a Bernoulli distribution. The probabilities are proportional to the likelihood, scaled by a constant to ensure they remain valid. Compared to the AL and GAL models' mixture representation, this approach increases computational demands.

    In the bi-exponential model, $\lambda_{1i}$ typically represents the initial, faster decay rate, while $\lambda_{2ij}$ reflects the slower, long-term decay. If $\lambda_{2ij}$ is estimated to be larger than $\lambda_{1i}$, their roles reverse, with $\lambda_{2ij}$ governing the initial phase and $\lambda_{1i}$ the later phase. This reversal also shifts $P_{1i}$ and $P_{2i}$, associating $P_{2i}$ with the initial decay and $P_{1i}$ with the long-term phase. To prevent this switch, setting initial values so that $\lambda_{1i}$ is larger than $\lambda_{2ij}$, along with appropriate values for $P_{1i}$ and $P_{2i}$, helps preserve the intended structure of the decay phases.

    All estimates are based on the posterior median, with 95\% highest posterior density (HPD) intervals representing the uncertainty around these estimates.

    We present $\bm\Sigma$ in its inverse form, $\bm\Omega = \bm\Sigma^{-1}$. The entries of $\bm\Omega$ are denoted by $\omega_{ij}$, for $1 \le i, j \le 4$.

    We present quantile-specific predictions for the viral load trajectory over time, setting random effects to zero alongside the fixed-effect parameter estimates and 95\% HPD intervals across quantiles under each model. CD4 count is included as a covariate, with predicted values over time based on a random intercept-slope model (intercept of approximately 2.25 and slope of 0.001). Predictions are generated for specific study days in line with the planned schedule of viral load and CD4 cell count measurements, including days~0, 2, 7, 10, 14, 21, 28, and weeks 8, 12, 24, and 48.

    The Kullback-Leibler (K-L) divergence is employed to assess the adequacy of the three models to identify influential observations and outliers \citep{WANG2016, TOMAZELLA2021}, as outlined in \autoref{sec:KL_METHOD}.

    We conducted a simulation-based residual analysis using the \verb|R| package \verb|DHARMa| \citep{HARTIG2021A, HARTIG2021B} to evaluate the adequacy of the three models. This method compares the observed data to the posterior predictive distribution, conditioned on the random effects for each observation, thereby generating \emph{scaled residuals} between 0 and~1. We used several tests provided by \verb|DHARMa| to assess these scaled residuals statistically. The Kolmogorov-Smirnov test evaluated the uniformity of the residual distribution; a parametric dispersion test checked for overdispersion or underdispersion, and a binomial test flagged observations significantly deviating from model expectations. We report \emph{p}-values from these tests to determine each model's adequacy of fit.

    The \texttt{loo} package \citep{vehtari2024LOO} in \verb|R| was used to perform leave-one-out (LOO) cross-validation to compare the models' predictive performance. To ensure numerical stability in the computation of the GAL and cGAL density functions, especially when dealing with exponentials of large magnitudes, we adopted a numerically stable implementation of the log-density function. The original GAL density function involves exponential and CDF terms that can cause overflow or underflow issues. To mitigate this, we reformulated the density computation to operate primarily on the logarithmic scale.

    The \verb|R| code implementing the methods described in this paper is available on GitHub in the repository \href{https://github.com/DABURGER1/Robust-Quantile-cGAL}{Robust-Quantile-cGAL}.

    In the next sections, we present and discuss the results of these models.

    \subsection{Model adequacy and comparison}

\begin{figure}
        \centering
        \begin{subfigure}{0.44\textwidth}
            \centering
            \includegraphics[width=\linewidth]{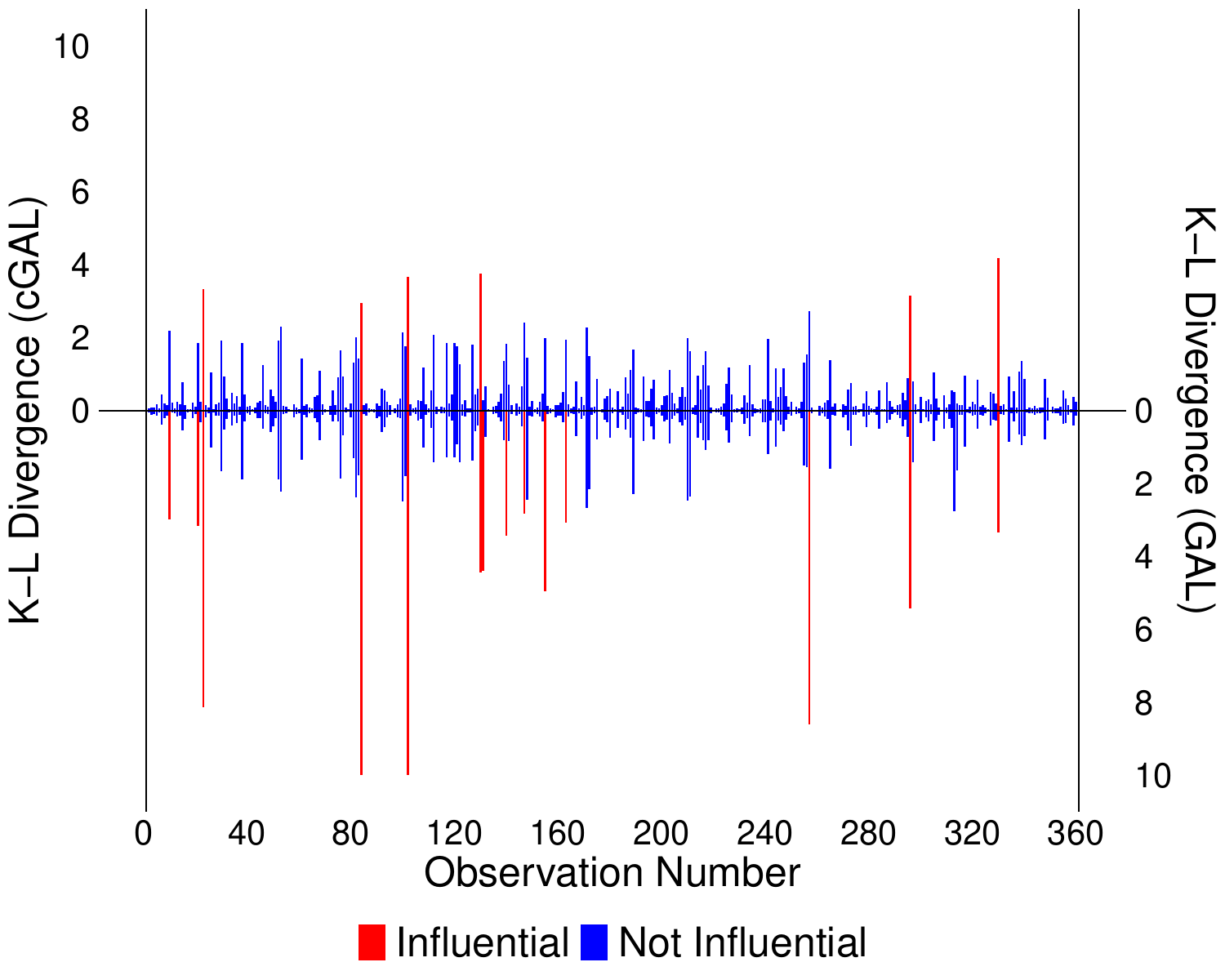}
            \caption{$p_0 = 0.05$}
        \end{subfigure}
        \begin{subfigure}{0.44\textwidth}
            \centering
            \includegraphics[width=\linewidth]{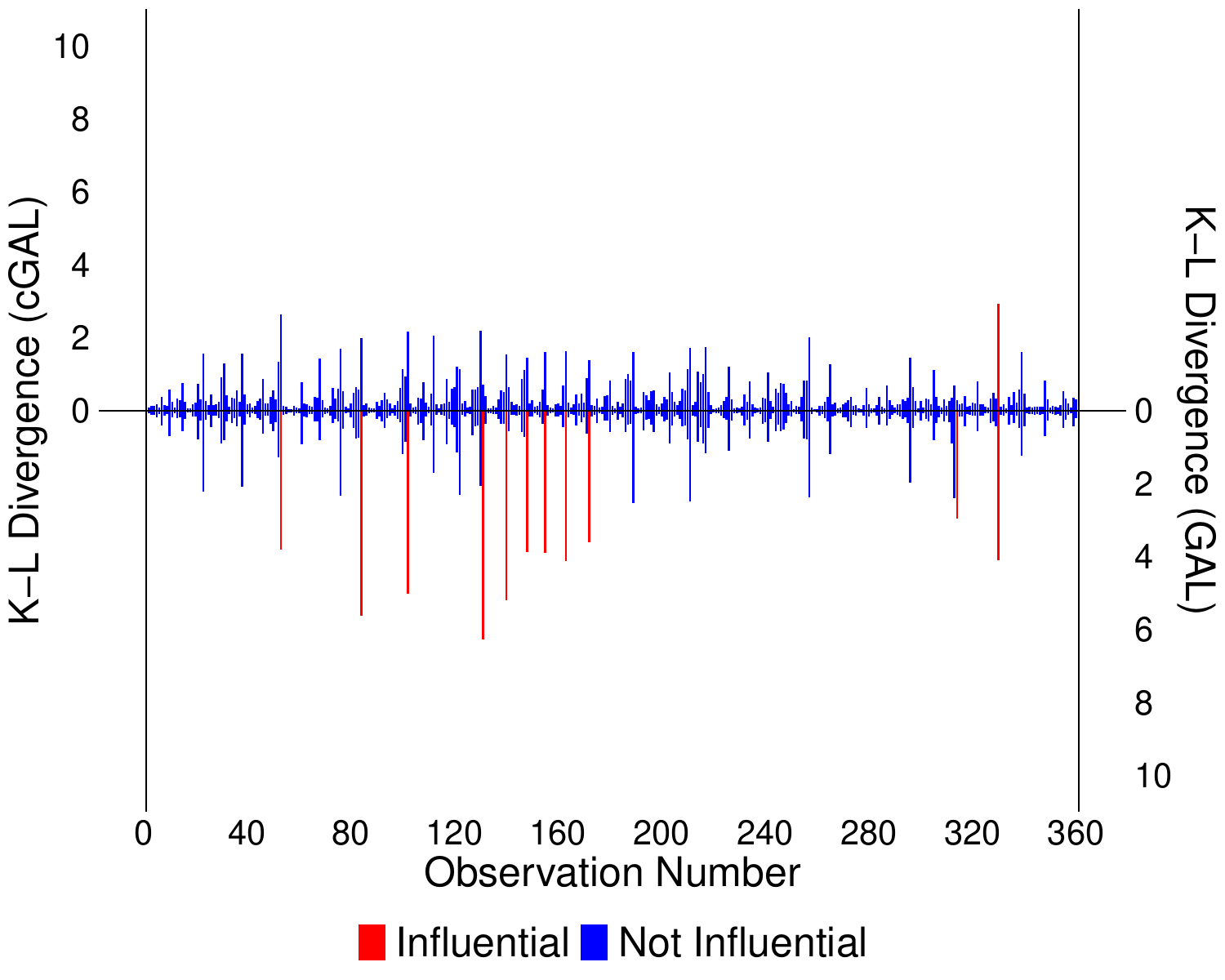}
            \caption{$p_0 = 0.25$}
        \end{subfigure}
        \\
        \begin{subfigure}{0.44\textwidth}
            \centering
            \includegraphics[width=\linewidth]{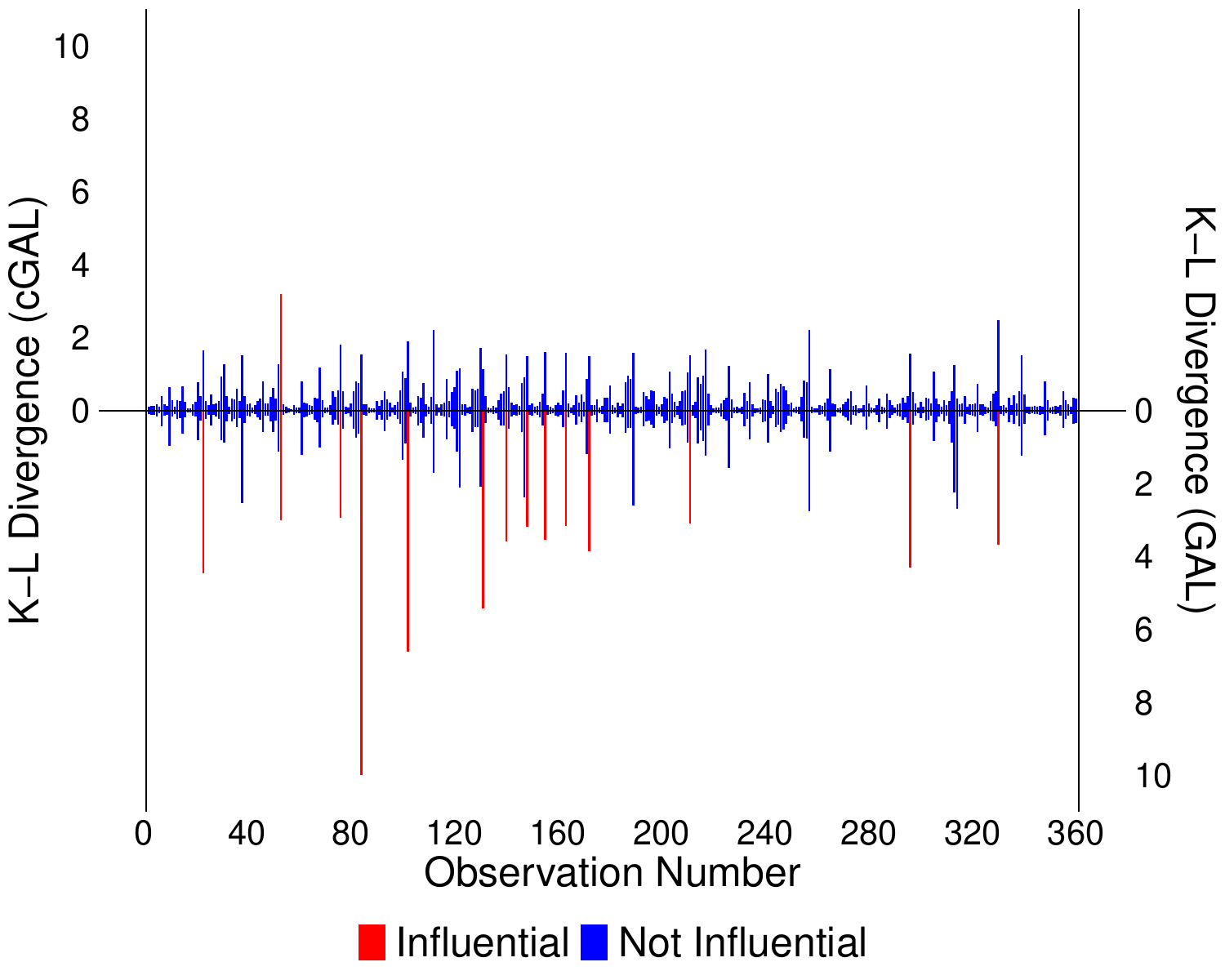}
            \caption{$p_0 = 0.5$}
        \end{subfigure}
        \begin{subfigure}{0.44\textwidth}
            \centering
            \includegraphics[width=\linewidth]{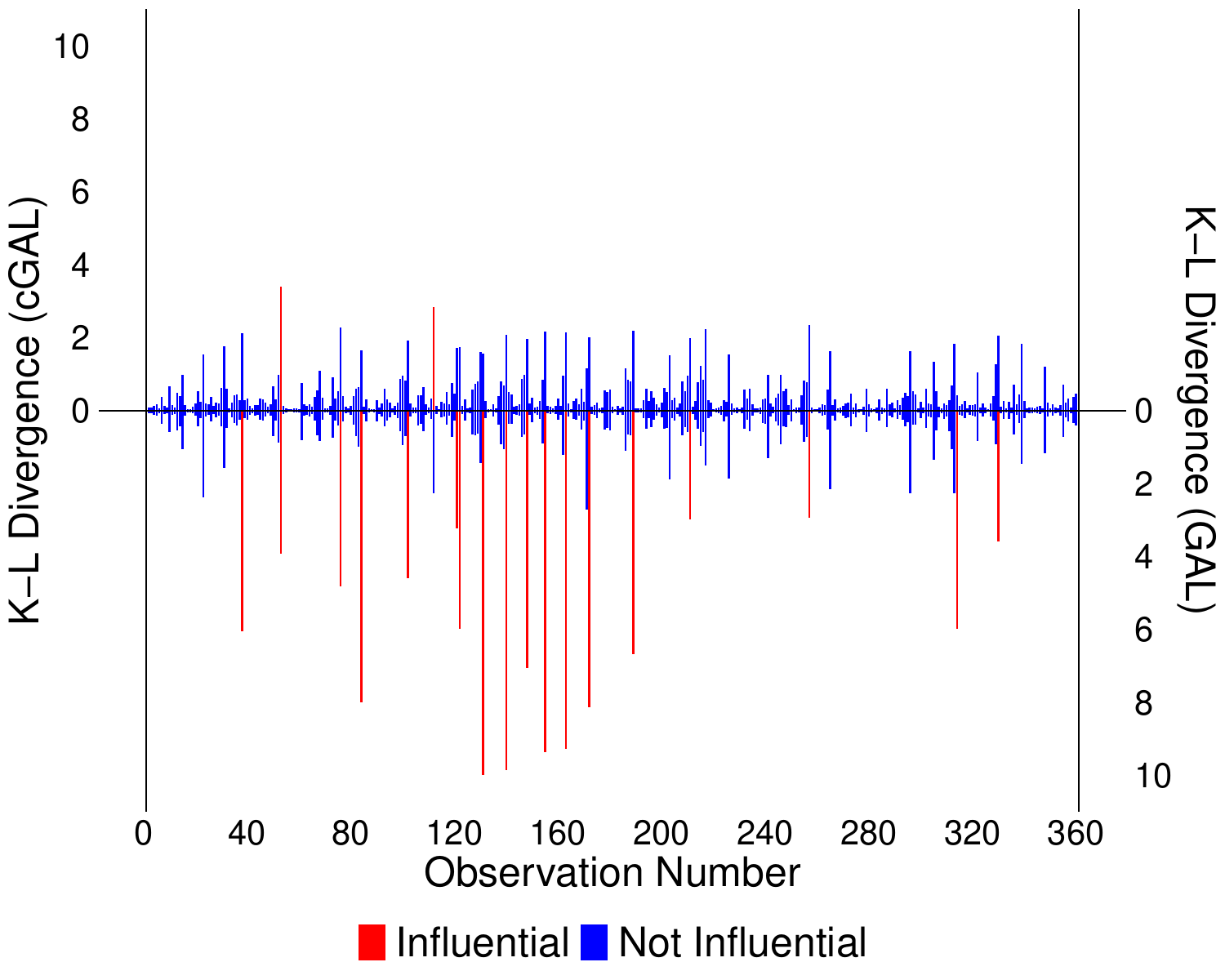}
            \caption{$p_0 = 0.75$}
        \end{subfigure}
        \\
        \begin{subfigure}{0.44\textwidth}
            \centering
            \includegraphics[width=\linewidth]{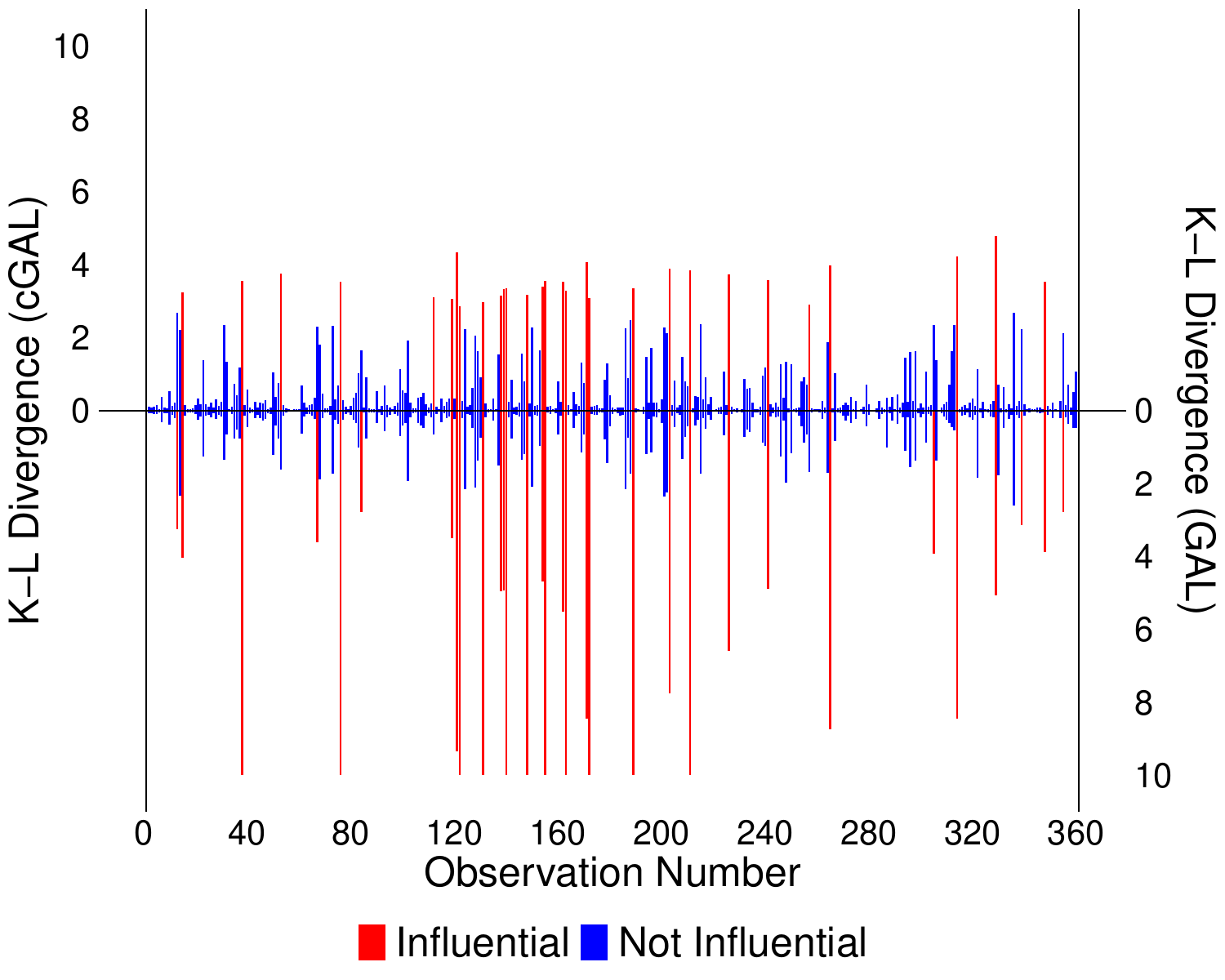}
            \caption{$p_0 = 0.95$}
        \end{subfigure}
        \caption{
            K-L divergence for the GAL and cGAL models across quantiles ($p_0 = 0.05, 0.25, 0.5, 0.75, 0.95$). The K-L divergence, capped at 10, is shown for all observations, with the GAL model on the bottom and the cGAL model on top. Influential observations (see \autoref{sec:KL_METHOD} for the definition) are highlighted in red, while non-influential ones are in blue. Dual y-axes represent the divergence for each model. Outliers appear to have a much larger impact under the GAL model compared to the cGAL model, indicating that the GAL model is more sensitive to extreme observations.
        }
        \label{fig:cGAL_GAL_QUANTILE_KL}
    \end{figure}

    \autoref{fig:cGAL_GAL_QUANTILE_KL}, Web \autoref{fig:cGAL_AL_QUANTILE_KL} (Supplementary Material), and Web \autoref{fig:GAL_AL_QUANTILE_KL} show pairwise comparisons of K-L divergence estimates among the models: \autoref{fig:cGAL_GAL_QUANTILE_KL} compares GAL vs. cGAL, Web \autoref{fig:cGAL_AL_QUANTILE_KL} compares AL vs. cGAL, and Web \autoref{fig:GAL_AL_QUANTILE_KL} compares AL vs. GAL. The results indicate that the cGAL consistently achieves the lowest divergence, demonstrating strong robustness to outliers across quantiles. While more resilient than AL, the GAL model still shows higher divergence values than cGAL, and AL exhibits the highest divergence overall, indicating its greater sensitivity to outliers.

    The \emph{p}-values from the standard DHARMa checks on scaled residuals---Kolmogorov--Smirnov for uniformity, a parametric dispersion test, and a binomial outlier test---are presented in \autoref{tab:LOO_COMPARE}. Generally, the models do not exhibit significant issues, with most \emph{p}-values remaining above the conventional 5\% threshold. However, for $p_0=0.75$ and $p_0=0.95$, the uniformity test indicates significant deviations, suggesting potential fit problems.

    \begin{table}[ht]
      \centering
      \begin{threeparttable}
        \caption{Comparison of LOOIC between AL, GAL, and cGAL models across quantiles. LOOIC is derived from full LOO cross-validation; it equals $-2\times$ the sum of the log-predictive densities of each observation given a model refit to the remaining data. Additionally, the table includes \emph{p}-values for the scaled residual diagnostics: a Kolmogorov-Smirnov test (uniformity), a dispersion test (under-/overdispersion), and a binomial test for outliers. In each test, higher \emph{p}-values (e.g., above 0.05) indicate that the data do not deviate substantially from model assumptions, suggesting a better fit.}
        \label{tab:LOO_COMPARE}
        
        \setlength{\tabcolsep}{7pt}
        \begin{tabular}{S[table-format=1.2]
                        l
                        S[table-format=1.3]
                        S[table-format=1.3]
                        r
                        S[table-format=4.3]}
          \toprule
          \textbf{Quantile}\tnote{a} &
          \textbf{Model}\tnote{b} &
          \textbf{Uniform}\tnote{c} &
          \textbf{Dispersion}\tnote{d} &
          \textbf{Outliers}\tnote{e} &
          \textbf{LOOIC}\tnote{f} \\
          \midrule
                      0.05 & AL & 0.052 & 0.264 & $>$0.999 & 2890.920 \\ 
                       & GAL & 0.100 & 0.220 & $>$0.999 & 749.032 \\ 
                       & cGAL & 0.076 & 0.754 & $>$0.999 & 643.851 \\ 
                      0.25 & AL & 0.161 & 0.251 & 0.165 & 624.925 \\ 
                       & GAL & 0.119 & 0.244 & $>$0.999 & 655.484 \\ 
                       & cGAL & 0.085 & 0.978 & $>$0.999 & 578.952 \\ 
                      0.5 & AL & 0.093 & 0.290 & 0.165 & 680.793 \\ 
                       & GAL & 0.107 & 0.228 & $>$0.999 & 672.699 \\ 
                       & cGAL & 0.058 & 0.820 & $>$0.999 & 601.527 \\ 
                      0.75 & AL & 0.012 & 0.474 & $>$0.999 & 1191.032 \\ 
                       & GAL & 0.056 & 0.288 & 0.165 & 818.534 \\ 
                       & cGAL & 0.029 & 0.835 & $>$0.999 & 640.377 \\ 
                      0.95 & AL & 0.029 & 0.550 & $>$0.999 & 6010.133 \\ 
                       & GAL & 0.026 & 0.412 & $>$0.999 & 1356.416 \\ 
                       & cGAL & 0.022 & 0.712 & $>$0.999 & 923.316 \\ 
          \bottomrule
        \end{tabular}
    
        \begin{tablenotes}
          \footnotesize
          \item[a] Quantile level being modeled.
          \item[b] AL (asymmetric Laplace), GAL (generalized asymmetric Laplace), cGAL (contaminated GAL).
          \item[c] $p$-value from the Kolmogorov-Smirnov test (uniformity).
          \item[d] $p$-value from the dispersion test (under-/overdispersion).
          \item[e] $p$-value from the binomial outlier test.
          \item[f] LOOIC (leave-one-out information criterion).
        \end{tablenotes}
    
      \end{threeparttable}
    \end{table}

    \autoref{tab:LOO_COMPARE} presents the LOO information criterion (LOOIC) comparison of the AL, GAL, and cGAL models across five quantiles. The cGAL model consistently shows the lowest LOOIC values, providing the best predictive performance and generalizing more effectively to unseen data. Between AL and GAL, the latter generally outperforms the former, as evidenced by lower LOOIC values in most cases. The enhanced performance of cGAL is consistent with its ability to handle outliers by introducing a second, inflated-scale component, making cGAL the strongest candidate across all studied quantiles.
 
    \subsection{Quantile fits}

\begin{figure}
        \centering
        \begin{subfigure}{0.49\textwidth}
            \centering
            \includegraphics[width=\linewidth]{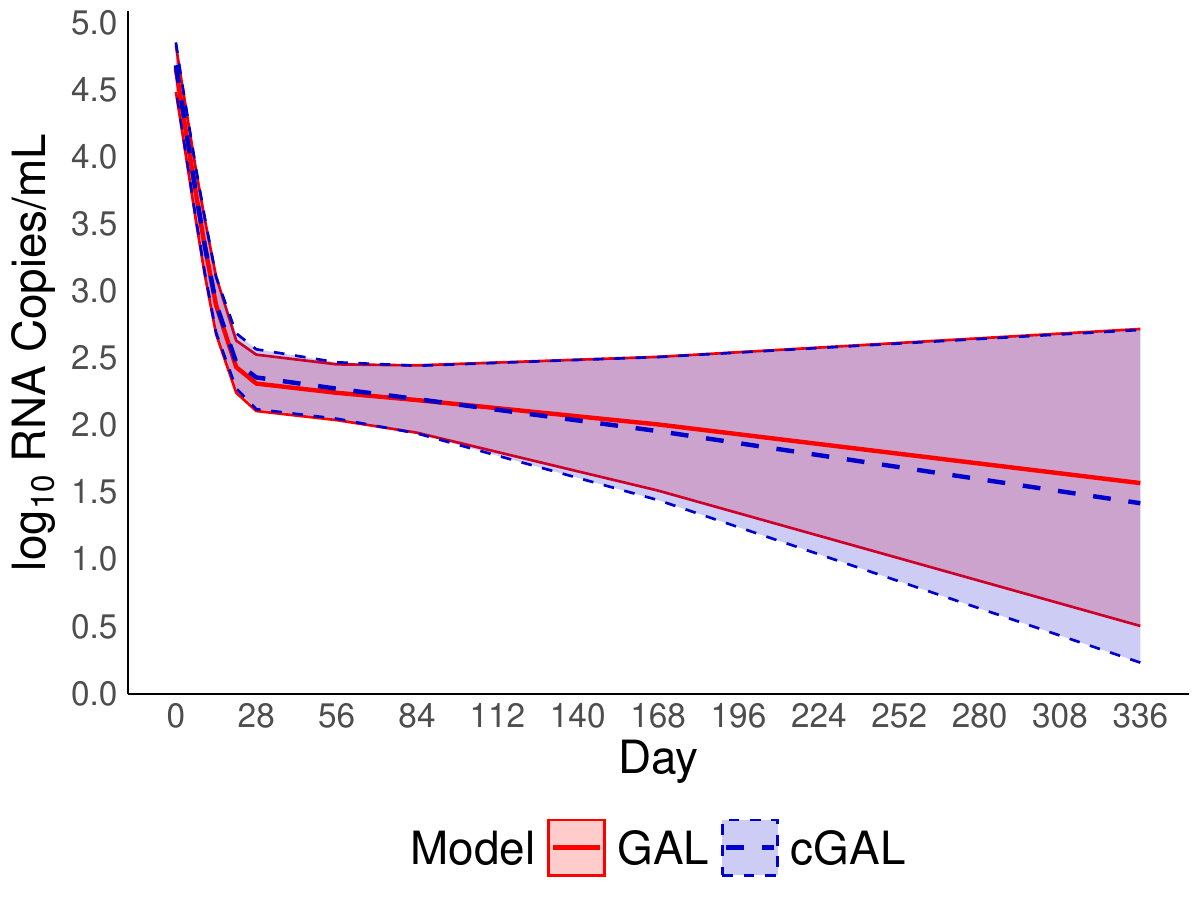}
            \caption{$p_0 = 0.05$}
        \end{subfigure}
        \begin{subfigure}{0.49\textwidth}
            \centering
            \includegraphics[width=\linewidth]{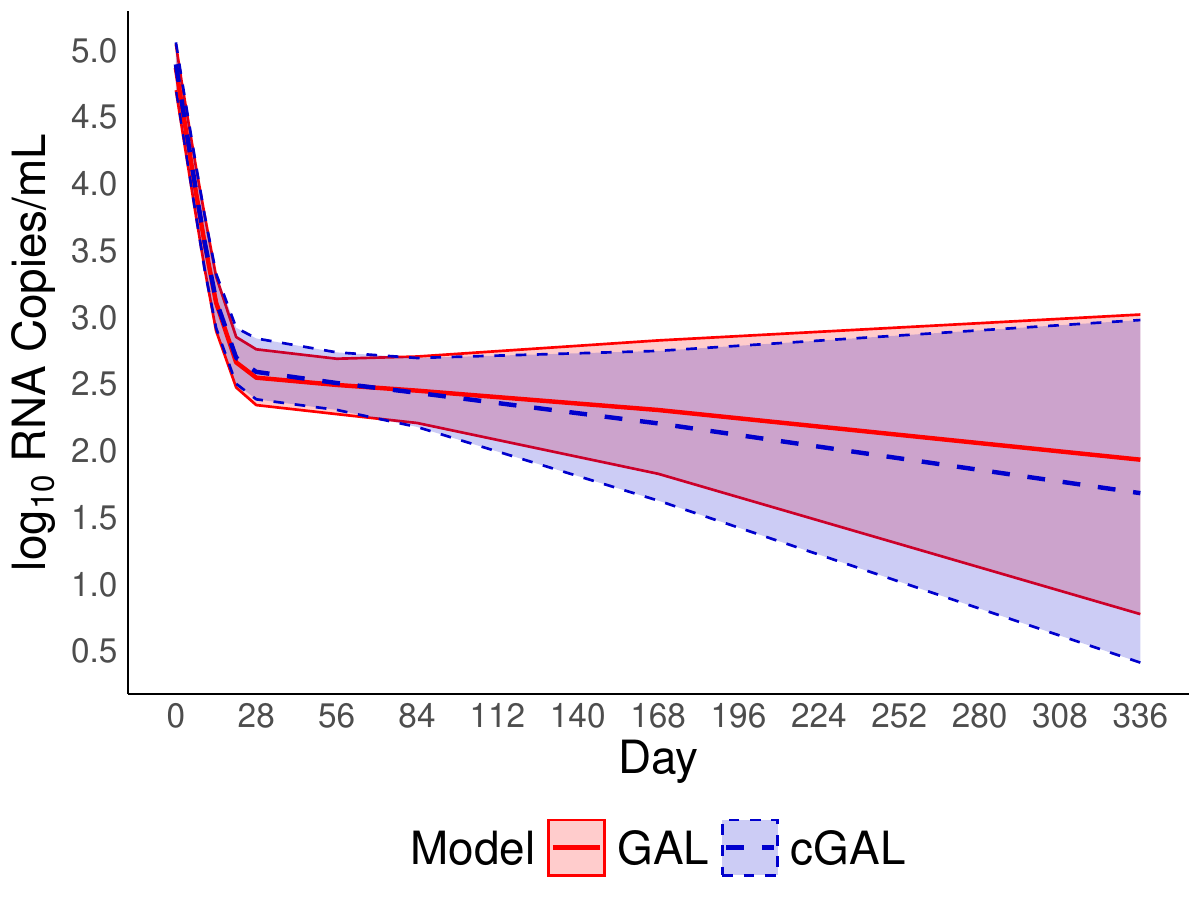}
            \caption{$p_0 = 0.25$}
        \end{subfigure}
        \\
        \begin{subfigure}{0.49\textwidth}
            \centering
            \includegraphics[width=\linewidth]{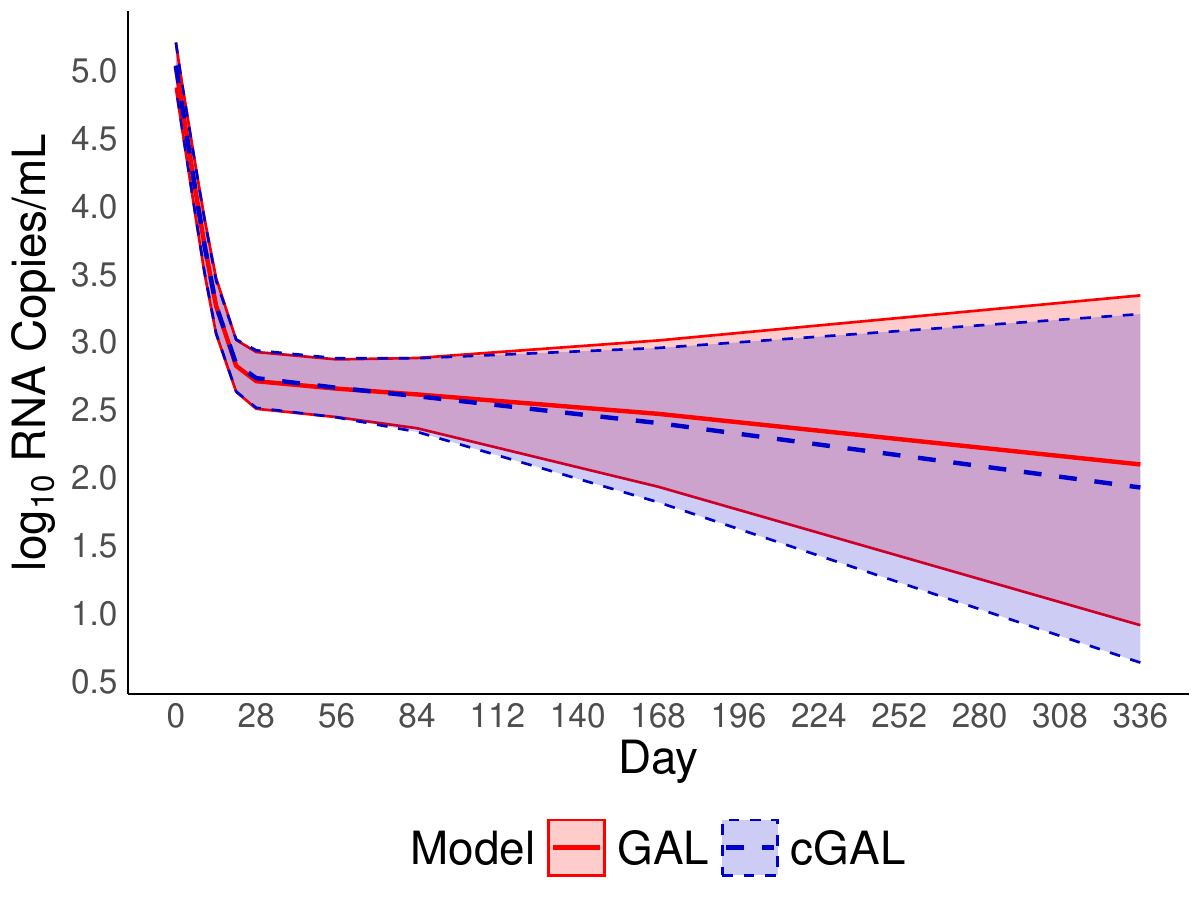}
            \caption{$p_0 = 0.5$}
        \end{subfigure}
        \begin{subfigure}{0.49\textwidth}
            \centering
            \includegraphics[width=\linewidth]{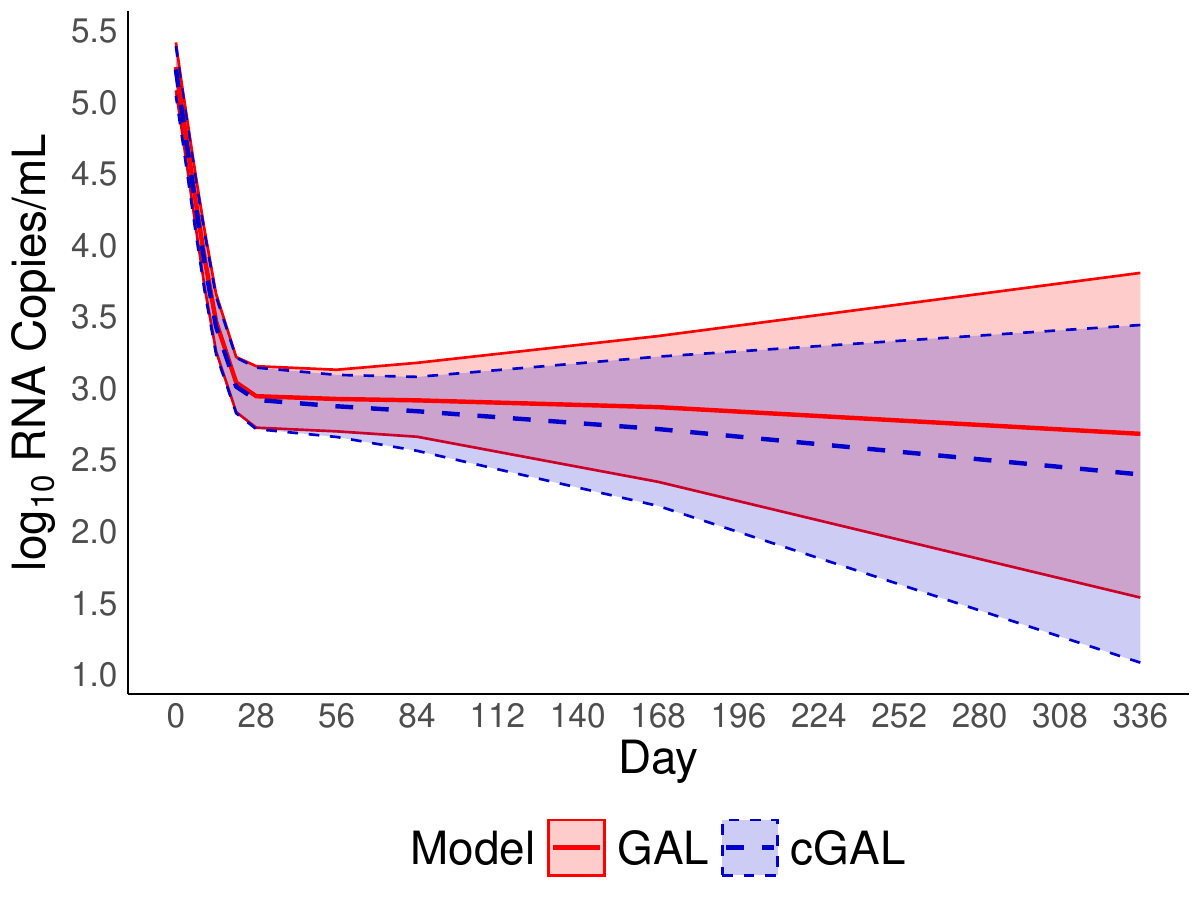}
            \caption{$p_0 = 0.75$}
        \end{subfigure}
        \\
        \begin{subfigure}{0.49\textwidth}
            \centering
            \includegraphics[width=\linewidth]{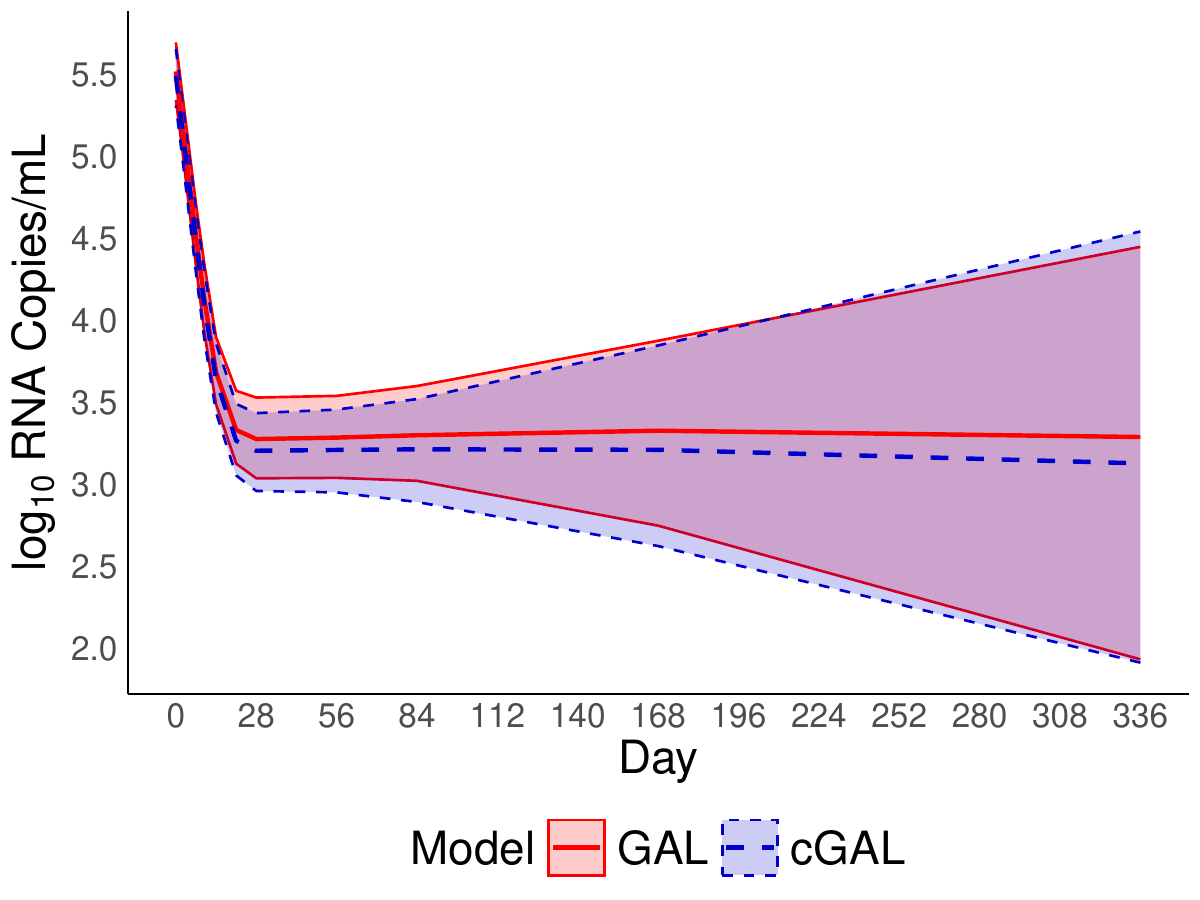}
            \caption{$p_0 = 0.95$}
        \end{subfigure}
        \caption{Comparison of quantile profiles for the GAL and cGAL models at five quantiles ($p_0 = 0.05, 0.25, 0.5, 0.75, 0.95$), depicting HIV viral load dynamics over time in the ACTG~315 study. The profiles illustrate the different decay trajectories across the two models, with the GAL model represented by the red solid lines and the cGAL model by the blue dashed lines. The shaded areas represent the 95\% HPD intervals around each profile.} \label{fig:CGALD_GALD_QUANTILE_PROFILES}
    \end{figure}

    \autoref{fig:CGALD_GALD_QUANTILE_PROFILES} presents the posterior estimates and 95\% HPD intervals for the GAL and cGAL models across five quantiles, modeling HIV viral load dynamics over time. In the Supplementary Material, Web \autoref{fig:CGALD_ALD_QUANTILE_PROFILES} provides the same quantile profiles for AL vs. cGAL. Web \autoref{fig:GALD_ALD_QUANTILE_PROFILES} compares AL vs. GAL.

    {\color{red} Figure~6} shows posterior estimates and 95\% HPD intervals for $\beta_1$ to $\beta_5$ across five quantiles, while Web \autoref{tab:POST_SUMMARY} summarizes parameter estimates and 95\% HPD intervals for all models. Web \autoref{tab:POST_SUMMARY} also reports results under markedly flatter priors; these are discussed in \autoref{sec:SENS}. From {\color{red} Figure~6}, we observe the following:
    \begin{itemize}
        \item As expected, $\beta_1$ and $\beta_3$ shift upward at higher quantiles since they capture higher intercepts of viral load for more extreme subpopulations.
        \item No major differences emerged among the models for $\beta_2$ and $\beta_5$, indicating that the short-term decay and CD4 effect were relatively stable across quantiles and robust to outliers.
        \item Estimates for $\beta_4$ suggest that the cGAL model generally predicts a faster second-phase decay than do the GAL and AL models, as also illustrated in \autoref{fig:CGALD_GALD_QUANTILE_PROFILES}. This can be attributed to the cGAL's robustness to outliers, particularly in the second phase, where the presence of irregular viral load trajectories influences the GAL and AL models, pulling their estimates upward and slowing the decay (see \autoref{fig:ACTG315_RNA}). In contrast, the cGAL model is less affected by these outliers, resulting in a more accurate representation of long-term viral load dynamics.
    \end{itemize}

    \afterpage{
        \begin{landscape}
            \begin{figure}
                \centering
                \includegraphics[width=1\linewidth, height=0.6\linewidth]{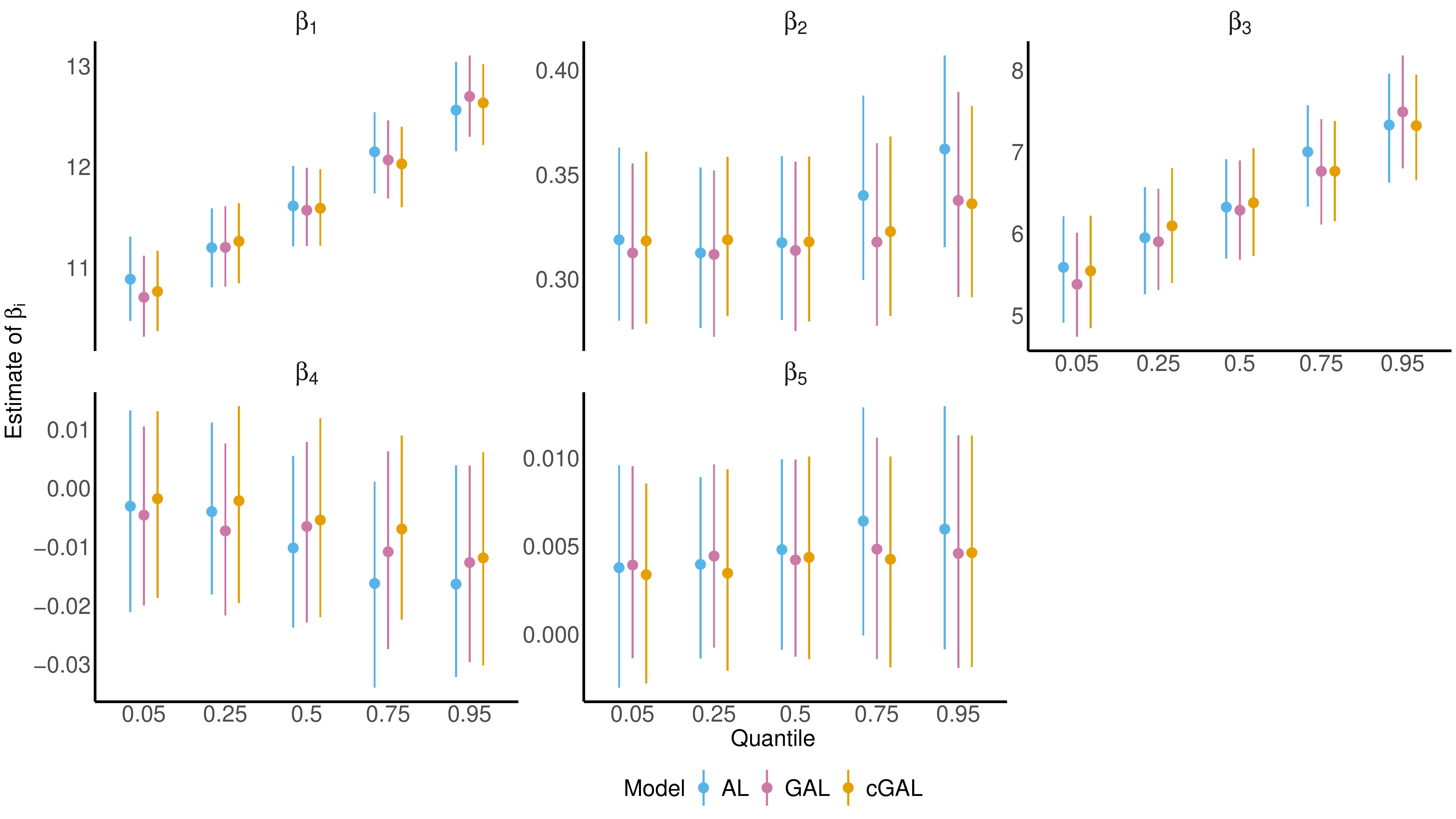}
                \caption{Posterior estimates and 95\% HPD intervals for the fixed effect parameters across quantiles $p_0 = 0.05, 0.25, 0.5, 0.75, 0.95$. The estimates are shown for three models: AL (blue), GAL (purple), and cGAL (orange) applied within an NLME framework to the HIV viral load in the ACTG~315 dataset. Here, $\beta_1$ and $\beta_3$ represent the initial viral load components for the short- and long-term decay phases, respectively. The short-term decay rate is captured by $\beta_2$. The long-term decay rate is represented by $\beta_4$ and incorporates CD4 as a covariate, with $\beta_5$ as its regression coefficient.}
                \label{fig:BETA_COMPARISON}
            \end{figure}        
        \end{landscape}
    }

    The parameter estimates align with our model comparison results, confirming that the cGAL's robust handling of outliers yields steeper long-term decay rates.

    \subsection{Sensitivity analyses} \label{sec:SENS}

    To confirm that our conclusions are not driven by the baseline priors, we re-estimated the cGAL model with a markedly flatter specification,
    \begin{equation}
      \sigma \sim t_{+}\left(0,100,1\right), \qquad
      B \sim \mathrm{Beta}\left(\tfrac12,\tfrac12\right), \qquad
      \alpha \sim \mathrm{Beta}\left(\tfrac12,\tfrac12\right).
    \end{equation}
    The half-Cauchy, denoted by $t_{+}\left(0,100,1\right)$, contributes virtually no information about the scale~$\sigma$. Jeffreys' $\mathrm{Beta}\!\left(\tfrac12,\tfrac12\right)$ places extra mass near the endpoints of the admissible intervals for $B$ (and, by transformation,~$\gamma$) and~$\alpha$ \citep{jeffreys1998theory}, letting the data rather than the prior determine where those parameters lie.

    We refer to the sensitivity-prior variant as $\text{cGAL}_{\text{sens}}$. Web \autoref{tab:POST_SUMMARY} shows that parameter estimates and 95\% HPD intervals under these flatter priors are almost unchanged from the baseline fit, confirming that our substantive conclusions are robust to prior choice.

    We re-estimated the model with $\tau_{0}\in\left\{6.\dot{6}\dot{7},20,40\right\}$. Web \autoref{tab:POST_SUMMARY_CONTAM} shows that all fixed-effect parameter estimates and their 95\% HPD intervals are very similar to the baseline fit ($\tau_{0}=10$), confirming that our results are not sensitive to the particular choice of $\tau_{0}$.

    \section{Simulation study} \label{sec:SIMULATION}

    \subsection{Model performance} \label{sec:PERFORM_CGAL}
    
    A simulation study was conducted to assess the performance of the cGAL model within the proposed NLME quantile regression framework in \autoref{sec:BIPHASIC_MODEL}. Data were generated under a two-phase viral load model for $N = 15$ subjects, each observed at $J_i = 9$ time points. Specifically, the model assumed:
    \begin{equation}
       y_{ij} =
       \log_{10}\left(
           \exp\left(\beta_1 + b_{i1} - \left(\beta_2 + b_{i2}\right) t_{ij}\right)
           +
           \exp\left(\beta_3 - \beta_4 t_{ij}\right)
   \right),
    \end{equation}
    where $y_{ij}$ denotes the outcome at the $j$\textsuperscript{th} measurement for cluster $i$, and $t_{ij} = j-1$ so that $t_{ij}$ ranges from 0 to 8. The parameters $\beta_1$ and $\beta_2$ were the main focus (representing the short-term viral load component), while $\beta_3$ and $\beta_4$ were fixed at their true values and were not estimated.
    
    Random effects $\bm b_i = \left(b_{i1}, b_{i2}\right)^\top$ were drawn from a bivariate normal distribution with zero mean and covariance matrix:
    \begin{equation}
       \bm{\Sigma} =
       \begin{pmatrix}
           0.95^2 & 0.05 \times 0.95 \times 0.95 \\[6pt]
           0.05 \times 0.95 \times 0.95 & 0.95^2
       \end{pmatrix}.
    \end{equation}
    The parameter values were set to $\beta_1 = 11.5$, $\beta_2 = 5.5$, $\beta_3 = 3.5$, $\beta_4 = 0.05$, $\sigma = 0.2$, and $\tau_0 = 10$. A shape parameter $\gamma = -0.3$ governed skewness.

    We report
    \begin{equation}
        \bm{\Sigma}^{-1}
        =
        \begin{pmatrix}
            \omega_{11} & \omega_{12} \\
            \omega_{12} & \omega_{22}
        \end{pmatrix}
    \end{equation}
    here to remain consistent with the approach used in the application section (see \autoref{sec:IMPLEMENT}).

    We assume a $\text{Uniform}\left(0,1\right)$ prior for $\alpha$, a more noninformative choice than the beta-distributed prior used in the application, which places greater mass on smaller values of $\alpha$.
    
    Four scenarios were created by varying the contamination proportion $\alpha$ (either $0.001$ or $0.05$) and the target quantile $p_0$ (either $0.50$ or $0.85$). In each scenario, $300$ datasets were generated. Model fitting and parameter inference were performed using the same NLME framework but substituting the appropriate likelihood (AL, GAL, or cGAL). Primary outcomes included bias and coverage probabilities of 95\% HPD intervals.
    
    The model performance results are presented in Web \autoref{tab:CGAL_PERFORM} of the Supplementary Material. Coverage probabilities for the model parameters range from acceptable (meeting nominal targets) to conservative, with none being overly lenient. However, higher contamination levels lead to increased bias in $\beta_2$, and the variance components also exhibit substantial bias, consistent with the findings by \citet{BURGER2020} for the covariance matrix under the MGH-$t$ prior distribution.

    \subsection{Robustness under data contamination}

    We conducted a data contamination simulation study to evaluate the robustness of the GAL model relative to the cGAL model, using the same scenarios described in \autoref{sec:PERFORM_CGAL} while sampling from the cGAL model. We report bias, coverage probabilities of 95\% HPD intervals, root mean square error (RMSE), and HPD interval width for $\beta_{1}$ and $\beta_{2}$.
    
    The results of the data contamination study are presented in Web \autoref{tab:CONTAMINATION}. Under minimal contamination ($\alpha=0.1\%$), both GAL and cGAL perform similarly. However, when 5\% contamination is introduced ($\alpha = 5\%$), both the RMSE and HPD interval length for $\beta_{2}$ are substantially larger for the GAL model than for the cGAL model. These findings suggest that the cGAL model is more robust to outliers than the GAL model.

    \section{Discussion} \label{sec:DISCUSSION}

    In this paper, we developed a robust mixed-effects quantile regression framework for modeling HIV viral load decay, comparing the AL, GAL, and cGAL distributions in an NLME setting. Our results indicate that the cGAL distribution, through its improved treatment of outliers in viral load data, more reliably captures the dynamics of HIV progression.

    An important practical consideration is that the cGAL model, while more robust to outliers than its AL and GAL counterparts, is computationally more demanding. Unlike AL and GAL, which have convenient mixture representations facilitating sampling, the cGAL distribution requires a custom likelihood formulation (via the ``ones trick" in \verb|JAGS|), substantially increasing runtime. This trade-off may be critical in large-scale studies or real-time clinical settings where computational efficiency is paramount. Nonetheless, the computational overhead may be justified when data contains influential outliers, given the cGAL's superior robustness and improved predictive performance in our results.
    
    We introduced $\tau_0$ as a scale inflation factor for the contaminated component of the cGAL distribution, fixing it to a relatively large value (in our case, $\tau_0=10$) to ensure sufficient coverage for potential heavy-tailed behavior. Although choosing a fixed value can work well in practice, especially if we suspect pronounced outliers, estimating $\tau_0$ from the data could present greater flexibility. In such an extended model, both the contamination proportion $\alpha$ and scale factor $\tau_0$ would be unknown parameters. This may present identifiability challenges, such as distinguishing whether extreme observations arise from a heavily scaled component or a small fraction of contamination. However, it arguably provides a clearer perspective on outlier behavior.

    Although we treated observed viral load data in a straightforward manner, future research may account for left-censored observations below the limit of quantification (LLOQ), which is common in HIV studies. In principle, the NLME model could handle this by treating sub-threshold observations as censored. However, implementing censoring under the GAL or cGAL distributions is nontrivial due to the lack of closed-form CDFs. A specialized sampling scheme or additional numerical integration might be required to accommodate censoring within the robust quantile framework. Similarly, incorporating measurement error in the CD4 covariate could refine the modeling of its effect on long-term viral decay \citep{wu2002joint}. Presently, we assume that CD4 was measured without error, but realistic clinical data often include substantial variability. Approaches such as Bayesian measurement-error models or joint modeling of CD4 and viral load could better characterize how immune status covaries with viral trajectory.

    Across the two contamination levels and target quantiles considered, the cGAL model's coverage often exceeds the 95\% target, indicating a conservative stance rather than an overly lenient one. The data contamination study further shows that the GAL is considerably affected by moderate contamination compared to the cGAL, particularly regarding RMSE and HPD interval width.

    In summary, the cGAL-based NLME quantile model constitutes a robust alternative to classical AL or GAL approaches by eliminating the need for explicit outlier identification and deletion. It exhibits stronger resistance to the effects of outliers while yielding more reliable model parameter estimates.
    
    \section*{Appendix}
    
    \begin{appendices}
    
    \section{L-statistic-based kurtosis} \label{sec:L_STAT_KUR}

    The L-statistic-based kurtosis is computed separately for a sorted sample's left and right tails to capture the distribution's tail behavior around the median. Given an ordered sample $\left\{X_{\left(1\right)}, X_{\left(2\right)}, \dots, X_{\left(n\right)} \right\}$, the sample median $F^{-1}_n\left(0.5\right)$ is defined as $X_{\left(\frac{n}{2}\right)}$ for an even sample size $n$. The left-side kurtosis $\phi_{n,L}$ is then calculated as:
    \begin{equation}
        \phi_{n,L} = \frac{\frac{2}{n} \sum_{i=1}^{n/2} \left(\frac{4i - 2}{n} - 1\right) X_{\left(i\right)}}{F^{-1}_n\left(0.5\right) - \frac{2}{n} \sum_{i=1}^{n/2} X_{\left(i\right)}}.
    \end{equation}    
    Similarly, the right-side kurtosis $\phi_{n,D}$ is computed as:    
    \begin{equation}
        \phi_{n,D} = \frac{\frac{2}{n} \sum_{i=n/2+1}^{n} \left(\frac{4i - 2}{n} - 3\right) X_{\left(i\right)}}{\frac{2}{n} \sum_{i=n/2+1}^{n} X_{\left(i\right)} - F^{-1}_n\left(0.5\right)}.
    \end{equation}

    \section{Kullback-Leibler divergence} \label{sec:KL_METHOD}
    
    To evaluate the influence of individual data points on the model, we use LOO to measure model adequacy following the approach detailed by \citet{WANG2016}. Specifically, we compute the K-L divergence, quantifying the difference between model fits when a particular observation is included versus excluded.

    Let $\bm{\eta}$ denote the collection of parameters. For the AL model, $\bm{\eta} = \left(\bm{\beta}^\top, \sigma\right)^\top$ includes the fixed effects vector $\bm{\beta}$ and scale parameter $\sigma$. For the GAL model, $\bm{\eta} = \left(\bm{\beta}^\top, \sigma, \gamma\right)^\top$, where $\gamma$ is the skewness parameter. In the cGAL model, $\bm{\eta} = \left(\bm{\beta}^\top, \sigma, \gamma, \alpha\right)^\top$, where $\alpha$ is the contamination parameter. Let $\bm{\Theta}$ represent the complete set of model parameters, including both fixed and random effects.
    
    Let $\bm{\eta}^{\left(k\right)}$ and $\bm{b}_{i}^{\left(k\right)}$ represent the $k$\textsuperscript{th} posterior sample of parameters $\bm{\eta}$ and random effects $\bm{b}_i$, respectively, for $k = 1, \ldots, K$. The vector $\bm{y}$ consists of all observations $y_{ij}$ for subjects $i = 1, \ldots, N$ and measurements $j = 1, \ldots, J_i$. The posterior distribution of the parameters $\bm{\Theta}$ given the full dataset is denoted by $P\left(\bm{\Theta} \left| \bm{y}\right.\right)$, while $P\left(\bm{\Theta} \left| \bm{y}_{\left[ij\right]}\right.\right)$ refers to the posterior distribution of $\bm{\Theta}$ after element-wise exclusion of the single observation $y_{ij}$. The Monte Carlo estimate of the K-L divergence between the posterior distributions under regression model $R$ is computed as:
    \begin{flalign}
        \text{KL}_R \left(P\left(\bm{\Theta} \left|\bm{y}\right.\right), P\left(\bm{\Theta} \left|\bm{y}_{\left[ij\right]}\right.\right)\right) & = \log \left\{\frac{1}{K} \sum_{k=1}^{K} \left[P\left(y_{ij} \left|\bm{\eta}^{\left(k\right)}, \bm{b}_i^{\left(k\right)} \right.\right)\right]^{-1} \right\} \\ \nonumber
        & + \frac{1}{K} \sum_{k=1}^{K} \log \left[P\left(y_{ij} \left|\bm{\eta}^{\left(k\right)}, \bm{b}_i^{\left(k\right)} \right.\right)\right].
    \end{flalign}
    This K-L divergence is calculated for each observation $y_{ij}$ to evaluate its potential influence under regression model $R$, identifying observations that may substantially impact the parameter estimates. Following \citet{TOMAZELLA2021}, we flag observation $y_{ij}$ as influential (``outlier") if it meets the following criterion:
    \begin{equation}
        0.5 \left(1 + \sqrt{1 - \exp \left[-2\text{KL}_R \left(P\left(\bm{\Theta} \left|\bm{y}\right.\right), P\left(\bm{\Theta} \left|\bm{y}_{\left[ij\right]}\right.\right)\right)\right]}\right) \geq 0.999.
    \end{equation}
    
    \end{appendices}

    \section*{Supplementary information}

    Supplementary material for this paper is available online. Example code for the model application is available on \href{https://github.com/DABURGER1/Robust-Quantile-cGAL}{Robust-Quantile-cGAL} on GitHub.

    \section*{Acknowledgements}
    
    This work is based on the research supported by the National Research Foundation (NRF) of South Africa (Grant number 132383). Opinions expressed and conclusions arrived at are those of the authors and are not necessarily to be attributed to the NRF.
    
    \section*{Declaration of conflicting interests}
    
    The authors declare no conflict of interest.

    \bibliographystyle{plainnat}

    \newpage
    \clearpage
    
    \setcounter{page}{1}
    \setcounter{figure}{0}
    \setcounter{table}{0}
    \renewcommand{\restoreapp}{}
    \renewcommand\appendixname{Web Appendix}
    \renewcommand{\figurename}{WEB FIGURE}
    \renewcommand{\tablename}{WEB TABLE}
    \renewcommand{\headrulewidth}{0pt}
    \titleformat{\section}{\large\bfseries}{\appendixname~\thesection .}{0.5em}{}
    
    \begin{appendices}
        
        {\Large\bf A robust mixed-effects quantile regression model using generalized Laplace mixtures to handle outliers and skewness}
        
        \vskip 1.0cm
        
        {\normalsize Divan A. Burger, Sean van der Merwe, and Emmanuel Lesaffre}
        
        \vskip 4.5truecm
        
        \begin{center}
            \noindent
            {\Large\bf Supporting Information}
        \end{center}

        \newpage
        \clearpage
    
        \begin{figure}
            \centering
            \begin{subfigure}{0.44\textwidth}
                \centering
                \includegraphics[width=\linewidth]{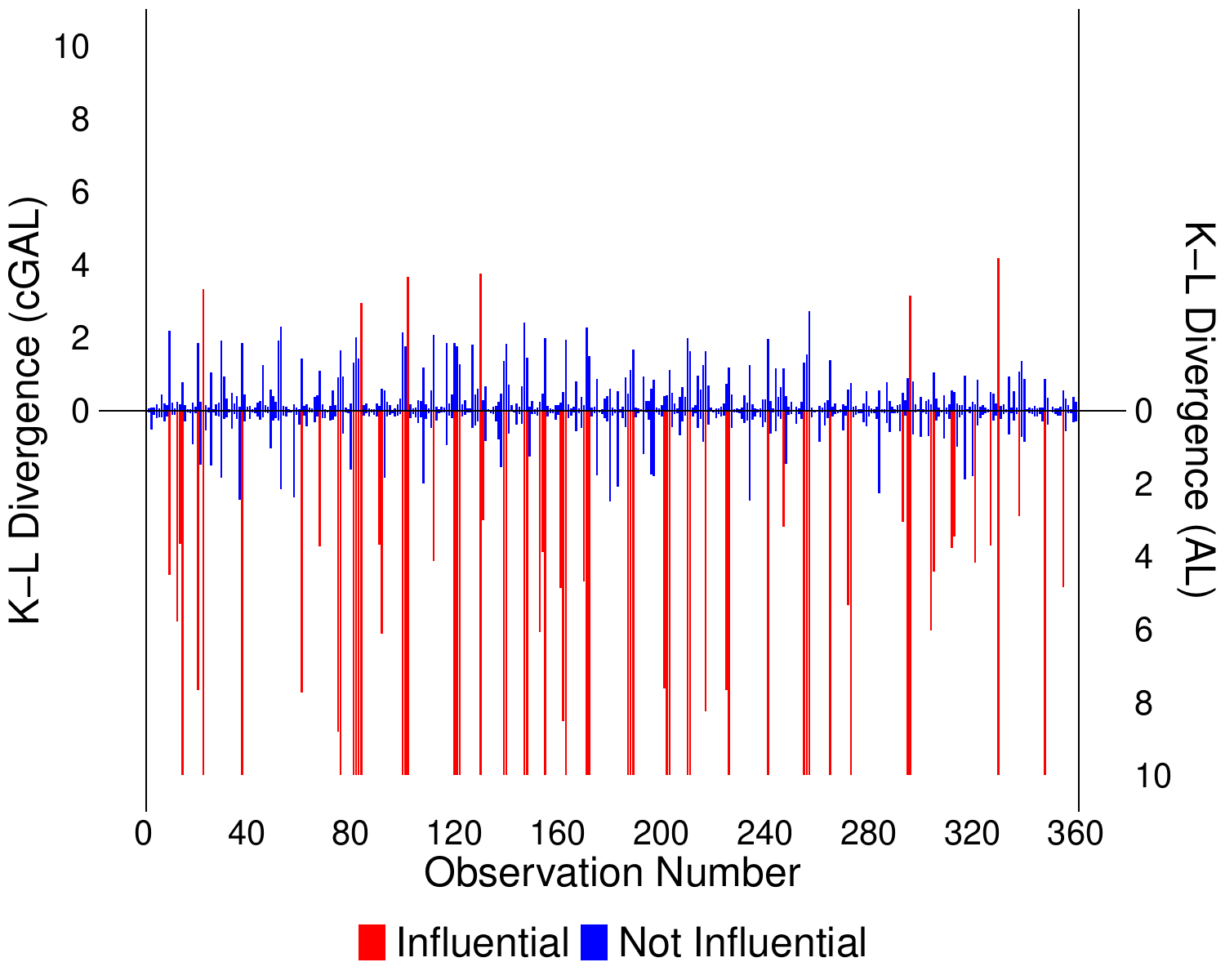}
                \caption{$p_0 = 0.05$}
            \end{subfigure}
            \begin{subfigure}{0.44\textwidth}
                \centering
                \includegraphics[width=\linewidth]{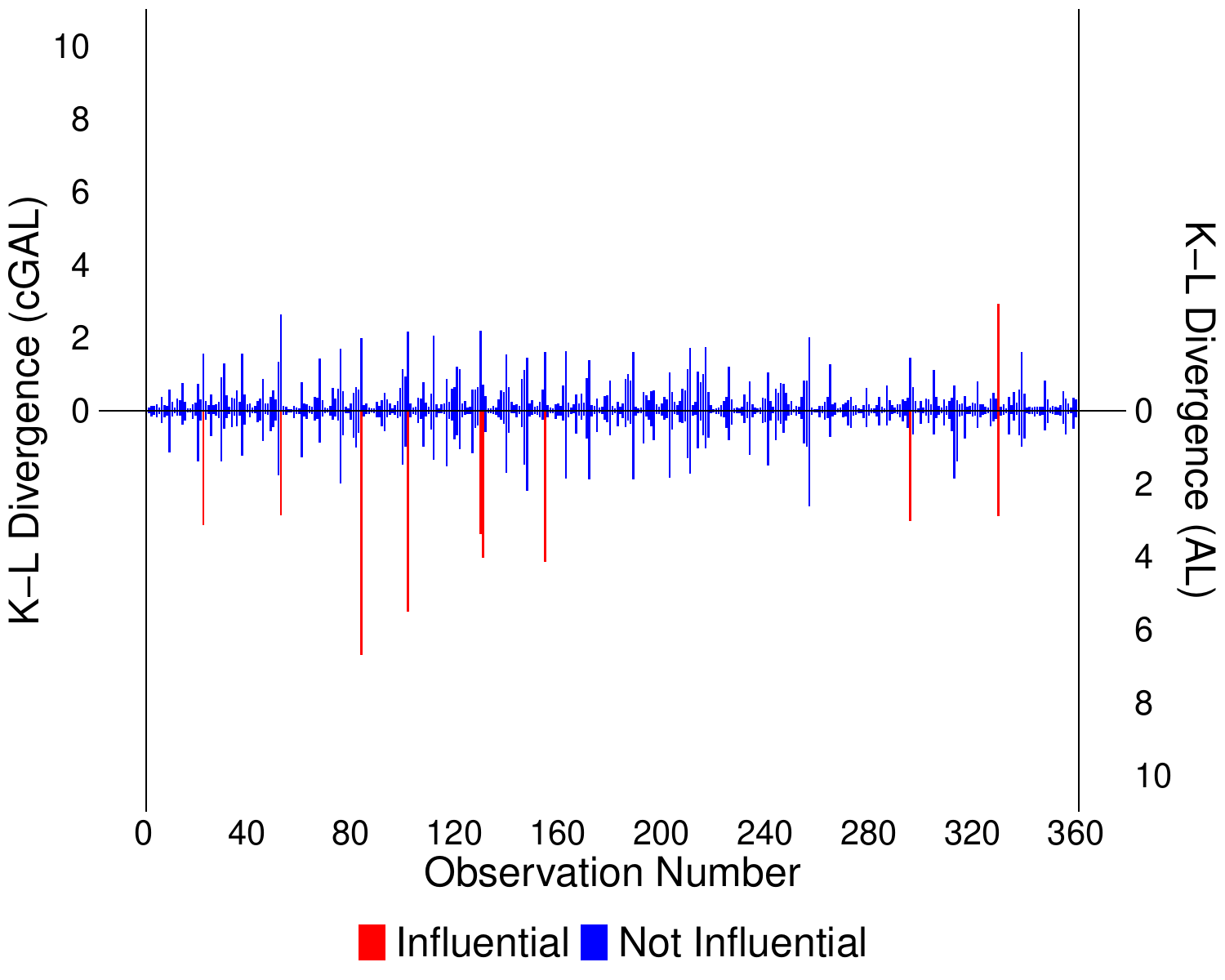}
                \caption{$p_0 = 0.25$}
            \end{subfigure}
            \\
            \begin{subfigure}{0.44\textwidth}
                \centering
                \includegraphics[width=\linewidth]{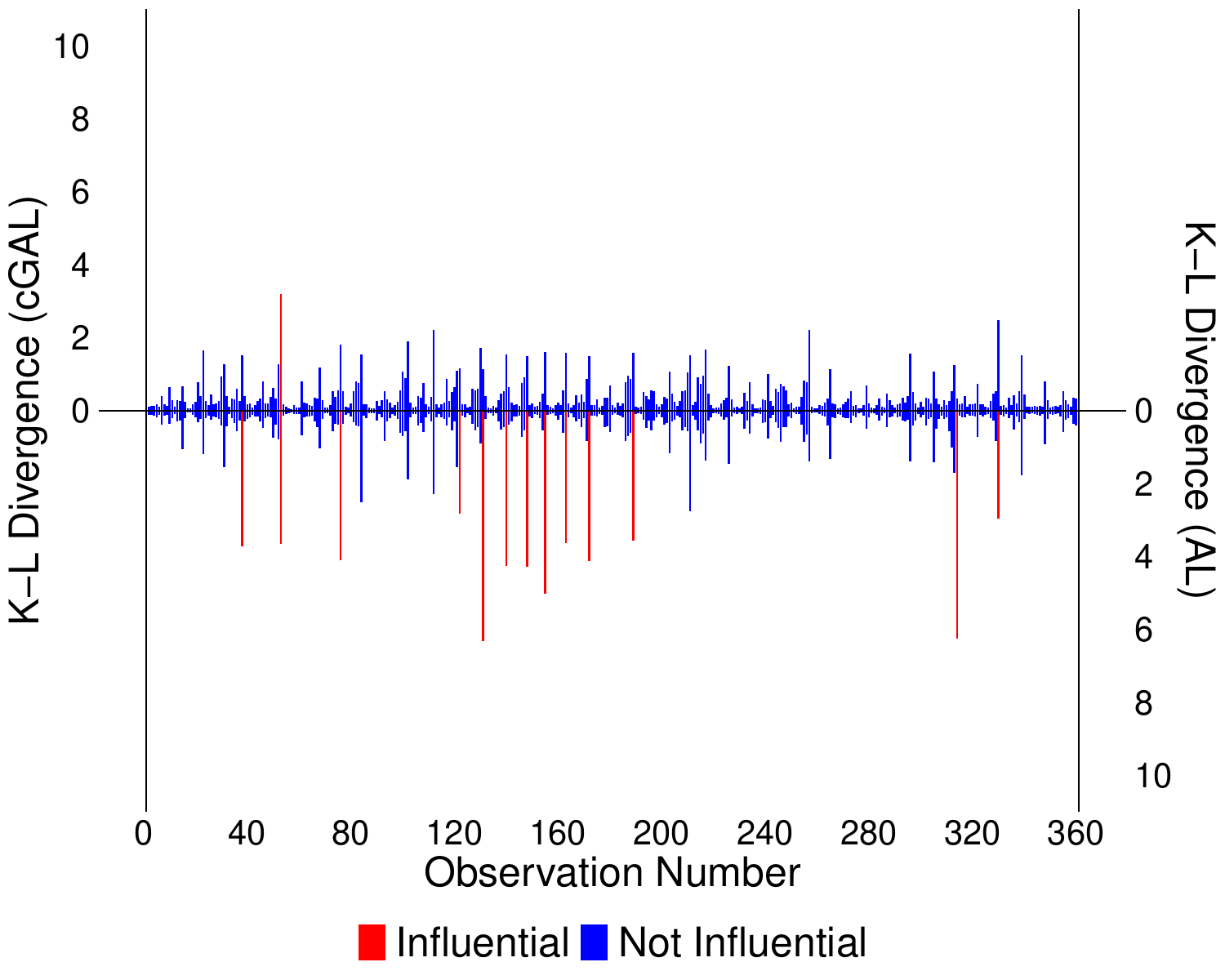}
                \caption{$p_0 = 0.5$}
            \end{subfigure}
            \begin{subfigure}{0.44\textwidth}
                \centering
                \includegraphics[width=\linewidth]{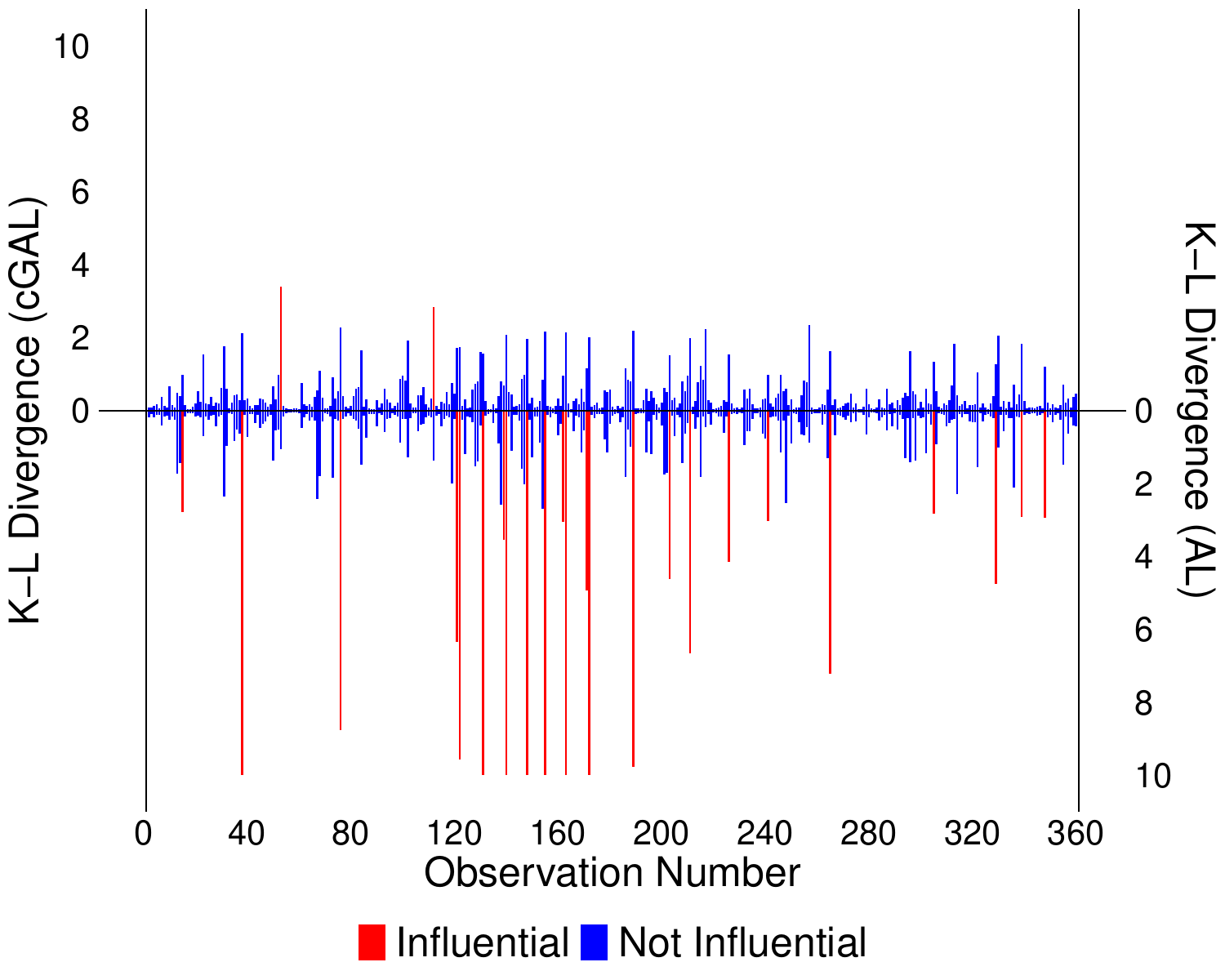}
                \caption{$p_0 = 0.75$}
            \end{subfigure}
            \\
            \begin{subfigure}{0.44\textwidth}
                \centering
                \includegraphics[width=\linewidth]{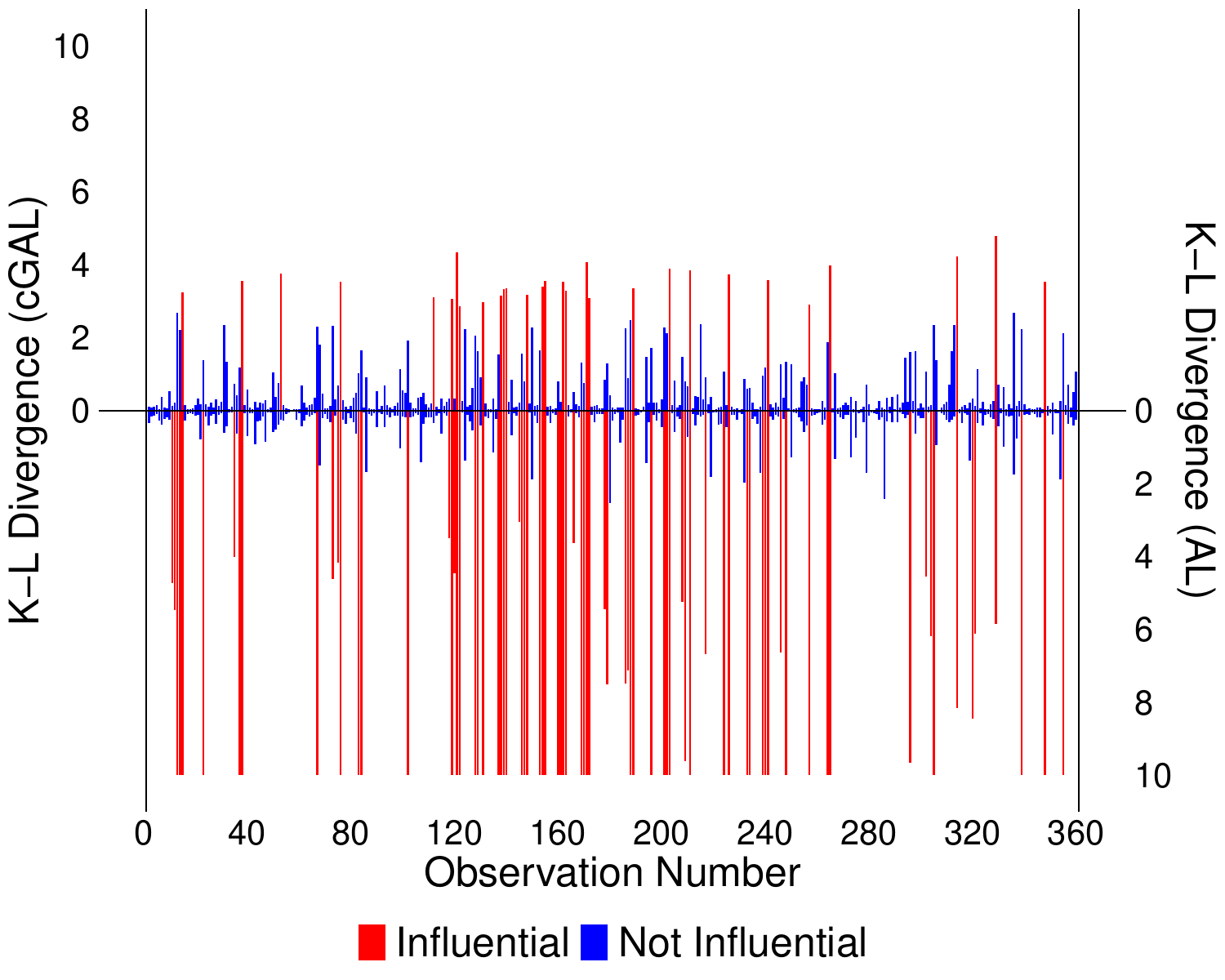}
                \caption{$p_0 = 0.95$}
            \end{subfigure}
            \caption{
                K-L divergence for the AL and cGAL models across quantiles ($p_0 = 0.05, 0.25, 0.5, 0.75, 0.95$). The K-L divergence, capped at 10, is shown for all observations, with the AL model on the bottom and the cGAL model on top. Influential observations (see \autoref{sec:KL_METHOD} for the definition) are highlighted in red, while non-influential ones are in blue. Dual y-axes represent the divergence for each model. Outliers appear to have a much larger impact under the AL model compared to the cGAL model, indicating that the AL model is more sensitive to extreme observations.
            }
            \label{fig:cGAL_AL_QUANTILE_KL}
        \end{figure}

        \newpage
        \clearpage
    
        \begin{figure}
            \centering
            \begin{subfigure}{0.44\textwidth}
                \centering
                \includegraphics[width=\linewidth]{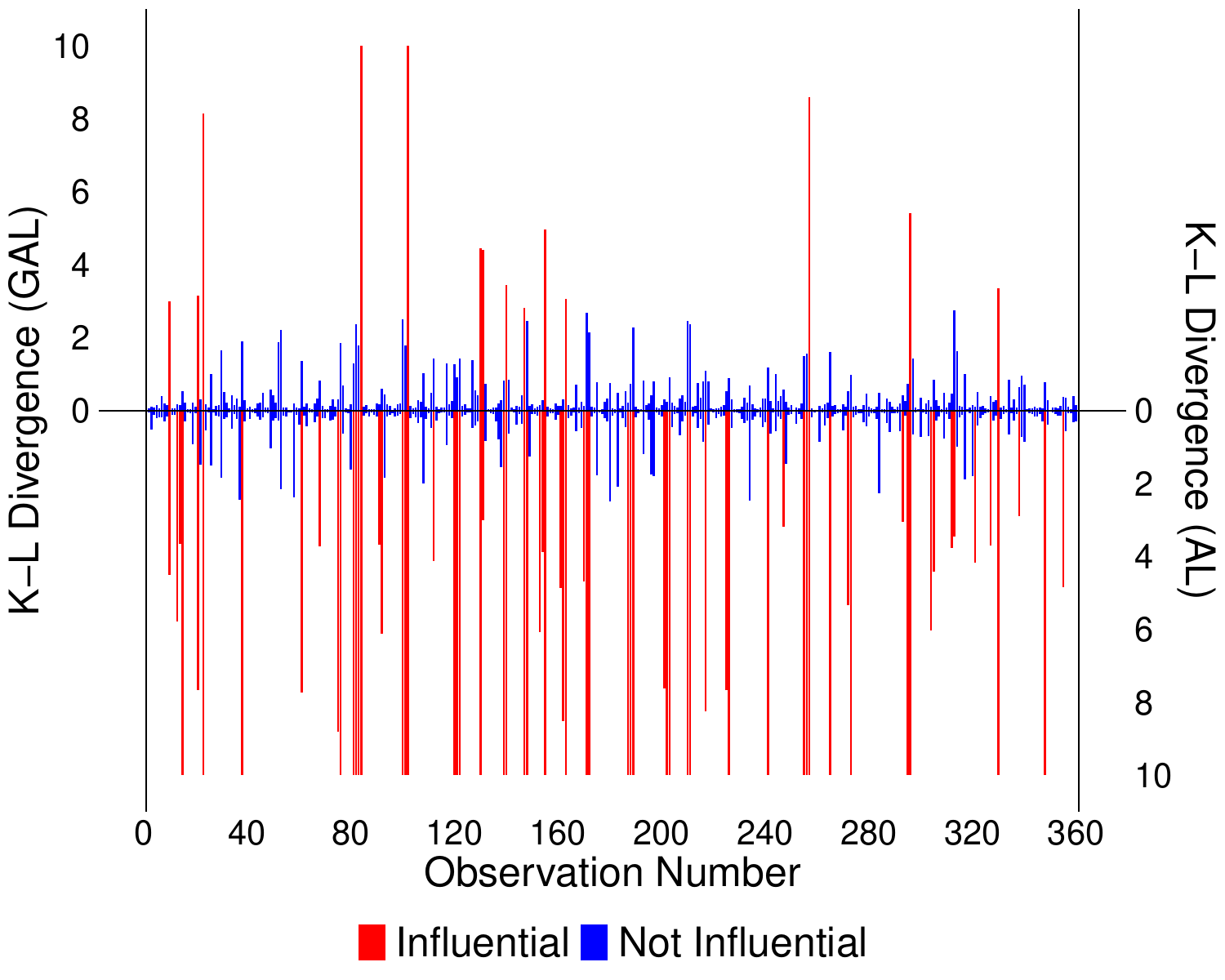}
                \caption{$p_0 = 0.05$}
            \end{subfigure}
            \begin{subfigure}{0.44\textwidth}
                \centering
                \includegraphics[width=\linewidth]{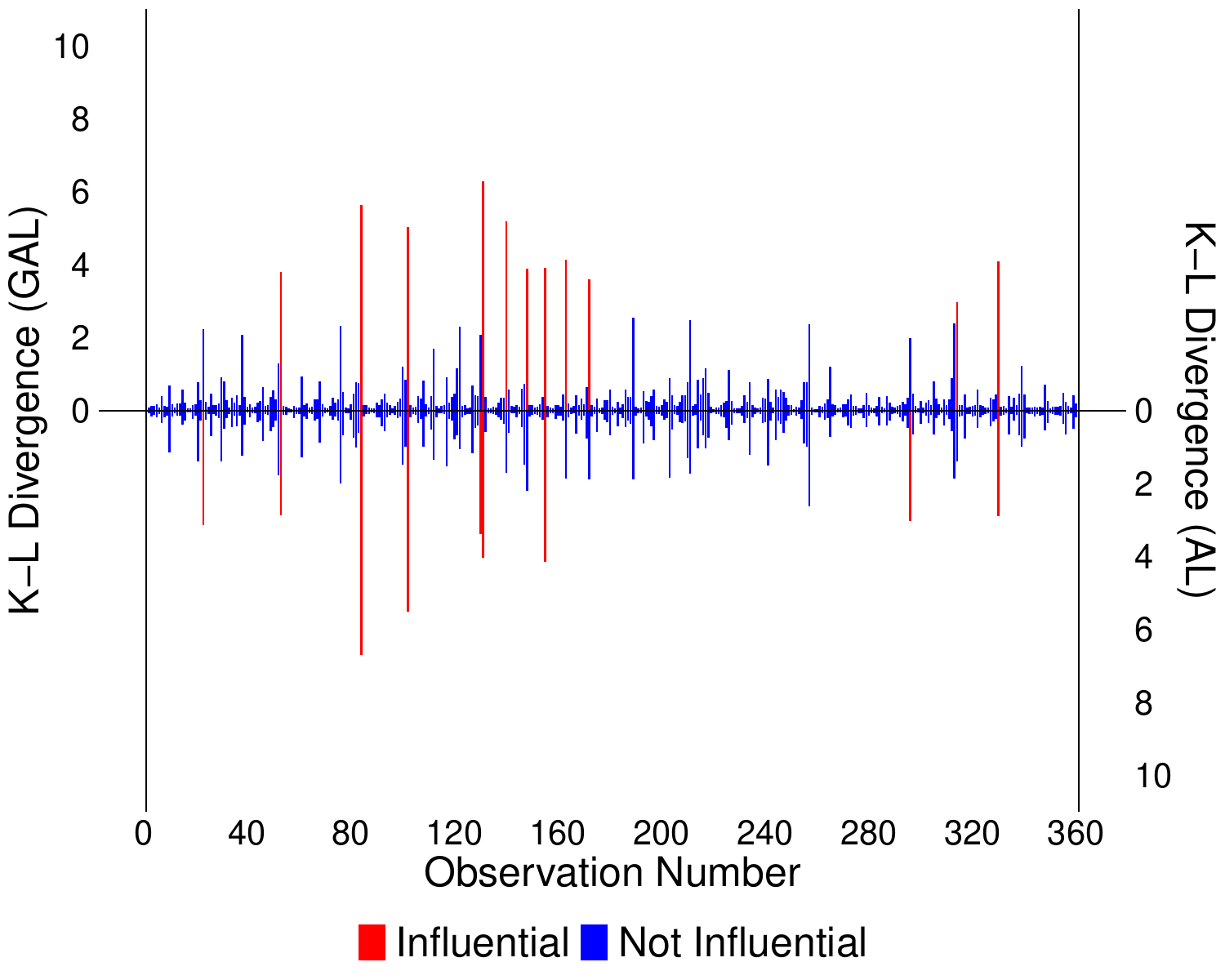}
                \caption{$p_0 = 0.25$}
            \end{subfigure}
            \\
            \begin{subfigure}{0.44\textwidth}
                \centering
                \includegraphics[width=\linewidth]{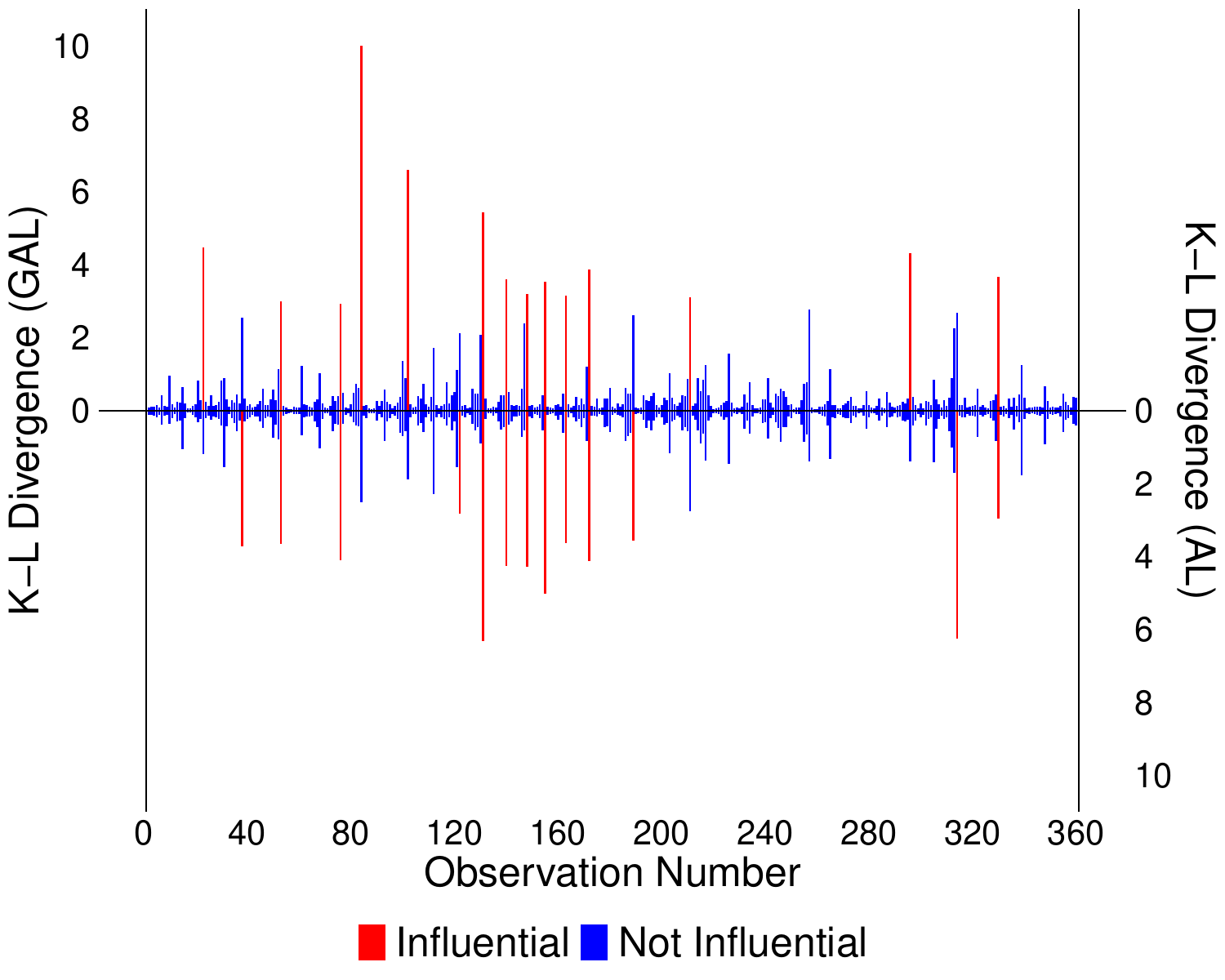}
                \caption{$p_0 = 0.5$}
            \end{subfigure}
            \begin{subfigure}{0.44\textwidth}
                \centering
                \includegraphics[width=\linewidth]{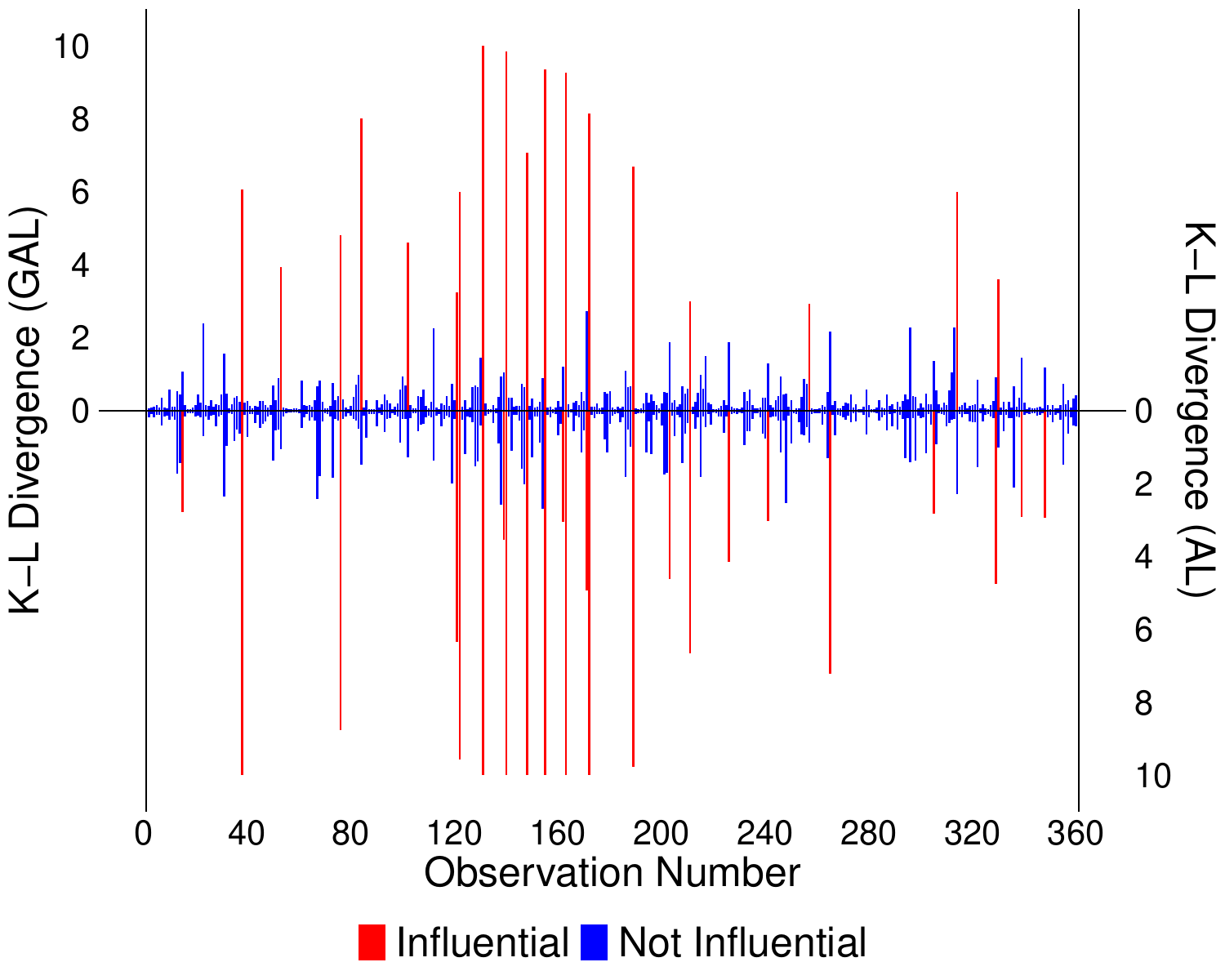}
                \caption{$p_0 = 0.75$}
            \end{subfigure}
            \\
            \begin{subfigure}{0.44\textwidth}
                \centering
                \includegraphics[width=\linewidth]{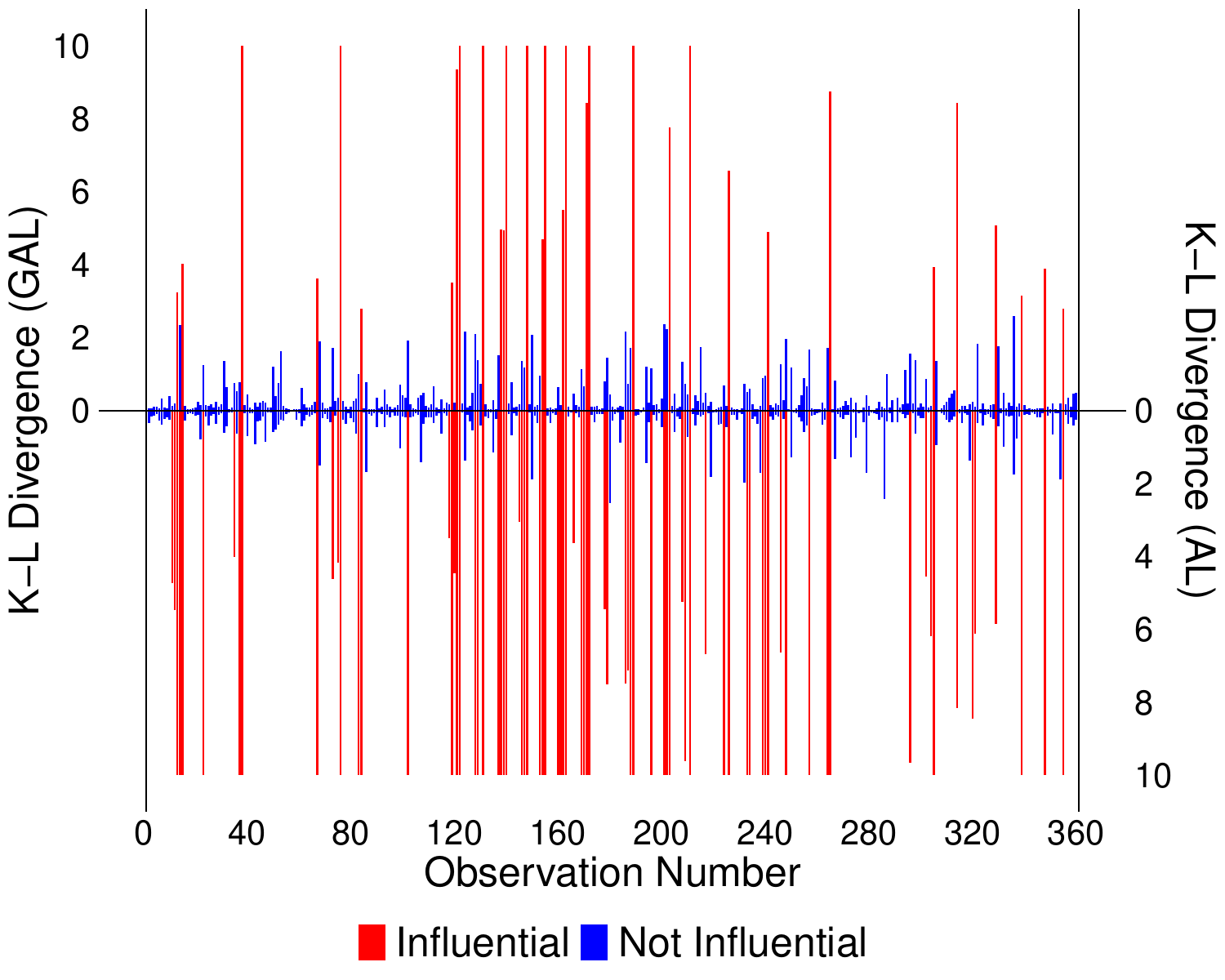}
                \caption{$p_0 = 0.95$}
            \end{subfigure}
            \caption{
                K-L divergence for the AL and GAL models across quantiles ($p_0 = 0.05, 0.25, 0.5, 0.75, 0.95$). The K-L divergence, capped at 10, is shown for all observations, with the AL model on the bottom and the GAL model on top. Influential observations (see \autoref{sec:KL_METHOD} for the definition) are highlighted in red, while non-influential ones are in blue. Dual y-axes represent the divergence for each model. Outliers generally have a larger impact under the AL model compared to the GAL model, suggesting that the AL model is more sensitive to extreme observations.
            }
            \label{fig:GAL_AL_QUANTILE_KL}
        \end{figure}

        \newpage
        \clearpage
        
        \begin{figure}
            \centering
            \begin{subfigure}{0.49\textwidth}
                \centering
                \includegraphics[width=\linewidth]{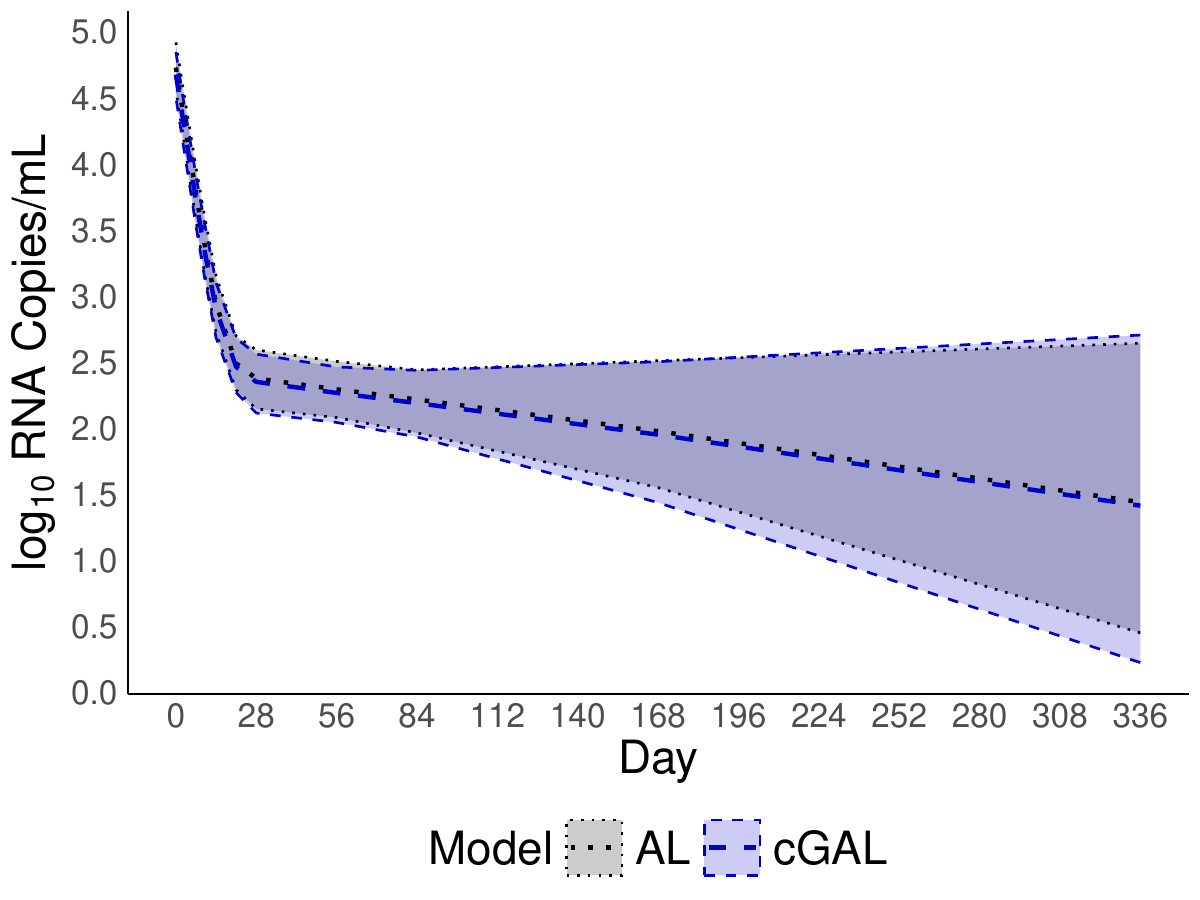}
                \caption{$p_0 = 0.05$}
            \end{subfigure}
            \begin{subfigure}{0.49\textwidth}
                \centering
                \includegraphics[width=\linewidth]{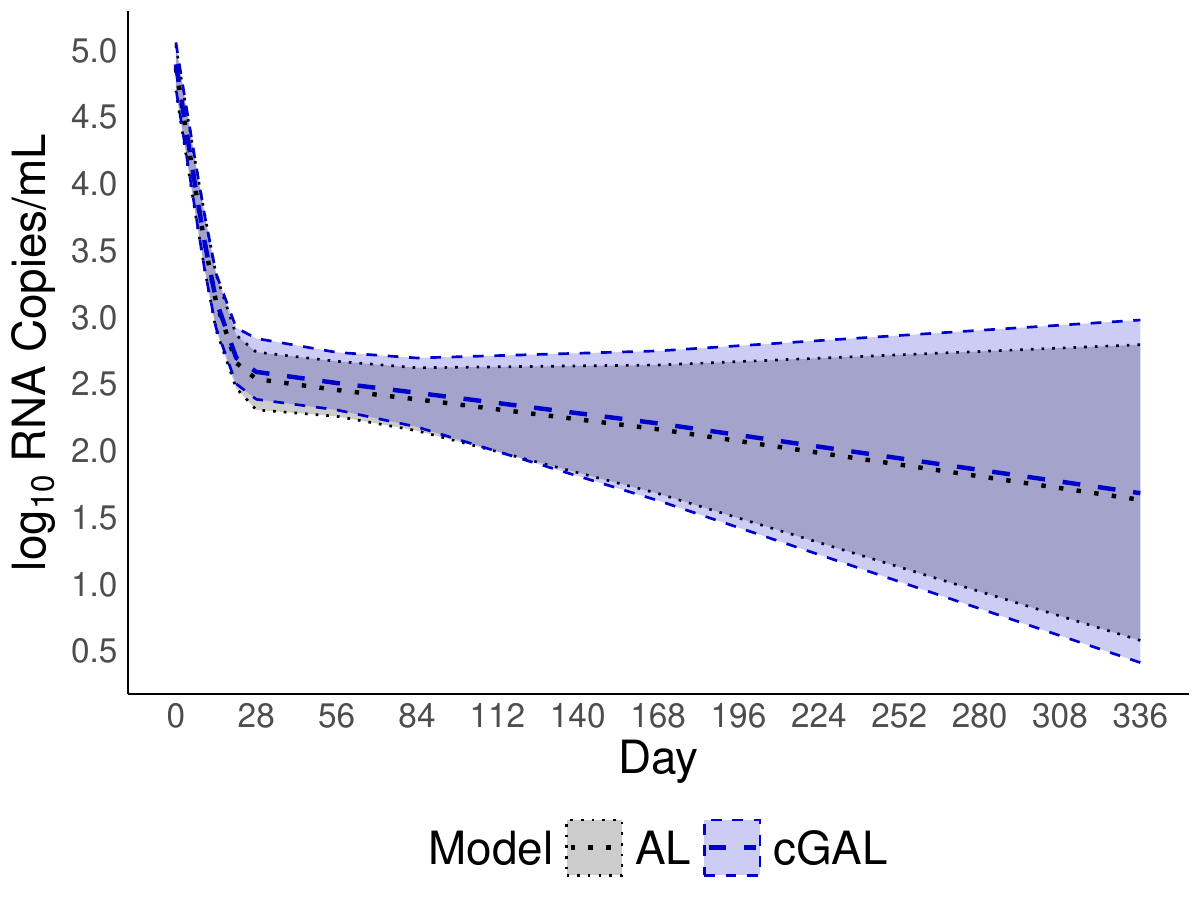}
                \caption{$p_0 = 0.25$}
            \end{subfigure}
            \\
            \begin{subfigure}{0.49\textwidth}
                \centering
                \includegraphics[width=\linewidth]{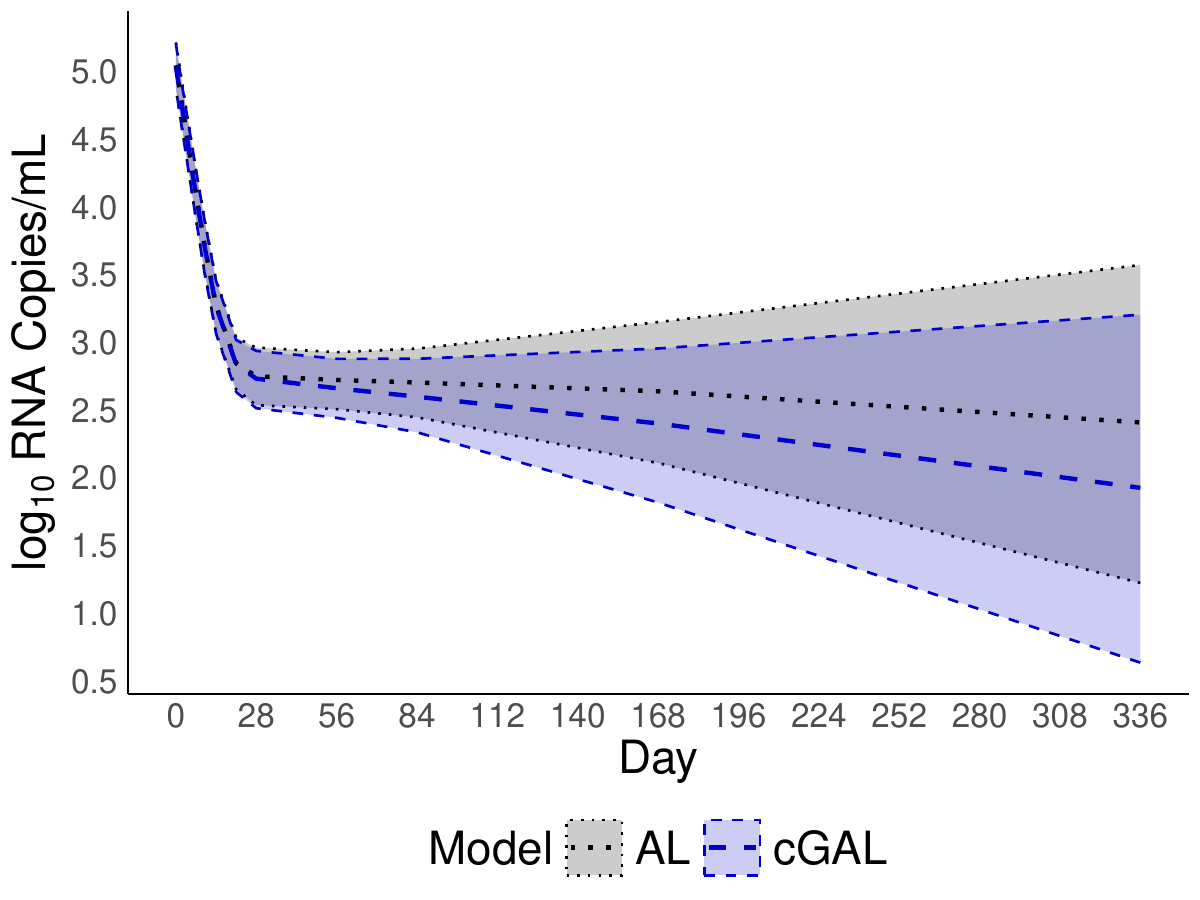}
                \caption{$p_0 = 0.5$}
            \end{subfigure}
            \begin{subfigure}{0.49\textwidth}
                \centering
                \includegraphics[width=\linewidth]{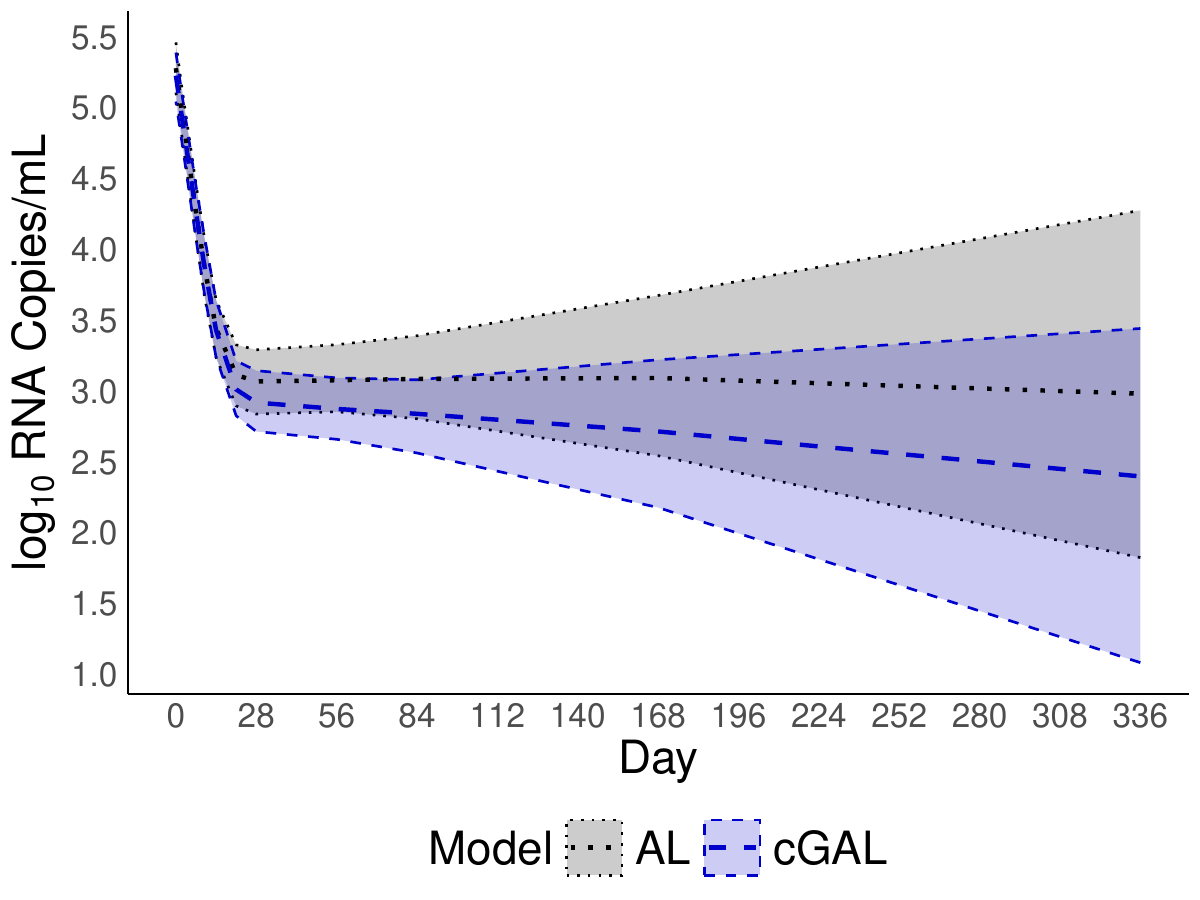}
                \caption{$p_0 = 0.75$}
            \end{subfigure}
            \\
            \begin{subfigure}{0.49\textwidth}
                \centering
                \includegraphics[width=\linewidth]{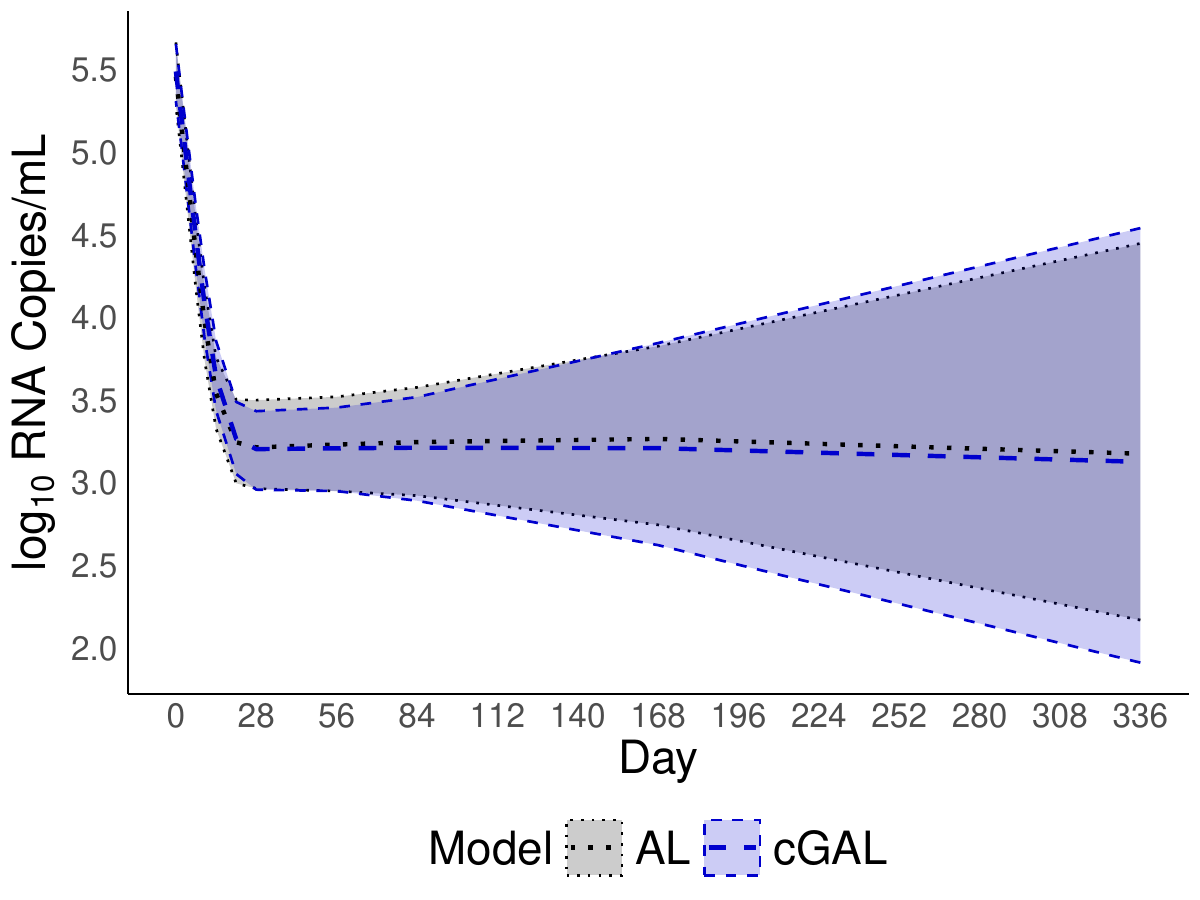}
                \caption{$p_0 = 0.95$}
            \end{subfigure}
            \caption{Comparison of quantile profiles for the AL and cGAL models at five quantiles ($p_0 = 0.05, 0.25, 0.5, 0.75, 0.95$), depicting HIV viral load dynamics over time in the ACTG~315 study. The profiles illustrate the different decay trajectories across the two models, with the AL model represented by the black dotted lines and the cGAL model by the blue dashed lines. The shaded areas represent the 95\% HPD intervals around each profile.}
            \label{fig:CGALD_ALD_QUANTILE_PROFILES}
        \end{figure}

        \newpage
        \clearpage
        
        \begin{figure}
            \centering
            \begin{subfigure}{0.49\textwidth}
                \centering
                \includegraphics[width=\linewidth]{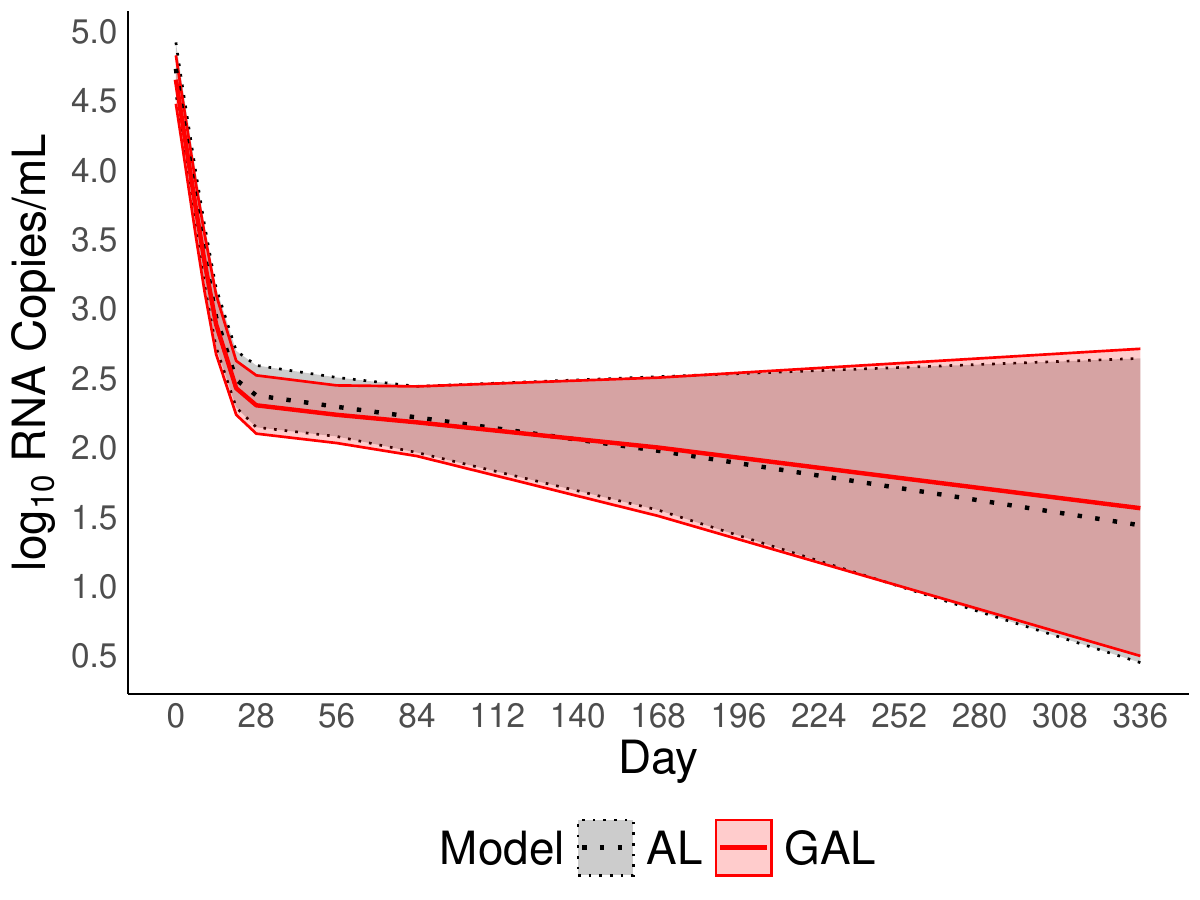}
                \caption{$p_0 = 0.05$}
            \end{subfigure}
            \begin{subfigure}{0.49\textwidth}
                \centering
                \includegraphics[width=\linewidth]{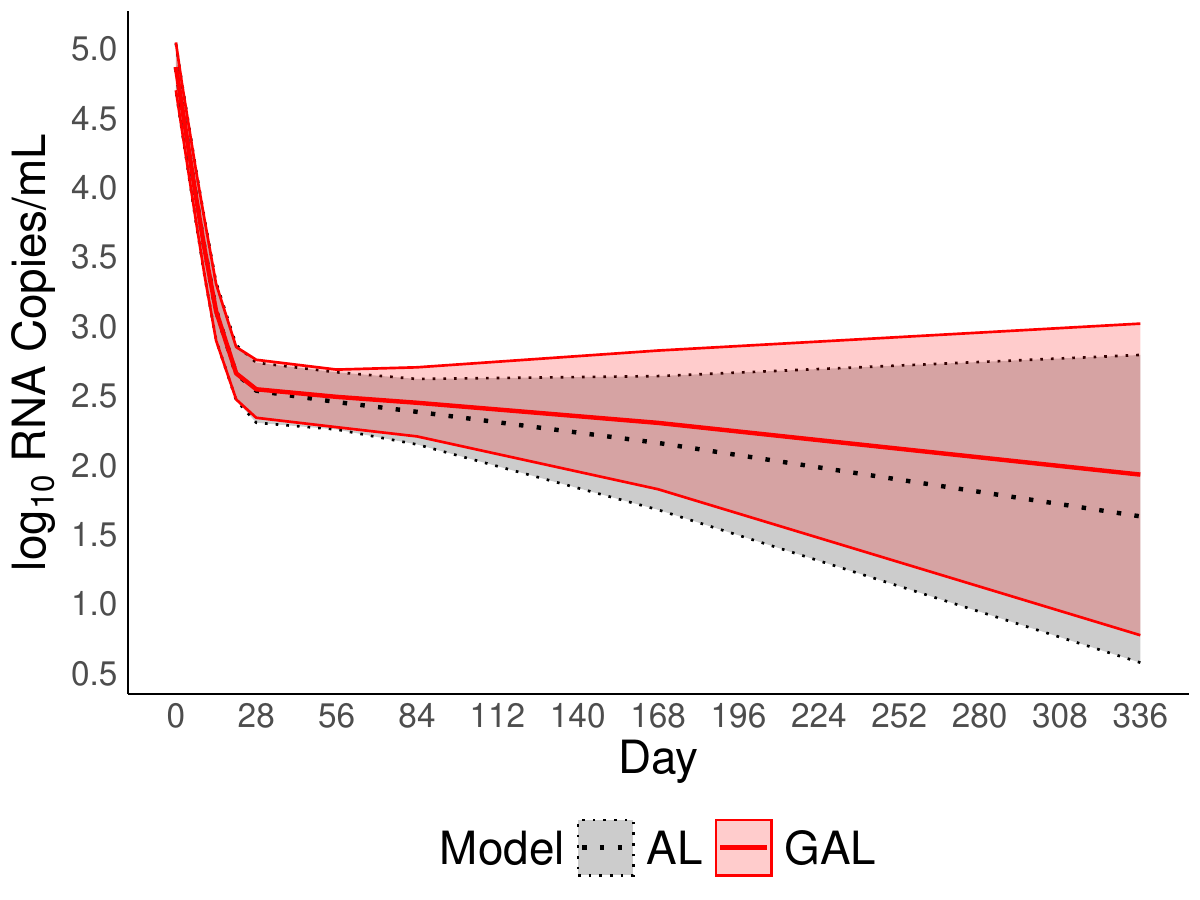}
                \caption{$p_0 = 0.25$}
            \end{subfigure}
            \\
            \begin{subfigure}{0.49\textwidth}
                \centering
                \includegraphics[width=\linewidth]{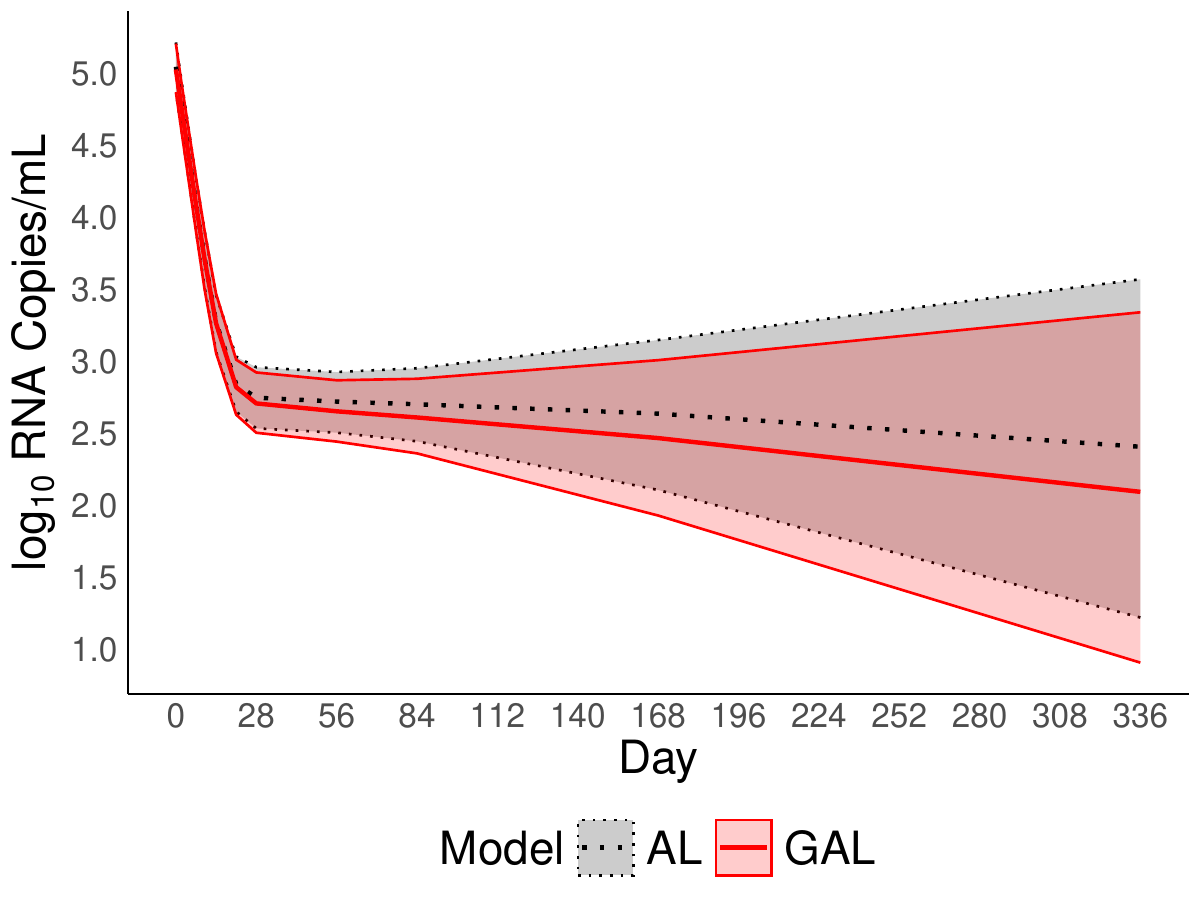}
                \caption{$p_0 = 0.5$}
            \end{subfigure}
            \begin{subfigure}{0.49\textwidth}
                \centering
                \includegraphics[width=\linewidth]{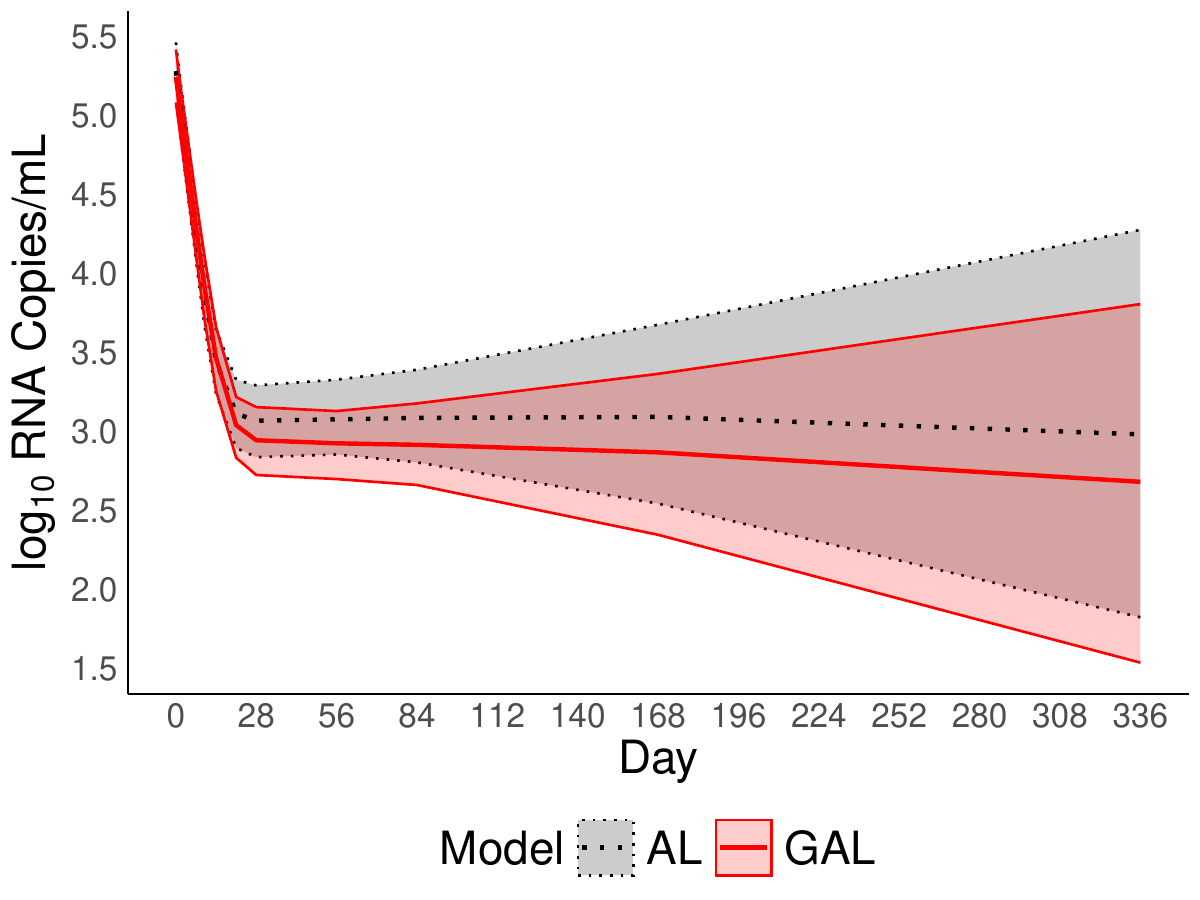}
                \caption{$p_0 = 0.75$}
            \end{subfigure}
            \\
            \begin{subfigure}{0.49\textwidth}
                \centering
                \includegraphics[width=\linewidth]{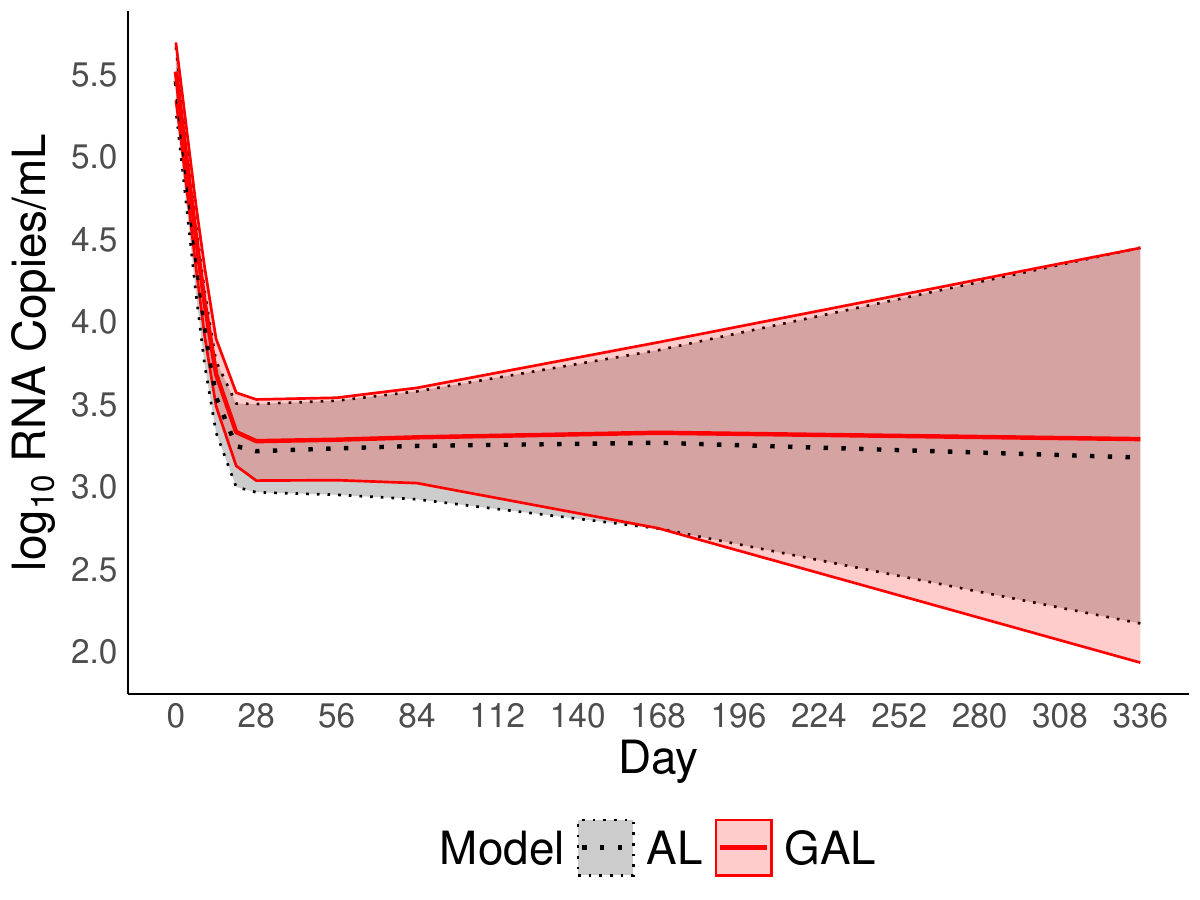}
                \caption{$p_0 = 0.95$}
            \end{subfigure}
            \caption{Comparison of quantile profiles for the AL and GAL models at five quantiles ($p_0 = 0.05, 0.25, 0.5, 0.75, 0.95$), depicting HIV viral load dynamics over time in the ACTG~315 study. The profiles illustrate the different decay trajectories across the two models, with the AL model represented by the black dotted lines and the GAL model by the red solid lines. The shaded areas represent the 95\% HPD intervals around each profile.}
            \label{fig:GALD_ALD_QUANTILE_PROFILES}
        \end{figure}

        \begin{landscape}

        \begin{footnotesize}
        
        \begin{ThreePartTable}
        
          \begin{TableNotes}[para, flushleft]
            \item[a] Quantile level being modeled. \\
            \item[b] AL (asymmetric Laplace), GAL (generalized asymmetric Laplace), cGAL (contaminated GAL), cGAL$_{\text{sens}}$ (sensitivity-analysis variant of cGAL).
          \end{TableNotes}
        
          \setlength{\tabcolsep}{3pt}
          \begin{longtable}{
            S[table-format=1.2]
            l
            S[table-format=4.3]
            S[table-format=5.3]
            S[table-format=4.3]
            c
            S[table-format=4.3]
            S[table-format=6.3]
            S[table-format=4.3]
            c
            S[table-format=4.3]
            S[table-format=6.3]
            S[table-format=4.3]
            c
            S[table-format=4.3]
            S[table-format=6.3]
            S[table-format=4.3]
          }
          \caption{Posterior summaries for each quantile level under the AL, GAL, cGAL, and cGAL$_{\text{sens}}$ models. Columns report each parameter's posterior estimate (Est.) and 95\% HPD interval. Results for higher quantiles continue on subsequent pages.}
          \label{tab:POST_SUMMARY}\\
        
          \toprule
          & & \multicolumn{15}{c}{\textbf{Model}\tnote{b}}\\
          \cline{3-17}
          \multicolumn{2}{c}{} &
          \multicolumn{3}{c}{\textbf{AL}} &
          \multicolumn{1}{c}{} &
          \multicolumn{3}{c}{\textbf{GAL}} &
          \multicolumn{1}{c}{} &
          \multicolumn{3}{c}{\textbf{cGAL}} &
          \multicolumn{1}{c}{} &
          \multicolumn{3}{c}{\textbf{cGAL$_{\text{sens}}$}}\\
          \cline{3-5}\cline{7-9}\cline{11-13}\cline{15-17}
          \textbf{Quantile}\tnote{a} & \textbf{Par.} &
          \textbf{Est.} & \multicolumn{2}{c}{\textbf{95\% HPD}} &
          &
          \textbf{Est.} & \multicolumn{2}{c}{\textbf{95\% HPD}} &
          &
          \textbf{Est.} & \multicolumn{2}{c}{\textbf{95\% HPD}} &
          &
          \textbf{Est.} & \multicolumn{2}{c}{\textbf{95\% HPD}}\\
          \midrule
          \endfirsthead
        
          \caption*{\textit{(continued)}}\\
          \toprule
          & & \multicolumn{15}{c}{\textbf{Model}\tnote{b}}\\
          \cline{3-17}
          \multicolumn{2}{c}{} &
          \multicolumn{3}{c}{\textbf{AL}} &
          \multicolumn{1}{c}{} &
          \multicolumn{3}{c}{\textbf{GAL}} &
          \multicolumn{1}{c}{} &
          \multicolumn{3}{c}{\textbf{cGAL}} &
          \multicolumn{1}{c}{} &
          \multicolumn{3}{c}{\textbf{cGAL$_{\text{sens}}$}}\\
          \cline{3-5}\cline{7-9}\cline{11-13}\cline{15-17}
          \textbf{Quantile}\tnote{a} & \textbf{Par.} &
          \textbf{Est.} & \multicolumn{2}{c}{\textbf{95\% HPD}} &
          &
          \textbf{Est.} & \multicolumn{2}{c}{\textbf{95\% HPD}} &
          &
          \textbf{Est.} & \multicolumn{2}{c}{\textbf{95\% HPD}} &
          &
          \textbf{Est.} & \multicolumn{2}{c}{\textbf{95\% HPD}}\\
          \midrule
          \endhead
        
          \midrule
          \multicolumn{17}{r}{\textit{Continued on next page}}\\
          \endfoot
        
          \bottomrule
          \insertTableNotes\\
          \endlastfoot
        
            0.05 & $\beta_{1}$ & 10.888 & [10.473; & 11.311] &  & 10.708 & [10.317; & 11.120] &  & 10.766 & [10.375; & 11.169] &  & 10.761 & [10.386; & 11.189] \\* 
               & $\beta_{2}$ & 0.319 & [0.280; & 0.363] &  & 0.313 & [0.276; & 0.355] &  & 0.318 & [0.279; & 0.361] &  & 0.319 & [0.282; & 0.363] \\* 
               & $\beta_{3}$ & 5.590 & [4.912; & 6.217] &  & 5.382 & [4.740; & 6.015] &  & 5.546 & [4.845; & 6.224] &  & 5.555 & [4.878; & 6.216] \\* 
               & $\beta_{4}$ & -0.003 & [-0.021; & 0.013] &  & -0.005 & [-0.020; & 0.011] &  & -0.002 & [-0.019; & 0.013] &  & -0.002 & [-0.018; & 0.014] \\* 
               & $\beta_{5}$ & 0.004 & [-0.003; & 0.010] &  & 0.004 & [-0.001; & 0.010] &  & 0.003 & [-0.003; & 0.009] &  & 0.003 & [-0.002; & 0.009] \\* 
               & $\sigma$ & 0.021 & [0.018; & 0.024] &  & 0.067 & [0.055; & 0.078] &  & 0.057 & [0.036; & 0.071] &  & 0.058 & [0.038; & 0.070] \\* 
               & $\gamma$ &  &  &  &  & 2.847 & [1.968; & 3.824] &  & 4.058 & [2.225; & 6.975] &  & 3.938 & [2.190; & 6.720] \\* 
               & $\alpha$ &  &  &  &  &  &  &  &  & 0.018 & [0.002; & 0.038] &  & 0.015 & [0.002; & 0.037] \\* 
               & $\omega_{11}$ & 0.929 & [0.452; & 1.600] &  & 1.139 & [0.515; & 2.210] &  & 1.094 & [0.484; & 1.987] &  & 1.097 & [0.451; & 1.998] \\* 
               & $\omega_{21}$ & 2.042 & [-2.583; & 8.240] &  & 3.393 & [-3.734; & 15.857] &  & 2.958 & [-3.758; & 13.770] &  & 2.866 & [-4.008; & 13.806] \\* 
               & $\omega_{22}$ & 126.647 & [45.138; & 273.301] &  & 182.982 & [41.391; & 495.406] &  & 175.010 & [57.743; & 442.609] &  & 175.704 & [60.605; & 448.345] \\* 
               & $\omega_{31}$ & -0.374 & [-0.807; & -0.056] &  & -0.469 & [-1.102; & -0.092] &  & -0.465 & [-1.045; & -0.087] &  & -0.463 & [-1.040; & -0.094] \\* 
               & $\omega_{32}$ & -0.377 & [-5.359; & 4.413] &  & -1.594 & [-10.095; & 5.187] &  & -1.945 & [-10.519; & 3.920] &  & -2.053 & [-10.437; & 4.425] \\* 
               & $\omega_{33}$ & 0.647 & [0.329; & 1.079] &  & 0.728 & [0.346; & 1.262] &  & 0.719 & [0.333; & 1.231] &  & 0.713 & [0.356; & 1.217] \\* 
               & $\omega_{41}$ & 14.600 & [-8.951; & 44.343] &  & 15.825 & [-15.627; & 51.615] &  & 13.007 & [-12.288; & 42.097] &  & 13.095 & [-12.239; & 40.346] \\* 
               & $\omega_{42}$ & -106.832 & [-534.535; & 273.684] &  & -165.956 & [-823.793; & 408.179] &  & -176.357 & [-774.993; & 198.428] &  & -186.963 & [-742.759; & 216.296] \\* 
               & $\omega_{43}$ & -24.531 & [-51.701; & -4.272] &  & -25.121 & [-51.457; & -2.802] &  & -22.975 & [-45.799; & -4.021] &  & -22.997 & [-45.336; & -3.144] \\* 
               & $\omega_{44}$ & 3550.850 & [1563.010; & 7617.490] &  & 3775.830 & [1668.130; & 7478.400] &  & 3171.280 & [1607.770; & 5613.480] &  & 3140.775 & [1540.320; & 5472.120] \\ 
              0.25 & $\beta_{1}$ & 11.200 & [10.806; & 11.592] &  & 11.204 & [10.813; & 11.612] &  & 11.264 & [10.846; & 11.642] &  & 11.243 & [10.843; & 11.646] \\* 
               & $\beta_{2}$ & 0.313 & [0.277; & 0.354] &  & 0.312 & [0.273; & 0.352] &  & 0.319 & [0.282; & 0.359] &  & 0.319 & [0.282; & 0.360] \\* 
               & $\beta_{3}$ & 5.953 & [5.261; & 6.573] &  & 5.904 & [5.313; & 6.553] &  & 6.098 & [5.396; & 6.806] &  & 6.071 & [5.440; & 6.758] \\* 
               & $\beta_{4}$ & -0.004 & [-0.018; & 0.011] &  & -0.007 & [-0.022; & 0.008] &  & -0.002 & [-0.020; & 0.014] &  & -0.002 & [-0.018; & 0.013] \\* 
               & $\beta_{5}$ & 0.004 & [-0.001; & 0.009] &  & 0.004 & [-0.001; & 0.010] &  & 0.003 & [-0.002; & 0.009] &  & 0.003 & [-0.002; & 0.010] \\* 
               & $\sigma$ & 0.093 & [0.081; & 0.107] &  & 0.111 & [0.095; & 0.131] &  & 0.097 & [0.041; & 0.121] &  & 0.098 & [0.052; & 0.123] \\* 
               & $\gamma$ &  &  &  &  & 0.473 & [0.063; & 0.833] &  & 0.745 & [0.129; & 2.113] &  & 0.698 & [0.053; & 1.876] \\* 
               & $\alpha$ &  &  &  &  &  &  &  &  & 0.022 & [0.002; & 0.051] &  & 0.018 & [0.000; & 0.044] \\* 
               & $\omega_{11}$ & 1.112 & [0.506; & 2.040] &  & 1.174 & [0.532; & 2.177] &  & 1.115 & [0.543; & 2.030] &  & 1.109 & [0.538; & 2.050] \\* 
               & $\omega_{21}$ & 2.896 & [-4.712; & 16.280] &  & 3.385 & [-5.005; & 19.242] &  & 2.688 & [-6.698; & 16.022] &  & 2.712 & [-5.798; & 15.642] \\* 
               & $\omega_{22}$ & 206.009 & [61.341; & 610.099] &  & 218.843 & [57.340; & 719.243] &  & 213.755 & [61.322; & 782.779] &  & 208.914 & [51.528; & 666.532] \\* 
               & $\omega_{31}$ & -0.447 & [-1.019; & -0.043] &  & -0.487 & [-1.108; & -0.079] &  & -0.465 & [-1.085; & -0.087] &  & -0.453 & [-1.056; & -0.076] \\* 
               & $\omega_{32}$ & -1.145 & [-11.137; & 7.147] &  & -1.869 & [-14.249; & 5.966] &  & -1.524 & [-12.826; & 10.053] &  & -1.748 & [-12.308; & 7.703] \\* 
               & $\omega_{33}$ & 0.729 & [0.339; & 1.250] &  & 0.753 & [0.344; & 1.353] &  & 0.755 & [0.330; & 1.328] &  & 0.745 & [0.342; & 1.304] \\* 
               & $\omega_{41}$ & 17.577 & [-10.986; & 50.696] &  & 18.502 & [-11.716; & 52.047] &  & 12.994 & [-16.221; & 40.612] &  & 13.213 & [-12.419; & 42.225] \\* 
               & $\omega_{42}$ & -181.731 & [-935.112; & 286.107] &  & -194.774 & [-1036.640; & 330.180] &  & -222.341 & [-1062.980; & 360.416] &  & -230.349 & [-1126.830; & 318.367] \\* 
               & $\omega_{43}$ & -27.220 & [-55.447; & -6.385] &  & -26.756 & [-55.040; & -5.211] &  & -24.384 & [-50.962; & -3.354] &  & -23.757 & [-47.201; & -1.912] \\* 
               & $\omega_{44}$ & 3734.935 & [1592.020; & 7133.150] &  & 3600.175 & [1598.020; & 6855.200] &  & 3220.385 & [1497.850; & 6166.040] &  & 3230.330 & [1495.320; & 5815.670] \\ 
              0.5 & $\beta_{1}$ & 11.614 & [11.216; & 12.011] &  & 11.572 & [11.216; & 11.993] &  & 11.591 & [11.220; & 11.980] &  & 11.595 & [11.187; & 11.977] \\* 
               & $\beta_{2}$ & 0.318 & [0.281; & 0.359] &  & 0.314 & [0.275; & 0.356] &  & 0.318 & [0.280; & 0.359] &  & 0.317 & [0.278; & 0.355] \\* 
               & $\beta_{3}$ & 6.327 & [5.698; & 6.913] &  & 6.289 & [5.684; & 6.897] &  & 6.380 & [5.729; & 7.047] &  & 6.366 & [5.723; & 7.056] \\* 
               & $\beta_{4}$ & -0.010 & [-0.024; & 0.006] &  & -0.006 & [-0.023; & 0.008] &  & -0.005 & [-0.022; & 0.012] &  & -0.006 & [-0.024; & 0.009] \\* 
               & $\beta_{5}$ & 0.005 & [-0.001; & 0.010] &  & 0.004 & [-0.001; & 0.010] &  & 0.004 & [-0.001; & 0.010] &  & 0.004 & [-0.001; & 0.010] \\* 
               & $\sigma$ & 0.126 & [0.109; & 0.144] &  & 0.112 & [0.020; & 0.137] &  & 0.102 & [0.039; & 0.129] &  & 0.102 & [0.022; & 0.127] \\* 
               & $\gamma$ &  &  &  &  & -0.326 & [-0.983; & -0.057] &  & -0.242 & [-0.836; & 0.097] &  & -0.260 & [-0.961; & 0.045] \\* 
               & $\alpha$ &  &  &  &  &  &  &  &  & 0.018 & [0.001; & 0.045] &  & 0.015 & [0.000; & 0.041] \\* 
               & $\omega_{11}$ & 1.201 & [0.559; & 2.336] &  & 1.180 & [0.504; & 2.259] &  & 1.112 & [0.543; & 2.066] &  & 1.131 & [0.527; & 2.006] \\* 
               & $\omega_{21}$ & 3.293 & [-5.487; & 17.590] &  & 3.526 & [-4.655; & 18.509] &  & 2.735 & [-6.302; & 16.953] &  & 2.668 & [-5.229; & 15.098] \\* 
               & $\omega_{22}$ & 206.829 & [64.149; & 684.648] &  & 218.613 & [61.656; & 657.225] &  & 214.907 & [58.781; & 738.424] &  & 206.373 & [59.435; & 591.464] \\* 
               & $\omega_{31}$ & -0.487 & [-1.140; & -0.071] &  & -0.491 & [-1.147; & -0.058] &  & -0.465 & [-1.055; & -0.069] &  & -0.459 & [-1.018; & -0.081] \\* 
               & $\omega_{32}$ & -1.710 & [-13.728; & 7.355] &  & -1.855 & [-12.912; & 6.737] &  & -1.912 & [-13.632; & 9.802] &  & -1.687 & [-11.740; & 6.298] \\* 
               & $\omega_{33}$ & 0.775 & [0.376; & 1.435] &  & 0.768 & [0.357; & 1.382] &  & 0.744 & [0.340; & 1.314] &  & 0.736 & [0.362; & 1.302] \\* 
               & $\omega_{41}$ & 18.944 & [-10.219; & 51.400] &  & 17.725 & [-13.623; & 50.222] &  & 15.522 & [-10.741; & 46.174] &  & 14.900 & [-11.623; & 42.636] \\* 
               & $\omega_{42}$ & -165.159 & [-949.283; & 332.340] &  & -202.437 & [-989.776; & 346.832] &  & -218.613 & [-1114.230; & 273.108] &  & -201.720 & [-850.471; & 286.951] \\* 
               & $\omega_{43}$ & -26.966 & [-56.662; & -6.934] &  & -27.142 & [-55.074; & -5.559] &  & -24.462 & [-49.770; & -4.232] &  & -24.282 & [-48.854; & -5.650] \\* 
               & $\omega_{44}$ & 3289.345 & [1451.350; & 5738.650] &  & 3568.830 & [1648.540; & 6723.000] &  & 3166.495 & [1543.930; & 5895.370] &  & 3174.795 & [1464.620; & 5535.030] \\ 
              0.75 & $\beta_{1}$ & 12.152 & [11.737; & 12.546] &  & 12.071 & [11.691; & 12.464] &  & 12.032 & [11.603; & 12.402] &  & 12.036 & [11.674; & 12.438] \\* 
               & $\beta_{2}$ & 0.340 & [0.300; & 0.388] &  & 0.318 & [0.278; & 0.365] &  & 0.323 & [0.282; & 0.368] &  & 0.326 & [0.284; & 0.370] \\* 
               & $\beta_{3}$ & 7.005 & [6.334; & 7.574] &  & 6.766 & [6.116; & 7.404] &  & 6.767 & [6.157; & 7.384] &  & 6.785 & [6.155; & 7.451] \\* 
               & $\beta_{4}$ & -0.016 & [-0.034; & 0.001] &  & -0.011 & [-0.027; & 0.006] &  & -0.007 & [-0.022; & 0.009] &  & -0.007 & [-0.023; & 0.011] \\* 
               & $\beta_{5}$ & 0.006 & [-0.000; & 0.013] &  & 0.005 & [-0.001; & 0.011] &  & 0.004 & [-0.002; & 0.010] &  & 0.004 & [-0.001; & 0.011] \\* 
               & $\sigma$ & 0.095 & [0.081; & 0.109] &  & 0.104 & [0.072; & 0.130] &  & 0.085 & [0.043; & 0.113] &  & 0.085 & [0.043; & 0.112] \\* 
               & $\gamma$ &  &  &  &  & -1.240 & [-1.804; & -0.741] &  & -1.408 & [-2.100; & -0.854] &  & -1.399 & [-2.066; & -0.788] \\* 
               & $\alpha$ &  &  &  &  &  &  &  &  & 0.015 & [0.001; & 0.035] &  & 0.013 & [0.001; & 0.033] \\* 
               & $\omega_{11}$ & 1.170 & [0.503; & 2.123] &  & 1.232 & [0.539; & 2.447] &  & 1.136 & [0.544; & 2.105] &  & 1.147 & [0.502; & 2.130] \\* 
               & $\omega_{21}$ & 1.698 & [-9.533; & 24.553] &  & 3.447 & [-4.343; & 17.750] &  & 2.972 & [-3.983; & 14.787] &  & 2.912 & [-3.855; & 14.449] \\* 
               & $\omega_{22}$ & 156.002 & [30.900; & 2301.640] &  & 193.226 & [48.892; & 570.078] &  & 186.693 & [59.594; & 472.967] &  & 180.619 & [57.655; & 461.854] \\* 
               & $\omega_{31}$ & -0.446 & [-1.040; & -0.029] &  & -0.498 & [-1.261; & -0.066] &  & -0.462 & [-1.064; & -0.075] &  & -0.473 & [-1.102; & -0.076] \\* 
               & $\omega_{32}$ & -0.271 & [-9.905; & 24.276] &  & -2.358 & [-12.522; & 7.086] &  & -2.335 & [-10.780; & 4.892] &  & -2.431 & [-12.540; & 3.656] \\* 
               & $\omega_{33}$ & 0.733 & [0.336; & 1.416] &  & 0.770 & [0.380; & 1.468] &  & 0.736 & [0.367; & 1.339] &  & 0.745 & [0.343; & 1.349] \\* 
               & $\omega_{41}$ & 18.636 & [-3.281; & 44.942] &  & 18.428 & [-8.155; & 53.426] &  & 14.099 & [-12.459; & 42.389] &  & 14.149 & [-11.200; & 42.412] \\* 
               & $\omega_{42}$ & -37.728 & [-730.643; & 916.398] &  & -159.418 & [-857.598; & 280.474] &  & -204.434 & [-837.495; & 225.430] &  & -197.066 & [-804.421; & 181.675] \\* 
               & $\omega_{43}$ & -18.554 & [-39.966; & -0.101] &  & -25.136 & [-51.927; & -4.386] &  & -21.977 & [-46.403; & -3.833] &  & -22.109 & [-44.681; & -2.863] \\* 
               & $\omega_{44}$ & 2311.940 & [1084.790; & 3975.270] &  & 3201.290 & [1538.090; & 5923.810] &  & 2989.335 & [1392.050; & 5168.770] &  & 2948.195 & [1547.980; & 5129.460] \\ 
              0.95 & $\beta_{1}$ & 12.566 & [12.157; & 13.044] &  & 12.701 & [12.300; & 13.108] &  & 12.638 & [12.221; & 13.021] &  & 12.663 & [12.259; & 13.090] \\* 
               & $\beta_{2}$ & 0.362 & [0.315; & 0.407] &  & 0.338 & [0.292; & 0.390] &  & 0.336 & [0.291; & 0.383] &  & 0.337 & [0.295; & 0.383] \\* 
               & $\beta_{3}$ & 7.333 & [6.626; & 7.963] &  & 7.494 & [6.803; & 8.185] &  & 7.325 & [6.656; & 7.951] &  & 7.456 & [6.740; & 8.075] \\* 
               & $\beta_{4}$ & -0.016 & [-0.032; & 0.004] &  & -0.013 & [-0.030; & 0.004] &  & -0.012 & [-0.030; & 0.006] &  & -0.011 & [-0.029; & 0.008] \\* 
               & $\beta_{5}$ & 0.006 & [-0.001; & 0.013] &  & 0.005 & [-0.002; & 0.011] &  & 0.005 & [-0.002; & 0.011] &  & 0.005 & [-0.002; & 0.011] \\* 
               & $\sigma$ & 0.020 & [0.018; & 0.023] &  & 0.070 & [0.059; & 0.080] &  & 0.060 & [0.042; & 0.075] &  & 0.062 & [0.043; & 0.077] \\* 
               & $\gamma$ &  &  &  &  & -3.734 & [-4.940; & -2.627] &  & -4.568 & [-6.571; & -2.968] &  & -4.436 & [-6.410; & -2.871] \\* 
               & $\alpha$ &  &  &  &  &  &  &  &  & 0.012 & [0.000; & 0.028] &  & 0.009 & [0.000; & 0.026] \\* 
               & $\omega_{11}$ & 0.839 & [0.441; & 1.318] &  & 1.096 & [0.501; & 2.071] &  & 1.044 & [0.519; & 1.945] &  & 1.059 & [0.474; & 1.931] \\* 
               & $\omega_{21}$ & -0.404 & [-4.923; & 4.207] &  & 1.508 & [-6.642; & 15.837] &  & 1.484 & [-4.387; & 9.463] &  & 1.624 & [-8.714; & 19.404] \\* 
               & $\omega_{22}$ & 94.944 & [28.917; & 246.147] &  & 132.469 & [25.499; & 888.207] &  & 128.435 & [37.715; & 312.930] &  & 145.248 & [28.271; & 1756.910] \\* 
               & $\omega_{31}$ & -0.289 & [-0.623; & -0.025] &  & -0.404 & [-0.941; & -0.013] &  & -0.389 & [-0.913; & -0.037] &  & -0.395 & [-0.937; & -0.014] \\* 
               & $\omega_{32}$ & -0.165 & [-4.611; & 5.048] &  & -1.314 & [-11.223; & 11.052] &  & -1.491 & [-8.969; & 3.656] &  & -1.134 & [-11.285; & 17.022] \\* 
               & $\omega_{33}$ & 0.548 & [0.268; & 0.935] &  & 0.676 & [0.283; & 1.267] &  & 0.658 & [0.337; & 1.181] &  & 0.677 & [0.315; & 1.305] \\* 
               & $\omega_{41}$ & 12.566 & [-0.376; & 28.142] &  & 16.897 & [-4.247; & 43.502] &  & 16.197 & [-3.971; & 43.321] &  & 15.422 & [-7.023; & 41.700] \\* 
               & $\omega_{42}$ & -50.011 & [-294.766; & 170.620] &  & -17.611 & [-524.265; & 456.264] &  & -78.953 & [-497.389; & 238.236] &  & -89.736 & [-844.415; & 747.349] \\* 
               & $\omega_{43}$ & -13.750 & [-26.763; & -1.535] &  & -18.478 & [-38.568; & -2.496] &  & -18.885 & [-39.459; & -2.917] &  & -18.189 & [-39.661; & -0.725] \\* 
               & $\omega_{44}$ & 1910.535 & [996.911; & 3076.930] &  & 2276.790 & [1068.020; & 3898.490] &  & 2522.015 & [1207.850; & 4309.600] &  & 2596.120 & [1208.850; & 4551.370] \\ 
        
        \end{longtable}
        
        \end{ThreePartTable}

        \end{footnotesize}

        \end{landscape}

        \begin{landscape}
        
        \begin{footnotesize}
        
        \begin{ThreePartTable}
        
          \begin{TableNotes}[para, flushleft]
            \item[a] Quantile level being modelled.
          \end{TableNotes}
        
          \setlength{\tabcolsep}{3pt}
          \begin{longtable}{
            S[table-format=1.2]
            l
            S[table-format=4.3]
            S[table-format=5.3]
            S[table-format=4.3]
            c
            S[table-format=4.3]
            S[table-format=5.3]
            S[table-format=4.3]
            c
            S[table-format=4.3]
            S[table-format=5.3]
            S[table-format=4.3]
          }
          \caption{Posterior summaries (cGAL model, $\tau_0$ sensitivity). For each quantile level and each scale-inflation constant $\tau_0 \in \left\{6.\dot{6}\dot{7}, 20, 40\right\}$, the table lists the posterior estimate (Est.) and 95\% HPD interval for all parameters. Results for higher quantiles continue on subsequent pages.}
          \label{tab:POST_SUMMARY_CONTAM}\\
        
          \toprule
          \multicolumn{2}{c}{} &
          \multicolumn{3}{c}{$\bm{\tau_0=6.\dot{6}\dot{7}}$} &
          \multicolumn{1}{c}{} &
          \multicolumn{3}{c}{$\bm{\tau_0=20}$} &
          \multicolumn{1}{c}{} &
          \multicolumn{3}{c}{$\bm{\tau_0=40}$}\\
          \cline{3-5}\cline{7-9}\cline{11-13}
          \textbf{Quantile}\textsuperscript{a} & \textbf{Par.} &
          \textbf{Est.} & \multicolumn{2}{c}{\textbf{95\% HPD}} &
          &
          \textbf{Est.} & \multicolumn{2}{c}{\textbf{95\% HPD}} &
          &
          \textbf{Est.} & \multicolumn{2}{c}{\textbf{95\% HPD}}\\
          \midrule
          \endfirsthead
        
          \caption*{\textit{(continued)}}\\
          \toprule
          \multicolumn{2}{c}{} &
          \multicolumn{3}{c}{$\bm{\tau_0=6.\dot{6}\dot{7}}$} &
          \multicolumn{1}{c}{} &
          \multicolumn{3}{c}{$\bm{\tau_0=20}$} &
          \multicolumn{1}{c}{} &
          \multicolumn{3}{c}{$\bm{\tau_0=40}$}\\
          \cline{3-5}\cline{7-9}\cline{11-13}
          \textbf{Quantile}\textsuperscript{a} & \textbf{Par.} &
          \textbf{Est.} & \multicolumn{2}{c}{\textbf{95\% HPD}} &
          &
          \textbf{Est.} & \multicolumn{2}{c}{\textbf{95\% HPD}} &
          &
          \textbf{Est.} & \multicolumn{2}{c}{\textbf{95\% HPD}}\\
          \midrule
          \endhead
        
          \midrule
          \multicolumn{13}{r}{\textit{Continued on next page}}\\
          \endfoot
        
          \bottomrule
          \insertTableNotes\\
          \endlastfoot
        
            0.05 & $\beta_{1}$ & 10.765 & [10.351; & 11.164] &  & 10.742 & [10.352; & 11.186] &  & 10.752 & [10.325; & 11.165] \\* 
               & $\beta_{2}$ & 0.318 & [0.279; & 0.361] &  & 0.319 & [0.279; & 0.361] &  & 0.318 & [0.276; & 0.359] \\* 
               & $\beta_{3}$ & 5.600 & [4.915; & 6.272] &  & 5.521 & [4.829; & 6.234] &  & 5.492 & [4.820; & 6.106] \\* 
               & $\beta_{4}$ & -0.000 & [-0.015; & 0.016] &  & -0.003 & [-0.019; & 0.013] &  & -0.003 & [-0.019; & 0.014] \\* 
               & $\beta_{5}$ & 0.003 & [-0.002; & 0.008] &  & 0.004 & [-0.002; & 0.010] &  & 0.003 & [-0.002; & 0.009] \\* 
               & $\sigma$ & 0.055 & [0.031; & 0.068] &  & 0.060 & [0.042; & 0.073] &  & 0.062 & [0.049; & 0.075] \\* 
               & $\gamma$ & 4.451 & [2.584; & 7.615] &  & 3.590 & [2.064; & 6.345] &  & 3.311 & [1.840; & 5.420] \\* 
               & $\alpha$ & 0.027 & [0.005; & 0.059] &  & 0.011 & [0.000; & 0.025] &  & 0.007 & [0.000; & 0.018] \\* 
               & $\omega_{11}$ & 1.110 & [0.491; & 1.976] &  & 1.082 & [0.487; & 1.991] &  & 1.085 & [0.477; & 2.071] \\* 
               & $\omega_{21}$ & 2.549 & [-4.491; & 14.749] &  & 2.982 & [-3.357; & 14.280] &  & 3.153 & [-3.258; & 13.810] \\* 
               & $\omega_{22}$ & 175.895 & [48.502; & 527.437] &  & 172.263 & [54.717; & 433.597] &  & 170.063 & [56.571; & 417.020] \\* 
               & $\omega_{31}$ & -0.469 & [-1.016; & -0.087] &  & -0.469 & [-1.053; & -0.079] &  & -0.468 & [-1.034; & -0.062] \\* 
               & $\omega_{32}$ & -1.539 & [-10.618; & 5.973] &  & -2.292 & [-11.448; & 3.306] &  & -2.238 & [-10.351; & 3.309] \\* 
               & $\omega_{33}$ & 0.728 & [0.366; & 1.248] &  & 0.725 & [0.365; & 1.285] &  & 0.726 & [0.309; & 1.230] \\* 
               & $\omega_{41}$ & 12.304 & [-14.146; & 40.045] &  & 13.343 & [-13.753; & 39.236] &  & 14.246 & [-11.271; & 43.967] \\* 
               & $\omega_{42}$ & -172.218 & [-775.830; & 353.575] &  & -186.886 & [-761.970; & 193.423] &  & -169.062 & [-746.004; & 174.997] \\* 
               & $\omega_{43}$ & -23.256 & [-46.723; & -4.300] &  & -22.979 & [-47.452; & -4.298] &  & -23.993 & [-47.123; & -4.752] \\* 
               & $\omega_{44}$ & 3320.605 & [1533.530; & 5997.970] &  & 3080.835 & [1480.750; & 5301.480] &  & 3117.705 & [1525.680; & 5563.090] \\ 
              0.25 & $\beta_{1}$ & 11.252 & [10.858; & 11.624] &  & 11.239 & [10.853; & 11.626] &  & 11.246 & [10.854; & 11.618] \\* 
               & $\beta_{2}$ & 0.319 & [0.282; & 0.360] &  & 0.316 & [0.281; & 0.355] &  & 0.317 & [0.277; & 0.357] \\* 
               & $\beta_{3}$ & 6.127 & [5.447; & 6.841] &  & 6.110 & [5.467; & 6.798] &  & 6.013 & [5.420; & 6.685] \\* 
               & $\beta_{4}$ & -0.003 & [-0.018; & 0.013] &  & -0.003 & [-0.020; & 0.011] &  & -0.005 & [-0.021; & 0.010] \\* 
               & $\beta_{5}$ & 0.004 & [-0.002; & 0.010] &  & 0.004 & [-0.002; & 0.009] &  & 0.004 & [-0.001; & 0.010] \\* 
               & $\sigma$ & 0.093 & [0.040; & 0.118] &  & 0.102 & [0.081; & 0.122] &  & 0.104 & [0.087; & 0.122] \\* 
               & $\gamma$ & 0.820 & [0.220; & 2.185] &  & 0.621 & [0.083; & 1.173] &  & 0.564 & [0.152; & 0.992] \\* 
               & $\alpha$ & 0.033 & [0.003; & 0.077] &  & 0.012 & [0.001; & 0.032] &  & 0.007 & [0.000; & 0.019] \\* 
               & $\omega_{11}$ & 1.123 & [0.551; & 2.033] &  & 1.119 & [0.549; & 2.000] &  & 1.114 & [0.518; & 2.051] \\* 
               & $\omega_{21}$ & 2.613 & [-8.442; & 18.120] &  & 2.822 & [-12.302; & 24.680] &  & 3.033 & [-4.894; & 15.046] \\* 
               & $\omega_{22}$ & 217.483 & [48.891; & 1076.700] &  & 268.924 & [46.422; & 2172.440] &  & 205.170 & [61.611; & 585.403] \\* 
               & $\omega_{31}$ & -0.468 & [-1.064; & -0.085] &  & -0.458 & [-1.106; & -0.031] &  & -0.462 & [-1.057; & -0.087] \\* 
               & $\omega_{32}$ & -1.166 & [-14.728; & 13.365] &  & -0.423 & [-14.681; & 43.440] &  & -1.910 & [-12.061; & 5.103] \\* 
               & $\omega_{33}$ & 0.759 & [0.365; & 1.380] &  & 0.803 & [0.325; & 1.831] &  & 0.741 & [0.358; & 1.271] \\* 
               & $\omega_{41}$ & 13.592 & [-14.227; & 45.509] &  & 14.227 & [-12.733; & 44.033] &  & 15.023 & [-10.319; & 44.864] \\* 
               & $\omega_{42}$ & -224.638 & [-1284.700; & 481.505] &  & -190.013 & [-1281.630; & 1275.230] &  & -222.940 & [-981.507; & 207.191] \\* 
               & $\omega_{43}$ & -24.804 & [-51.145; & -3.490] &  & -23.840 & [-50.514; & 3.075] &  & -24.753 & [-49.743; & -5.291] \\* 
               & $\omega_{44}$ & 3377.915 & [1463.660; & 6259.400] &  & 3254.420 & [1578.080; & 6085.090] &  & 3186.470 & [1465.790; & 5779.780] \\ 
              0.5 & $\beta_{1}$ & 11.579 & [11.190; & 11.982] &  & 11.595 & [11.232; & 11.990] &  & 11.597 & [11.201; & 11.966] \\* 
               & $\beta_{2}$ & 0.318 & [0.283; & 0.360] &  & 0.318 & [0.279; & 0.358] &  & 0.321 & [0.282; & 0.361] \\* 
               & $\beta_{3}$ & 6.392 & [5.773; & 7.025] &  & 6.365 & [5.746; & 7.027] &  & 6.381 & [5.762; & 7.009] \\* 
               & $\beta_{4}$ & -0.004 & [-0.019; & 0.010] &  & -0.006 & [-0.021; & 0.011] &  & -0.005 & [-0.021; & 0.011] \\* 
               & $\beta_{5}$ & 0.004 & [-0.001; & 0.010] &  & 0.004 & [-0.002; & 0.010] &  & 0.004 & [-0.002; & 0.010] \\* 
               & $\sigma$ & 0.099 & [0.029; & 0.126] &  & 0.105 & [0.040; & 0.135] &  & 0.108 & [0.054; & 0.136] \\* 
               & $\gamma$ & -0.270 & [-0.909; & 0.058] &  & -0.235 & [-0.870; & 0.082] &  & -0.229 & [-0.776; & 0.112] \\* 
               & $\alpha$ & 0.028 & [0.002; & 0.067] &  & 0.010 & [0.000; & 0.027] &  & 0.007 & [0.000; & 0.020] \\* 
               & $\omega_{11}$ & 1.115 & [0.532; & 2.033] &  & 1.112 & [0.524; & 2.038] &  & 1.126 & [0.545; & 2.088] \\* 
               & $\omega_{21}$ & 2.613 & [-5.481; & 16.121] &  & 2.951 & [-4.092; & 14.968] &  & 3.387 & [-3.877; & 16.112] \\* 
               & $\omega_{22}$ & 214.692 & [60.897; & 704.507] &  & 208.184 & [63.386; & 543.967] &  & 218.523 & [67.660; & 557.683] \\* 
               & $\omega_{31}$ & -0.459 & [-0.999; & -0.051] &  & -0.461 & [-1.033; & -0.099] &  & -0.464 & [-1.094; & -0.085] \\* 
               & $\omega_{32}$ & -1.330 & [-12.396; & 8.858] &  & -1.944 & [-11.300; & 5.678] &  & -2.139 & [-11.370; & 5.211] \\* 
               & $\omega_{33}$ & 0.753 & [0.354; & 1.321] &  & 0.738 & [0.360; & 1.290] &  & 0.752 & [0.376; & 1.298] \\* 
               & $\omega_{41}$ & 15.101 & [-10.463; & 46.551] &  & 14.460 & [-11.468; & 44.061] &  & 13.312 & [-13.967; & 42.605] \\* 
               & $\omega_{42}$ & -218.077 & [-1178.910; & 296.574] &  & -222.335 & [-945.746; & 222.452] &  & -259.053 & [-1025.470; & 200.366] \\* 
               & $\omega_{43}$ & -24.769 & [-51.602; & -4.059] &  & -23.909 & [-48.239; & -4.441] &  & -23.865 & [-47.171; & -4.612] \\* 
               & $\omega_{44}$ & 3278.055 & [1559.320; & 6139.130] &  & 3168.355 & [1513.260; & 5823.790] &  & 3250.930 & [1575.220; & 5918.410] \\ 
              0.75 & $\beta_{1}$ & 12.027 & [11.627; & 12.409] &  & 12.038 & [11.646; & 12.438] &  & 12.055 & [11.650; & 12.437] \\* 
               & $\beta_{2}$ & 0.323 & [0.283; & 0.367] &  & 0.324 & [0.282; & 0.365] &  & 0.323 & [0.281; & 0.369] \\* 
               & $\beta_{3}$ & 6.765 & [6.116; & 7.326] &  & 6.798 & [6.149; & 7.427] &  & 6.761 & [6.124; & 7.438] \\* 
               & $\beta_{4}$ & -0.007 & [-0.022; & 0.008] &  & -0.005 & [-0.022; & 0.011] &  & -0.008 & [-0.024; & 0.010] \\* 
               & $\beta_{5}$ & 0.004 & [-0.002; & 0.009] &  & 0.004 & [-0.003; & 0.010] &  & 0.004 & [-0.002; & 0.011] \\* 
               & $\sigma$ & 0.084 & [0.045; & 0.112] &  & 0.088 & [0.046; & 0.117] &  & 0.092 & [0.050; & 0.117] \\* 
               & $\gamma$ & -1.394 & [-2.055; & -0.855] &  & -1.364 & [-2.039; & -0.792] &  & -1.317 & [-1.994; & -0.800] \\* 
               & $\alpha$ & 0.020 & [0.002; & 0.047] &  & 0.010 & [0.001; & 0.026] &  & 0.007 & [0.000; & 0.020] \\* 
               & $\omega_{11}$ & 1.149 & [0.510; & 2.137] &  & 1.143 & [0.537; & 2.172] &  & 1.162 & [0.543; & 2.184] \\* 
               & $\omega_{21}$ & 2.884 & [-3.858; & 14.781] &  & 3.062 & [-3.923; & 15.313] &  & 3.281 & [-3.647; & 16.413] \\* 
               & $\omega_{22}$ & 184.734 & [49.599; & 473.156] &  & 185.568 & [58.234; & 480.389] &  & 187.017 & [55.066; & 501.216] \\* 
               & $\omega_{31}$ & -0.466 & [-1.091; & -0.060] &  & -0.477 & [-1.087; & -0.069] &  & -0.474 & [-1.100; & -0.046] \\* 
               & $\omega_{32}$ & -2.269 & [-11.070; & 4.253] &  & -2.366 & [-12.101; & 4.082] &  & -2.524 & [-11.676; & 4.609] \\* 
               & $\omega_{33}$ & 0.747 & [0.344; & 1.319] &  & 0.741 & [0.361; & 1.347] &  & 0.734 & [0.349; & 1.349] \\* 
               & $\omega_{41}$ & 15.473 & [-10.767; & 46.464] &  & 12.684 & [-13.921; & 40.890] &  & 13.184 & [-14.583; & 44.545] \\* 
               & $\omega_{42}$ & -186.032 & [-832.360; & 246.156] &  & -206.357 & [-836.341; & 209.837] &  & -214.059 & [-846.871; & 197.812] \\* 
               & $\omega_{43}$ & -22.684 & [-46.743; & -3.723] &  & -21.581 & [-43.592; & -0.669] &  & -21.579 & [-45.878; & -3.143] \\* 
               & $\omega_{44}$ & 3017.280 & [1361.650; & 5218.110] &  & 2966.075 & [1418.280; & 5190.710] &  & 3012.040 & [1415.220; & 5311.600] \\ 
              0.95 & $\beta_{1}$ & 12.630 & [12.213; & 13.022] &  & 12.653 & [12.200; & 13.025] &  & 12.634 & [12.226; & 13.035] \\* 
               & $\beta_{2}$ & 0.338 & [0.292; & 0.388] &  & 0.340 & [0.297; & 0.393] &  & 0.341 & [0.295; & 0.389] \\* 
               & $\beta_{3}$ & 7.293 & [6.581; & 7.911] &  & 7.409 & [6.802; & 8.061] &  & 7.364 & [6.744; & 8.000] \\* 
               & $\beta_{4}$ & -0.013 & [-0.033; & 0.004] &  & -0.009 & [-0.026; & 0.008] &  & -0.007 & [-0.025; & 0.011] \\* 
               & $\beta_{5}$ & 0.005 & [-0.002; & 0.012] &  & 0.004 & [-0.002; & 0.011] &  & 0.004 & [-0.003; & 0.010] \\* 
               & $\sigma$ & 0.059 & [0.041; & 0.074] &  & 0.059 & [0.041; & 0.075] &  & 0.059 & [0.035; & 0.074] \\* 
               & $\gamma$ & -4.619 & [-6.639; & -3.061] &  & -4.624 & [-6.868; & -3.221] &  & -4.620 & [-7.084; & -2.839] \\* 
               & $\alpha$ & 0.015 & [0.001; & 0.034] &  & 0.010 & [0.000; & 0.024] &  & 0.009 & [0.000; & 0.022] \\* 
               & $\omega_{11}$ & 1.063 & [0.544; & 1.860] &  & 1.021 & [0.478; & 1.764] &  & 0.996 & [0.499; & 1.776] \\* 
               & $\omega_{21}$ & 1.293 & [-4.357; & 8.460] &  & 1.373 & [-4.021; & 9.592] &  & 1.289 & [-3.891; & 8.452] \\* 
               & $\omega_{22}$ & 116.540 & [37.604; & 274.975] &  & 130.710 & [39.166; & 330.028] &  & 124.546 & [40.959; & 294.064] \\* 
               & $\omega_{31}$ & -0.406 & [-0.893; & -0.055] &  & -0.377 & [-0.869; & -0.035] &  & -0.373 & [-0.835; & -0.044] \\* 
               & $\omega_{32}$ & -1.738 & [-8.127; & 4.375] &  & -1.386 & [-9.119; & 4.411] &  & -1.578 & [-8.362; & 3.743] \\* 
               & $\omega_{33}$ & 0.650 & [0.322; & 1.172] &  & 0.653 & [0.301; & 1.180] &  & 0.638 & [0.325; & 1.147] \\* 
               & $\omega_{41}$ & 17.664 & [-0.889; & 44.938] &  & 13.540 & [-9.317; & 36.100] &  & 12.267 & [-8.363; & 36.518] \\* 
               & $\omega_{42}$ & -45.124 & [-391.948; & 230.798] &  & -125.369 & [-599.248; & 205.060] &  & -140.773 & [-565.293; & 166.391] \\* 
               & $\omega_{43}$ & -19.370 & [-39.120; & -2.639] &  & -18.086 & [-37.685; & -1.444] &  & -17.714 & [-36.401; & -1.210] \\* 
               & $\omega_{44}$ & 2413.385 & [1084.930; & 4178.740] &  & 2631.625 & [1303.870; & 4550.270] &  & 2638.600 & [1129.980; & 4393.670] \\ 
        
          \end{longtable}
        
        \end{ThreePartTable}
        
        \end{footnotesize}
        
        \end{landscape}

        \setlength{\tabcolsep}{6pt}
        
        \begin{ThreePartTable}
        
          \begin{TableNotes}[flushleft]
            \footnotesize
            \item[a] $p_0$ is the target quantile level.
            \item[b] $\alpha$ is the contamination level.
            \item[c] “True” denotes the true parameter value.
            \item[d] “Bias” is the mean difference between the estimated and true values.
            \item[e] “CP” is the coverage probability of the 95\% HPD intervals.
          \end{TableNotes}
        
          \begin{longtable}{
            S[table-format=1.3]
            S[table-format=1.3]
            l
            S[table-format=2.3]
            S[table-format=1.3]
            S[table-format=1.3]
          }
        \caption{
          Model performance characteristics for the cGAL model under various scenarios.
          Columns ``$p_0$'' and ``$\alpha$'' correspond to the target quantile and
          contamination level, respectively. The column ``True'' denotes the true
          parameter value, ``Bias'' is the mean difference between the estimated and
          true values, and ``CP'' represents the coverage probability of the 95\% HPD
          intervals.
        }
        \label{tab:CGAL_PERFORM} \\
        \toprule
        ${\bm {p_0}}$\tnote{a} & $\bm \alpha$\tnote{b} & {\bf Parameter} & {\bf True}\tnote{c} & {\bf Bias}\tnote{d} & {\bf CP}\tnote{e} \\
        \midrule
        \endfirsthead
            
        \caption*{\textit{(continued)}}\\
        \toprule
        ${\bm {p_0}}$\tnote{a} & $\bm \alpha$\tnote{b} & {\bf Parameter} & {\bf True}\tnote{c} & {\bf Bias}\tnote{d} & {\bf CP}\tnote{e} \\
        \midrule
        \endhead
            
        \midrule
        \multicolumn{6}{r}{\textit{Continued on next page}}\\
        \endfoot
            
        \bottomrule
        \insertTableNotes
        \endlastfoot
            
        0.50 & 0.001 & $\beta_{1}$ & 11.500 & 0.023 & 0.970 \\* 
           &  & $\beta_{2}$ & 5.500 & 0.311 & 0.980 \\* 
           &  & $\sigma$ & 0.200 & -0.024 & 0.940 \\* 
           &  & $\gamma$ & -0.300 & -0.060 & 0.973 \\* 
           &  & $\alpha$ & 0.001 & 0.016 & 0.980 \\* 
           &  & $\omega_{11}$ & 1.111 & 1.348 & 0.983 \\* 
           &  & $\omega_{12}$ & -0.056 & 0.001 & 1.000 \\* 
           &  & $\omega_{22}$ & 1.111 & 0.622 & 0.997 \\ 
          0.50 & 0.050 & $\beta_{1}$ & 11.500 & 0.005 & 0.967 \\* 
           &  & $\beta_{2}$ & 5.500 & 0.570 & 0.973 \\* 
           &  & $\sigma$ & 0.200 & -0.019 & 0.957 \\* 
           &  & $\gamma$ & -0.300 & -0.016 & 0.973 \\* 
           &  & $\alpha$ & 0.050 & 0.015 & 0.930 \\* 
           &  & $\omega_{11}$ & 1.111 & 1.247 & 0.977 \\* 
           &  & $\omega_{12}$ & -0.056 & 0.012 & 1.000 \\* 
           &  & $\omega_{22}$ & 1.111 & 0.529 & 0.993 \\ 
          0.85 & 0.001 & $\beta_{1}$ & 11.500 & -0.019 & 0.993 \\* 
           &  & $\beta_{2}$ & 5.500 & 0.614 & 0.997 \\* 
           &  & $\sigma$ & 0.200 & -0.001 & 0.987 \\* 
           &  & $\gamma$ & -0.300 & -0.232 & 0.963 \\* 
           &  & $\alpha$ & 0.001 & 0.030 & 0.977 \\* 
           &  & $\omega_{11}$ & 1.111 & 0.991 & 0.983 \\* 
           &  & $\omega_{12}$ & -0.056 & 0.001 & 1.000 \\* 
           &  & $\omega_{22}$ & 1.111 & 0.414 & 0.993 \\ 
          0.85 & 0.050 & $\beta_{1}$ & 11.500 & -0.015 & 0.970 \\* 
           &  & $\beta_{2}$ & 5.500 & 1.088 & 0.980 \\* 
           &  & $\sigma$ & 0.200 & 0.004 & 0.983 \\* 
           &  & $\gamma$ & -0.300 & -0.294 & 0.947 \\* 
           &  & $\alpha$ & 0.050 & 0.015 & 0.950 \\* 
           &  & $\omega_{11}$ & 1.111 & 1.000 & 0.983 \\* 
           &  & $\omega_{12}$ & -0.056 & 0.003 & 1.000 \\* 
           &  & $\omega_{22}$ & 1.111 & 0.272 & 0.993 \\ 
        
          \end{longtable}
        
        \end{ThreePartTable}

        \begin{ThreePartTable}
        
          \begin{TableNotes}[flushleft]
            \footnotesize
            \item[a] $p_0$ is the target quantile level.
            \item[b] $\alpha$ is the contamination level.
            \item[c] GAL (generalized asymmetric Laplace), cGAL (contaminated GAL).
            \item[d] “True” denotes the true parameter value.
            \item[e] “Bias” is the mean difference between the estimated and true values.
            \item[f] “RMSE” is the root mean square error.
            \item[g] “CP” is the coverage probability of the 95\% HPD intervals.
            \item[h] “HPD Len.” is the mean width of the HPD interval.
          \end{TableNotes}
        
          \setlength{\tabcolsep}{6pt}
          \begin{longtable}{
            S[table-format=1.3]
            S[table-format=1.4]
            l
            l
            S[table-format=2.3]
            S[table-format=1.3]
            S[table-format=1.3]
            S[table-format=1.3]
            S[table-format=2.3]
          }
            \caption{
              Model performance characteristics for GAL and cGAL fixed effects parameters under various scenarios. Columns ``$p_0$'' and ``$\alpha$'' correspond to the target quantile and contamination level, respectively. The column ``True'' denotes the true parameter value, ``Bias'' is the mean difference between the estimated and true values, ``RMSE'' is the root mean square error, ``CP'' is the coverage probability of the 95\% HPD intervals, and ``HPD Len.'' is the mean width of the HPD interval.
            }
            \label{tab:CONTAMINATION}\\
        
            \toprule
            ${\bm {p_0}}$\tnote{a} & $\bm \alpha$\tnote{b} & {\bf Model}\tnote{c} & {\bf Parameter} & {\bf True}\tnote{d} & {\bf Bias}\tnote{e} & {\bf RMSE}\tnote{f} & {\bf CP}\tnote{g} & {\bf HPD Len.}\tnote{h} \\
            \midrule
            \endfirsthead
            
            \caption*{\textit{(continued)}}\\
            \toprule
            ${\bm {p_0}}$\tnote{a} & $\bm \alpha$\tnote{b} & {\bf Model}\tnote{c} & {\bf Parameter} & {\bf True}\tnote{d} & {\bf Bias}\tnote{e} & {\bf RMSE}\tnote{f} & {\bf CP}\tnote{g} & {\bf HPD Len.}\tnote{h} \\
            \midrule
            \endhead
            
            \midrule
            \multicolumn{9}{r}{\textit{Continued on next page}} \\
            \endfoot
            
            \bottomrule
            \insertTableNotes
            \endlastfoot
        
            0.50 & 0.001 & GAL & $\beta_{1}$ & 11.500 & 0.031 & 0.420 & 0.967 & 1.822 \\* 
               &  &  & $\beta_{2}$ & 5.500 & 0.319 & 0.857 & 0.970 & 6.238 \\ 
              0.50 & 0.001 & cGAL & $\beta_{1}$ & 11.500 & 0.023 & 0.416 & 0.970 & 1.813 \\* 
               &  &  & $\beta_{2}$ & 5.500 & 0.311 & 0.846 & 0.980 & 6.058 \\ 
              0.50 & 0.050 & GAL & $\beta_{1}$ & 11.500 & 0.072 & 0.483 & 0.973 & 2.159 \\* 
               &  &  & $\beta_{2}$ & 5.500 & 1.141 & 3.441 & 0.987 & 14.892 \\ 
              0.50 & 0.050 & cGAL & $\beta_{1}$ & 11.500 & 0.005 & 0.459 & 0.967 & 1.939 \\* 
               &  &  & $\beta_{2}$ & 5.500 & 0.570 & 2.068 & 0.973 & 7.811 \\ 
              0.85 & 0.001 & GAL & $\beta_{1}$ & 11.500 & 0.004 & 0.494 & 0.993 & 2.413 \\* 
               &  &  & $\beta_{2}$ & 5.500 & 0.604 & 2.047 & 0.997 & 7.065 \\ 
              0.85 & 0.001 & cGAL & $\beta_{1}$ & 11.500 & -0.019 & 0.498 & 0.993 & 2.411 \\* 
               &  &  & $\beta_{2}$ & 5.500 & 0.614 & 2.068 & 0.997 & 7.108 \\ 
              0.85 & 0.050 & GAL & $\beta_{1}$ & 11.500 & 0.219 & 0.605 & 0.963 & 2.745 \\* 
               &  &  & $\beta_{2}$ & 5.500 & 1.252 & 3.552 & 0.977 & 11.498 \\ 
              0.85 & 0.050 & cGAL & $\beta_{1}$ & 11.500 & -0.015 & 0.522 & 0.970 & 2.486 \\* 
               &  &  & $\beta_{2}$ & 5.500 & 1.088 & 3.179 & 0.980 & 9.722 \\ 
        
          \end{longtable}
        
        \end{ThreePartTable}

    \end{appendices}
    
\end{document}